\documentclass[preprint,11pt]{aastex}
\newcommand{\ee}[1]{\mbox{${} \times 10^{#1}$}}
\newcommand{\eten}[1]{\mbox{$10^{#1}$}}

\newcommand{\kms}{\mbox{km s$^{-1}$}}
\newcommand\cmv{\mbox{cm$^{-3}$}}

\newcommand{\sfr }{\mbox{$\dot M_{\star}$}}
\newcommand{\tff }{\mbox{$t_{ff}$}}

\newcommand{\lsun}{\mbox{L$_\odot$}}
\newcommand{\msun}{\mbox{M$_\odot$}}
\newcommand{\ta}{{$T_A^*$}}

\newcommand{\tr}{\mbox{$T_R$}}

\newcommand{\lir}{\mbox{$L_{IR}$}} 
\newcommand{\lfir}{\mbox{$L_{IR}$}} 
\newcommand{\lmol}{\mbox{$L'_{mol}$}} 
\newcommand{\tbol}{\mbox{$T_{bol}$}} 

\newcommand{\nbar}{\mbox{$\overline{n}$}}
\newcommand{\nlvg}{\mbox{$n_{LVG}$}}
\newcommand{\mvir}{\mbox{$M_{Vir}$}} 
\newcommand{\mean}[1]{\mbox{$\langle#1\rangle$}} 

\newcommand{\hh}{\mbox{{\rm H}$_2$}}
\newcommand{\form}{H$_2$CO}

\newcommand{\ammonia}{\mbox{{\rm NH}$_3$}}

\newcommand{\cooo}{C$^{18}$O}

\newcommand{\hcop}{HCO$^+$}

\newcommand{\cs}{CS}

\input{epsf}

\newcommand{\jj}[2]{\mbox{$J = #1\rightarrow#2$}}
\newcommand{\jkkjkk}[6]{\mbox{$J_{K_{-1}K_{+1}}
                        = #1_{#2#3}\rightarrow#4_{#5#6}$}}

\shorttitle{Massive Dense Clumps}
\shortauthors{Wu \& Evans }
\begin{document}
\title {\bf The Properties of Massive, Dense, Clumps:
Mapping Surveys of HCN and CS }
\author {Jingwen Wu}
\affil{Harvard-Smithsonian Center for Astrophysics, 60 Garden st.,MS78,
 Cambridge, MA, 02138}
\email{jwu@cfa.harvard.edu}

\author {Neal J. Evans II}
\affil{Department of Astronomy, The University of Texas at Austin,
  1 University Station, C1400, Austin, Texas 78712}
\email{nje@astro.as.utexas.edu}

\author{Yancy L. Shirley}
\affil{Steward Observatory, University of Arizona, Tucson, AZ 85721.}
\email{yshirley@as.arizona.edu}

\author {Claudia Knez}
\affil{Department of Astronomy, University of Maryland, College Park, MD 20742}


\begin{abstract}
We have mapped over 50 massive, dense clumps with four dense gas tracers:
HCN $J=1-0$ and $3-2$; and CS $J=2-1$ and $7-6$ transitions.
Spectral lines of
optically thin H$^{13}$CN 3-2 and C$^{34}$S 5-4 were also obtained towards the
map centers. These maps usually demonstrate single well-peaked distributions
at our resolution, even with higher J transitions.
The size, virial mass, surface density, and mean volume density
within a well-defined
angular size (FWHM) were calculated from the contour maps for each transition.
We found that transitions with higher effective density usually trace the more
compact, inner part of the clumps but have larger linewidths, leading to an
inverse linewidth-size relation using different tracers. The mean surface
densities are 0.29, 0.33, 0.78, 1.09 g cm$^{-2}$ within FWHM contours of CS
2-1, HCN 1-0, HCN 3-2 and CS 7-6, respectively. We find no correlation of
\lir\ with surface density and a possible inverse correlation with
mean volume density, contrary to some theoretical expectations.
Molecular line luminosities
$L'_{mol}$ were derived for each transition.
We see no evidence in the data for the relation between \lmol\ and
mean density posited by modelers.
 The correlation between $L'_{mol}$
and the virial mass is roughly linear for each dense gas tracer.
No obvious correlation was found between the line luminosity ratio and
infrared luminosity, bolometric temperature, or the $L_{IR}/M_{Vir}$ ratio.
A nearly linear correlation was found between the infrared luminosity and the
line luminosity of all dense gas tracers for these massive, dense clumps, with a
lower cutoff in luminosity at $\lir=\eten{4.5}$ \lsun. The
$L_{IR}$-$L'_{HCN1-0}$ correlation agrees well with the one found in galaxies.
These correlations indicate a constant star formation rate per unit mass from
the scale of dense clumps to that of distant galaxies when the mass is
measured for dense gas. These results support the suggestion that
starburst galaxies may be understood as having a large fraction of gas
in dense clumps.

\end{abstract}

\section{Introduction}

A full understanding of how stars form, both in the Milky Way and in
other galaxies, requires an improved picture of the formation of massive
stars.  Most of the stars in our Galaxy form in regions of clustered
star formation, and nearly all massive stars appear to form in dense
clusters (Elmegreen 1985, Carpenter 2000, Lada \& Lada 2003). Massive forming
stars provide the only obvious signposts of star formation in distant regions
of our Galaxy and certainly in other galaxies. The star formation prescription
most widely used in simulations of galaxy formation and evolution, the
Kennicutt-Schmidt law (Schmidt 1959, Kennicutt 1998) is based on
tracers like H$\alpha$ (e.g., Calzetti 2008),
which trace the formation only of relatively massive stars. The signs of
formation of lower mass stars are mostly invisible at the distances of
other galaxies. For all these reasons, a better understanding of the
formation of massive stars is important.

However, our knowledge of the massive star formation process is still
limited, in both observational and theoretical aspects, in contrast to
the progress made in the study of low-mass star formation.
Observational progress has been limited by distance, complexity, and the
rapid evolution of young massive stars.  Most high-mass star-forming regions
are very distant from us, normally at least an order of magnitude farther
than the nearby low-mass star forming regions. When they are close enough
to study in detail, such as the Orion molecular cloud, we find an
extremely complex
environment marked by high turbulence, high density and opacity (e.g.,
O$'$Dell 2005), and multiple large-scale motions, including possible
explosive events. The fast evolution of the new-born massive stars leads
to a harsh radiative environment. Together with the jets, winds, and outflows,
the radiation field rapidly destroys the evidence that we need.

The confusing observational picture allows quite a variety of theoretical
models to exist. Recently, the models have tended to fall into one of two
camps.
Bonnell et al. (1997, 1998) and Bonnell \& Bate (2005)
have begun with a clump containing many Jeans masses, which fragments rapidly
into small stars that compete to accrete the remaining gas. The other camp,
represented by McKee \& Tan (2002, 2003) and Tan, Krumholz \& McKee (2006),
argue
that such a clump will be composed of a spectrum of turbulent cores,
essentially scaled-up versions of the well-studied cores that form low-mass
stars. The higher mass stars then form because turbulent pressure produces
a much higher accretion rate. For an interesting discussion of these
different approaches, see Krumholz \& Bonnell (2007). Recent simulations
including radiative feedback from forming protostars found that the extreme
fragmentation found in earlier, isothermal simulations was greatly suppressed,
allowing more massive stars to form (Krumholz et al. 2007, Urban et al. 2010,
Krumholz et al. 2010).

In this situation, one positive step to study massive star formation is to make
statistical studies based on systematic surveys towards a large and
well-characterized sample of massive star-forming regions. Using suitable
tracers, which can reveal the physical and dynamical properties of the regions,
we can provide the basic constraints that successful theories must satisfy.

Star formation is not uniformly distributed over molecular clouds, but
is restricted to regions within molecular clouds where the density
exceeds some threshold, roughly $n > \eten4$ or $\eten5$ cm$^{-3}$
(e.g., Evans, 2008). There have been many detailed studies of
individual regions of massive star formation and surveys of likely
regions in tracers of dense gas or dust (e.g., Plume et al. 1992, 1997;
Zinchenko, Mattila, \& Toriseva, 1995; Anglada et al. 1996;
Sridharan et al. 2002; Beuther et al. 2002; Wu \& Evans 2003).
These surveys have mostly used signposts of
massive star formation, such as masers, HII regions, or infrared sources with
particular colors. Less biased surveys, using infrared dark clouds or
blind surveys for millimeter wavelength dust continuum, are beginning to
become available (e.g., Price et al. 2001, Williams et al. 2007, Schuller et
al. 2009, Aguirre et al. 2009, Rosolowsky et al. 2009).
These will demand follow-up studies, and the question
arises of what can be learned from different tracers.

This work will focus on  a large sample of massive star forming regions
that are in an early phase of star formation, as indicated originally
by the association with water masers. We have in the past referred to these
regions as ``cores", but McKee and co-workers prefer to restrict the
use of ``core" to a smaller region that will form a single or small star
group, and they refer to regions that will form clusters as ``clumps."
The expectation is that star-forming clumps will contain multiple cores.
We adopt the McKee terminology here and refer to these objects as clumps.
We emphasize however, that their properties are very different from the
objects usually going by this name. As described by Williams et al. (2000),
clumps are traced by extinction or \cooo\ peaks and have modest mean
densities of $n \sim \eten3$. Most may not be gravitationally bound.
 The clumps we are  studying
here are very dense ($\mean{n} \sim 10^{6}$, e.g. Plume et al. 1997),
on average 100 times denser than the ``cores" in low-mass star forming
regions (e.g. Mueller et al. 2002).  To keep this distinction in mind,
we refer to them as ``dense clumps." Since most also have high masses,
we often refer to them as ``massive dense clumps."

The sample has been well studied by multiple CS line surveys
(Plume et al. 1992, 1997). A subsample of these sources has been
mapped in the CS 5-4 transition (Shirley et al. 2003) and in 350
$\mu$m dust continuum emission (Mueller et al. 2002).
Based on the dust continuum data, a 1D model was constructed to
determine the mass, bolometric luminosity, and density distribution of those
massive dense clumps; a power-law density distribution
($n(r) \propto r^{-p}$ with $\mean{p} = 1.8$) was found to fit
the data well (Mueller et al. 2002).
The CS 5-4 maps have provided further constraints on the
size and virial mass of the dense clumps (Shirley et al. 2003).

While the maps of CS 5-4 emission and dust continuum emission provided
valuable statistical data on sizes and masses, systematic studies of other
commonly used tracers can provide a valuable comparison set for surveys
in these tracers, both in distant regions of our Galaxy and in other galaxies.
In particular, HCN has been used much more widely in surveys of starburst
galaxies, and we need a set of fiducial objects in order to understand
 those surveys.

Additional molecular lines that can trace
dense clumps, as well as their maps, are needed for this purpose. We selected
the CS 7-6, CS 2-1, HCN 3-2, and HCN 1-0 transitions for further mapping.
CS 7-6 and CS 2-1 maps can constrain models of the clumps that have been
studied with
CS 5-4 at different excitation levels. HCN 3-2 is a
possible indicator of infall in massive star forming regions (Wu \&
Evans 2003), and it may be able to constrain mass inflow rates needed in
theories of massive star formation.

Recent work (Gao \& Solomon 2004a,b) shows that the luminosity of HCN 1-0
has a tight and linear correlation with infrared luminosity, for both nearby
normal galaxies and distant starburst galaxies,
suggesting that star formation efficiency may be constant in these different
systems when measured against tracers of the dense gas. We can directly test
this
$L'_{HCN}$-L$_{IR}$ correlation in Galactic dense clumps, by mapping HCN 1-0
towards massive dense clumps. Our results, based on the data in this survey,
did indeed suggest that the ratio of massive star formation to the massive
dense gas is similar in Galactic dense clumps as in starburst galaxies
(Wu et al. 2005).  That result suggests that  HCN can be used to connect star
formation on scales ranging from individual dense clumps in our Galaxy
 to distant galaxies. Then we may be
able to apply our knowledge of Galactic massive star formation obtained
from CS and HCN studies to star formation in other galaxies.

For all these purposes, and for building a complete dataset to
constrain models of individual massive dense clumps, we have carried out
mapping surveys using HCN 3-2, 1-0, CS 7-6 and 2-1. In this paper we report
the basic results of the survey.

\section{Observations}

\subsection{The Sample}

The sample in the survey is a subset of the larger sample of dense clumps
associated with H$_{2}$O masers that have been surveyed with several CS
transitions (Plume et al. 1992, 1997), most of which have been mapped with
CS 5-4 and dust emission (Shirley et al. 2003, Mueller et al. 2002).
Table \ref{sourcelist} lists the information on sources that have been mapped in this survey.
The sources mapped in CS 5-4 (Shirley et al. 2003) have virial masses within
the nominal core radius (R$_{CS}$) ranging from 30 M$_{\odot}$ to 2750
M$_{\odot}$,
with a mean of 920 M$_{\odot}$. The sources in this category have infrared
luminosities ranging from 10$^{3}$ - 10$^{7}$ L$_{\odot}$
and most contain compact or ultracompact  H $\amalg$ (UCH$\amalg$) regions.
These dense clumps are mostly quite massive and have been well studied by
several tracers; we
will focus on these sources for statistical studies of the properties of
massive dense clumps.
To extend the sample towards lower luminosities for the study of
$L'_{HCN}$-L$_{IR}$ correlation, we selected 14 IRAS sources from outflow
surveys (Zhang et al. 2005; Wu et al. 2004) and a few lower mass clumps from
other publications.  These added sources are listed in part 2 of
Table \ref{sourcelist}, separated by a line from the well studied massive
clumps in Table \ref{sourcelist}.
These clumps have various luminosities; in this paper
they are not included in our statistics of the properties of massive
dense clumps.

We mapped massive clumps in the HCN 3-2, HCN 1-0, CS 7-6, and CS 2-1
transitions. Because of observing constraints, maps of a particular transition
are not available for all sources in our sample,
but we have attempted to map as many sources as possible in all the tracers.
In addition to mapping the clumps in the main transitions, we also observed
the optically thin isotopes of HCN and CS (H$^{13}$CN 3-2 and C$^{34}$S 5-4)
at the center position of most of these massive clumps to constrain
models and to study infall in massive clumps.  Observations were taken
between 1996 to 2007, with information presented in Table \ref{information}.

\subsection{Observations with the FCRAO}

Maps of HCN 1-0 and CS 2-1 were made with the
14-m telescope of the Five College Radio Astronomy Observatory (FCRAO).
The 16-element focal plane array (SEQUOIA) was used, with typical system
temperatures 100-200K. A velocity resolution of 0.1 \kms\ was achieved with
the 25 MHz bandwidth on the dual channel correlator (DCC).
We converted the measured
\ta\ to \tr\ via $\tr = T_{A}^{*}/(\eta_{FSS}\eta_c)$,
with
$\eta_{FSS}=0.7$. The value of $\eta_{c}$ depends on source size; for
the typical map in this study ($\sim 10\arcmin$), $\eta_{C}=0.7$.
The map size was extended until the edge of the HCN \jj10\ and CS \jj21\
 emission was reached, typically at the $2\sigma$ level
(mean $\sigma$ $\sim$ 0.3 K \kms), so we could obtain the total line luminosity.

\subsection{Observations with the CSO}

All the other observations were made at the 10.4-m telescope of the
Caltech Submillimeter Observatory\footnote[1]
{The CSO is operated by the California Institute of
Technology under funding from the National Science Foundation, contract
AST 90-15755.} (CSO).
The observing date, line frequency, beam size, and main beam efficiency of
each line are listed in Table 2.  The rest frequencies of
HCN $3-2$, H$^{13}$CN $3-2$, and C$^{34}$S  5-4 lines have been updated
(Ahrens et al. 2002, Gottlieb et al. 2003)
after some of our observations. We have corrected our data to the new
frequencies listed in Table 2. The position switching mode was used;
reference positions were checked when necessary. This is important for tracers
that
may show absorption features in the spectra (for example, HCN 3-2). We
checked all the reference positions for HCN 3-2 emission and found them to be
free from emission ($T_A^* < 0.5$ K).
Pointing was checked periodically using planets and CO-bright stars.
The pointing accuracy was usually better than 6\arcsec\ for all the runs.
Map sizes of HCN 3-2 and CS 7-6 were extended
to at least the point where the emission was one third of the peak strength,
but usually maps extend to as low as one
tenth of the peak strength (typically the $3\sigma$ level, with the
mean $\sigma$ $\sim$ 1 K \kms). The beam sizes and efficiencies were
fairly constant for all transitions except for the CS 7-6, for which
various different receiver optics were in place at different times.
All data were scaled by the efficiency for the run on which they were
taken to obtain the values of $T_R^*$ given in tables and used in the
analysis.

\section{Observational Results}

Maps were obtained for 56 clumps with HCN 1-0, 43 clumps with HCN 3-2,
56 clumps with CS 2-1, and 51 clumps with CS 7-6.
The spectral line information at the central (0,0) position of each map
is listed in Table \ref{cstable} to Table \ref{hcn10table}.

\subsection{CS Results}

Table \ref{cstable} presents the peak temperature of the line
(T$_{R}^{*}$), integrated intensity ($\int$T$_{R}^{*}$d$\nu$), LSR velocity,
and FWHM linewidth of CS 2-1, CS 7-6, and C$^{34}$S 5-4 transitions for massive
dense clumps. Central position spectra of CS 2-1 and CS 7-6 have also been
observed by Plume et al. (1992, 1997), who obtained CS 2-1 lines with the
IRAM 30-m telescope at Pico Veleta, and CS 7-6 lines from the CSO. Comparing to
the results in Plume et al. (1992, 1997), the observed parameters of CS 7-6 are
consistent for the two CSO observations. The mean ratio of FWHM linewidth
in the current observations to those from Plume's earlier data
is 1.1$\pm$0.4, with median ratio of 1.1. The mean ratio for
$\int$T$_{R}^{*}$d$\nu$ is 1.1$\pm$0.7, with a median ratio of 1.0.
For the results of CS 2-1 observations, the IRAM data give a
stronger T$_{R}^{*}$ and larger $\int$T$_{R}^{*}$d$\nu$; the ratio
of $\int$T$_{R}^{*}$d$\nu$ for the IRAM
data to current FCRAO data is 3.3$\pm$1.8.
This difference comes from the different
telescope sizes and beam sizes of the IRAM and the FCRAO.
But the linewidth from the
two observations are close: the FWHM ratio of IRAM CS 2-1 data to FCRAO data is
1.1$\pm$0.2, and the median is 1.1.

\subsection{HCN Results}

Table \ref{hcntable} presents the line information for HCN 3-2 and its isotopologue H$^{13}$CN
3-2.  The mean integrated intensity ratio of HCN 3-2 to H$^{13}$CN 3-2 is
9.7$\pm$6.8.
This ratio is smaller than the usual $^{12}C/^{13}C$ isotopic abundance ratio, partly due to the large
fraction of self-absorption
features presented in HCN 3-2 lines. About half of the sources in the massive
core sample show double peaks. The description of line profiles is
in Table \ref{hcntable}.

Table \ref{hcn10table} lists the observational results for HCN 1-0. HCN
1-0 has three hyperfine components (F=1-0, F=2-1, F=1-1), separated
by $-7.064$ \kms\ (F=1-0 to F=2-1) and 4.842 \kms (F=1-1 to F=2-1).
We tried to fit
the spectra with 3 gaussian components with the same line width. The
peak temperatures of every component, the LSR velocity of
the main component (F=2-1), the linewidth, and the integrated intensity
are presented in Table \ref{hcn10table}. Massive clumps usually have large
linewidths; sometimes the line width is wider than the separation of hyperfine
components, so that two or three of these components will overlap.
Sometimes one or two hyperfine components are optically thick
enough to show absorption.  We have noted these features in Table \ref{hcn10table}.
If the HCN 1-0 line is optically thin and the hyperfine levels are populated
according to LTE, the line ratios between the three hyperfine components are
I(F=1-0):I(F=2-1):I(F=1-1)=1:5:3.
The observed ratios for most of our clumps (Table \ref{hcn10table}) do not
agree with these
optically thin ratios, as expected for lines that are almost surely optically
thick. This hyperfine anomaly has also been reported in other massive star
forming regions (e.g., Pirogov et al. 1996, Pirogov 1999).

\subsection{Morphology and Multiplicity}

\begin{figure}[hbt!]
\epsscale{0.90}
\plotone{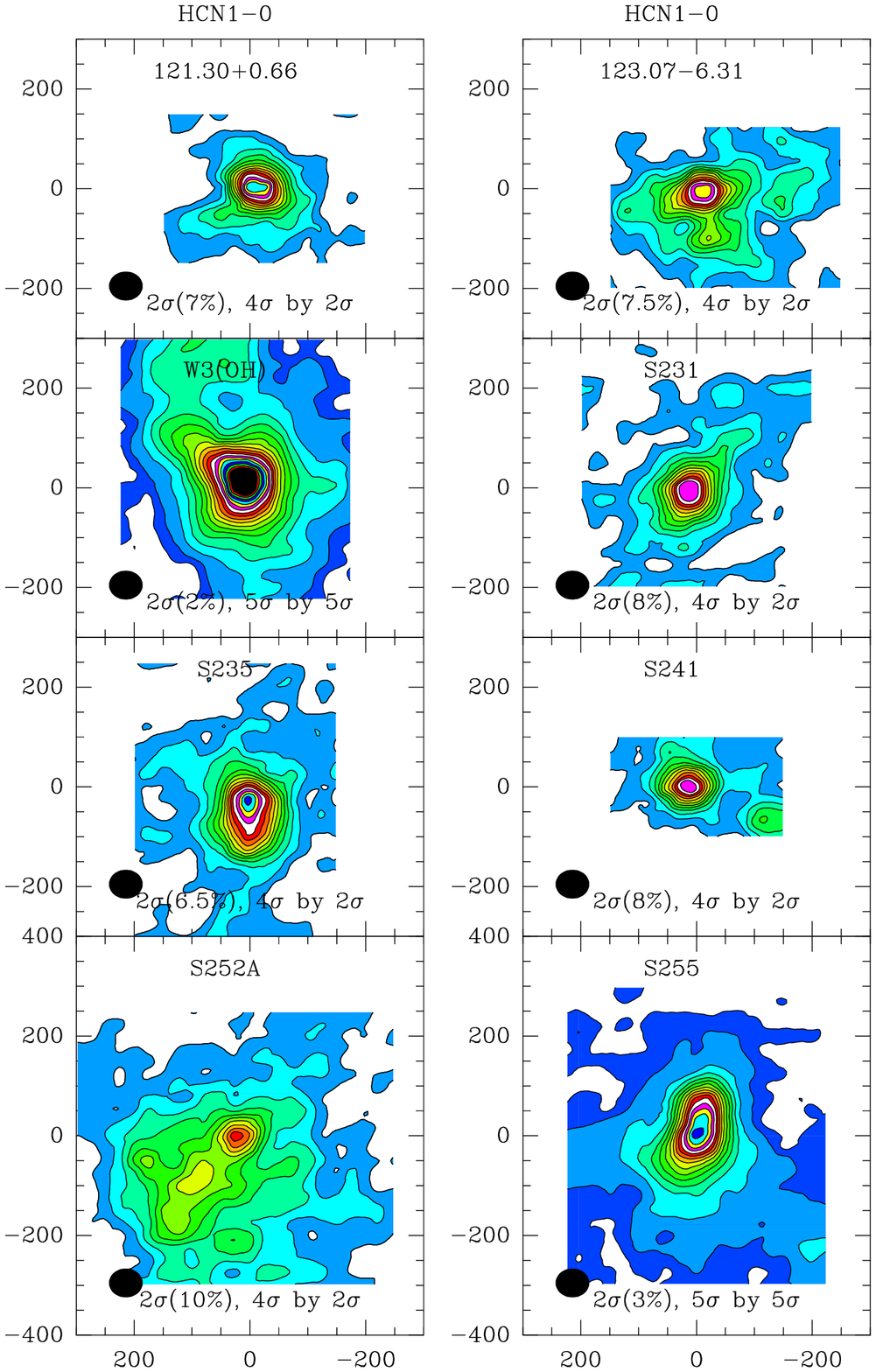}
\caption{\label{hcn1}HCN 1-0 contour maps of massive clumps. The lowest contour
level and
increasing step of contours are indicated in the plot. The beam size is
shown at the lower left of each map. }
!\label{f1}
\end{figure}

\begin{figure}[hbt!]
\epsscale{0.90}
\plotone{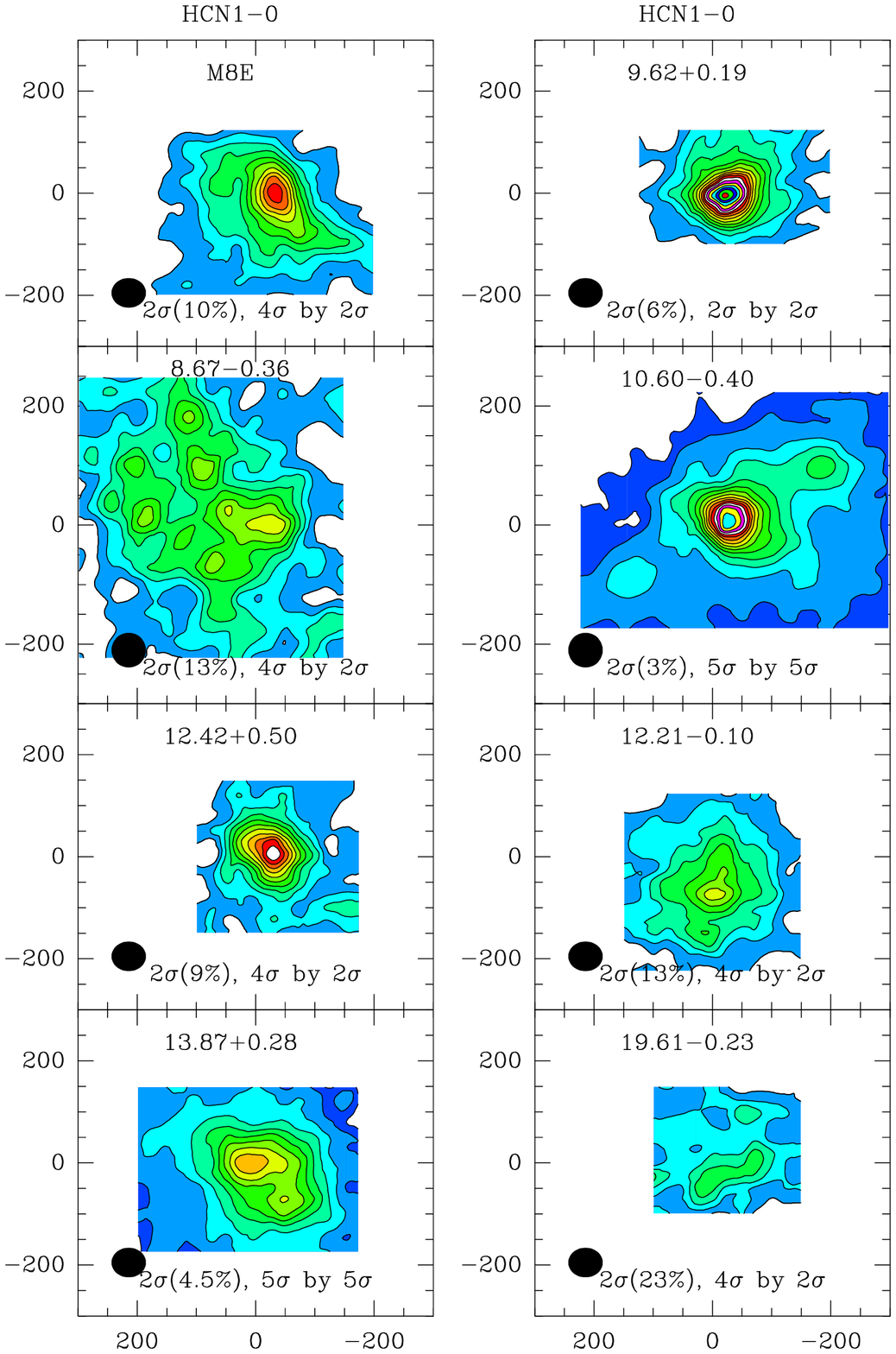}
\caption{\label{hcn2}HCN 1-0 contour maps continue... }
\end{figure}

\begin{figure}[hbt!]
\epsscale{0.90}
\plotone{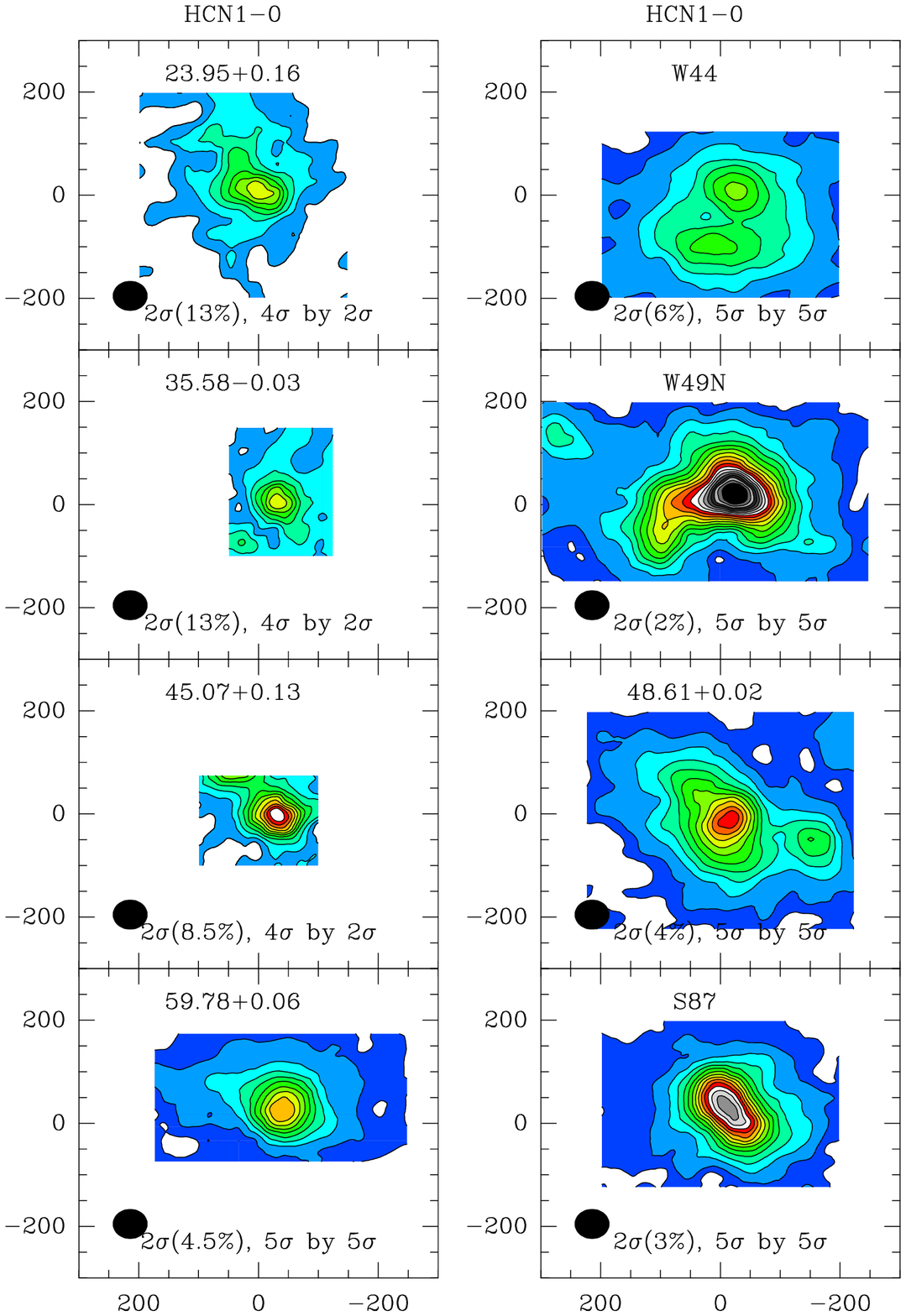}
\caption{\label{hcn3}HCN 1-0 contour maps continue... }
\end{figure}

\begin{figure}[hbt!]
\epsscale{0.90}
\plotone{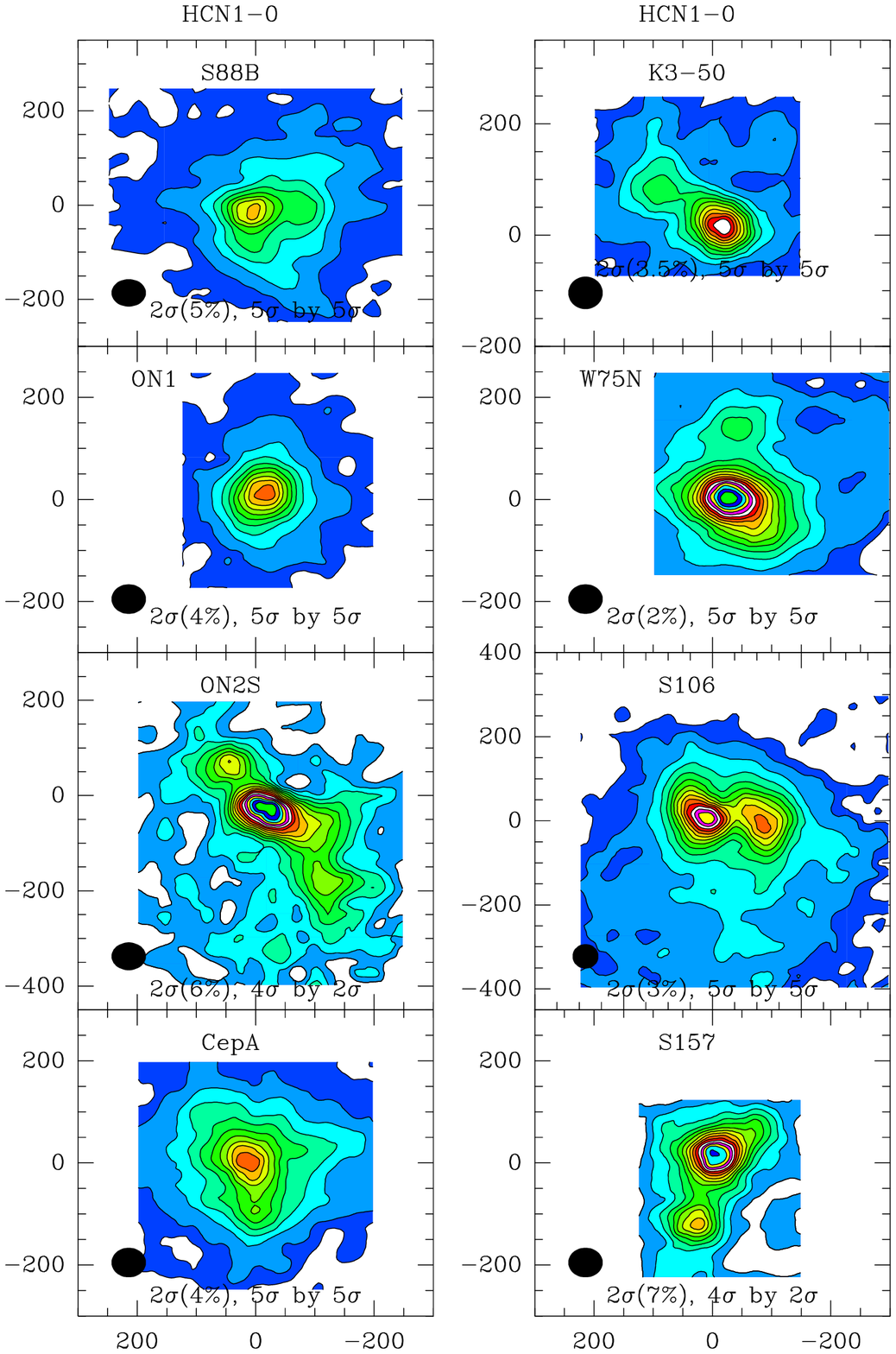}
\caption{\label{hcn4}HCN 1-0 contour maps continue... }
\end{figure}

\begin{figure}[hbt!]
\epsscale{0.90}
\plotone{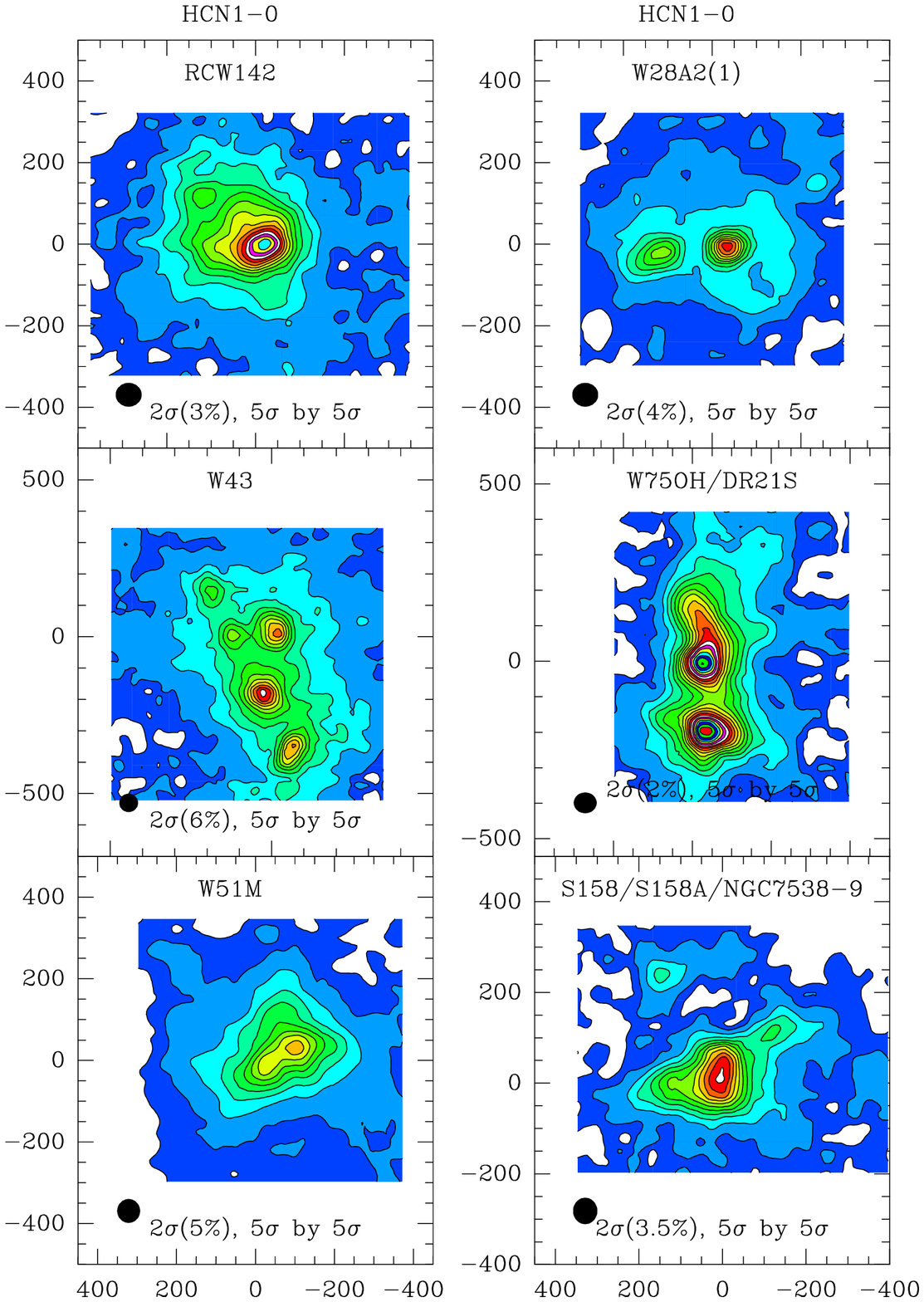}
\caption{\label{hcn5}HCN 1-0 contour maps continue... }
\end{figure}

\begin{figure}[hbt!]
\epsscale{0.90}
\plotone{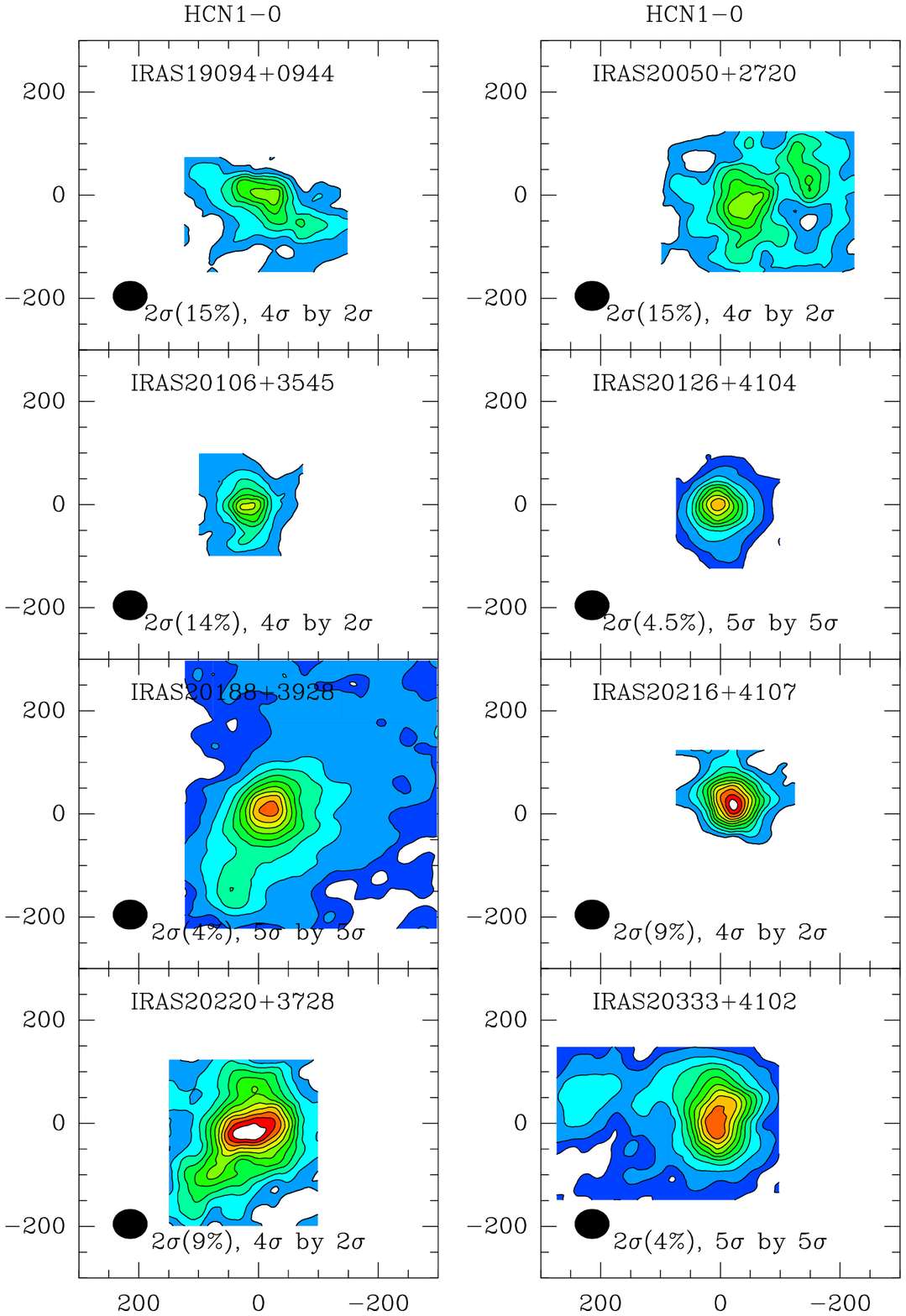}
\caption{\label{hcn6}HCN 1-0 contour maps continue... }
\end{figure}

\begin{figure}[hbt!]
\epsscale{0.90}
\plotone{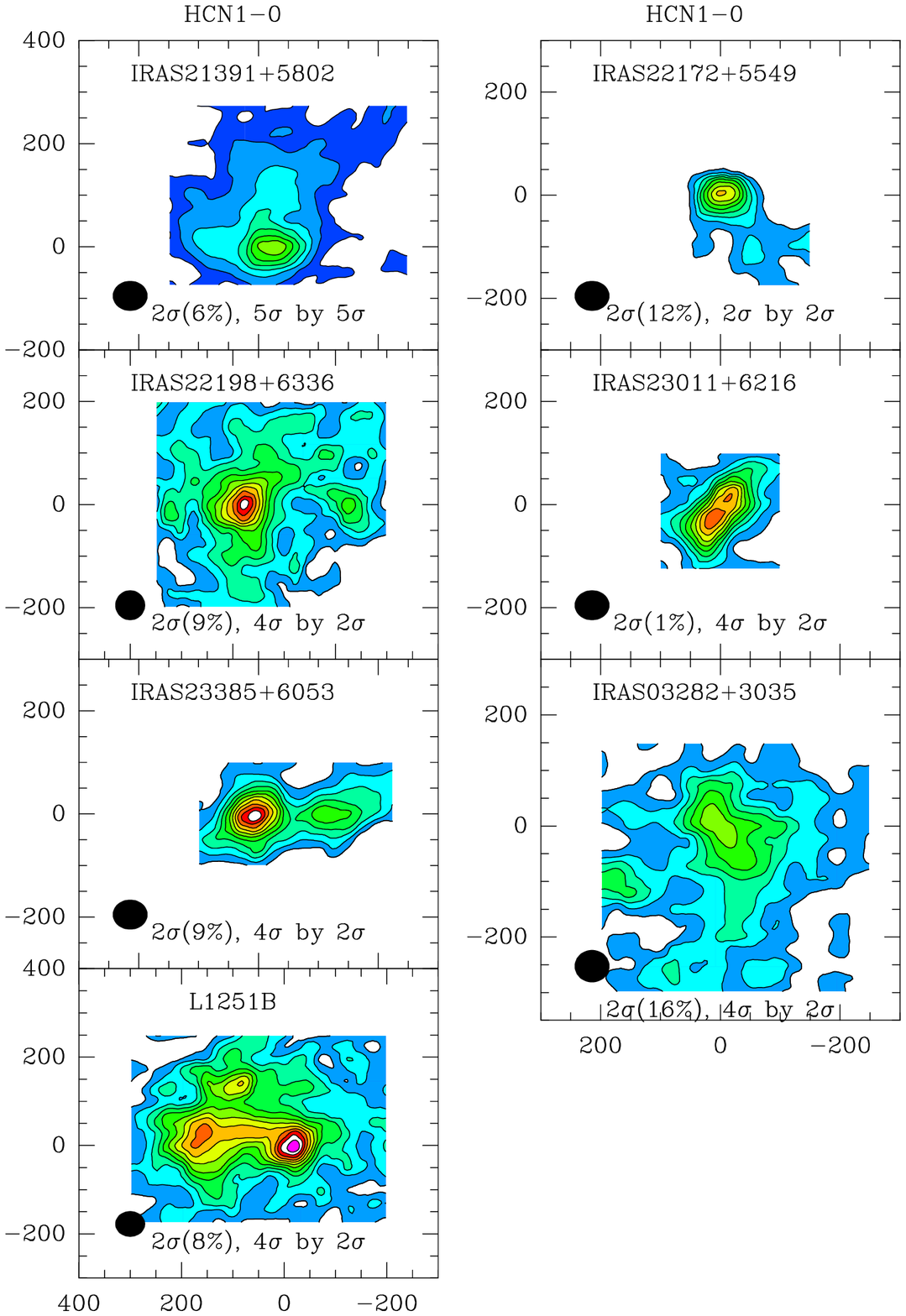}
\caption{\label{hcn7}HCN 1-0 contour maps continue... }
\end{figure}

\begin{figure}[hbt!]
\epsscale{0.90}
\plotone{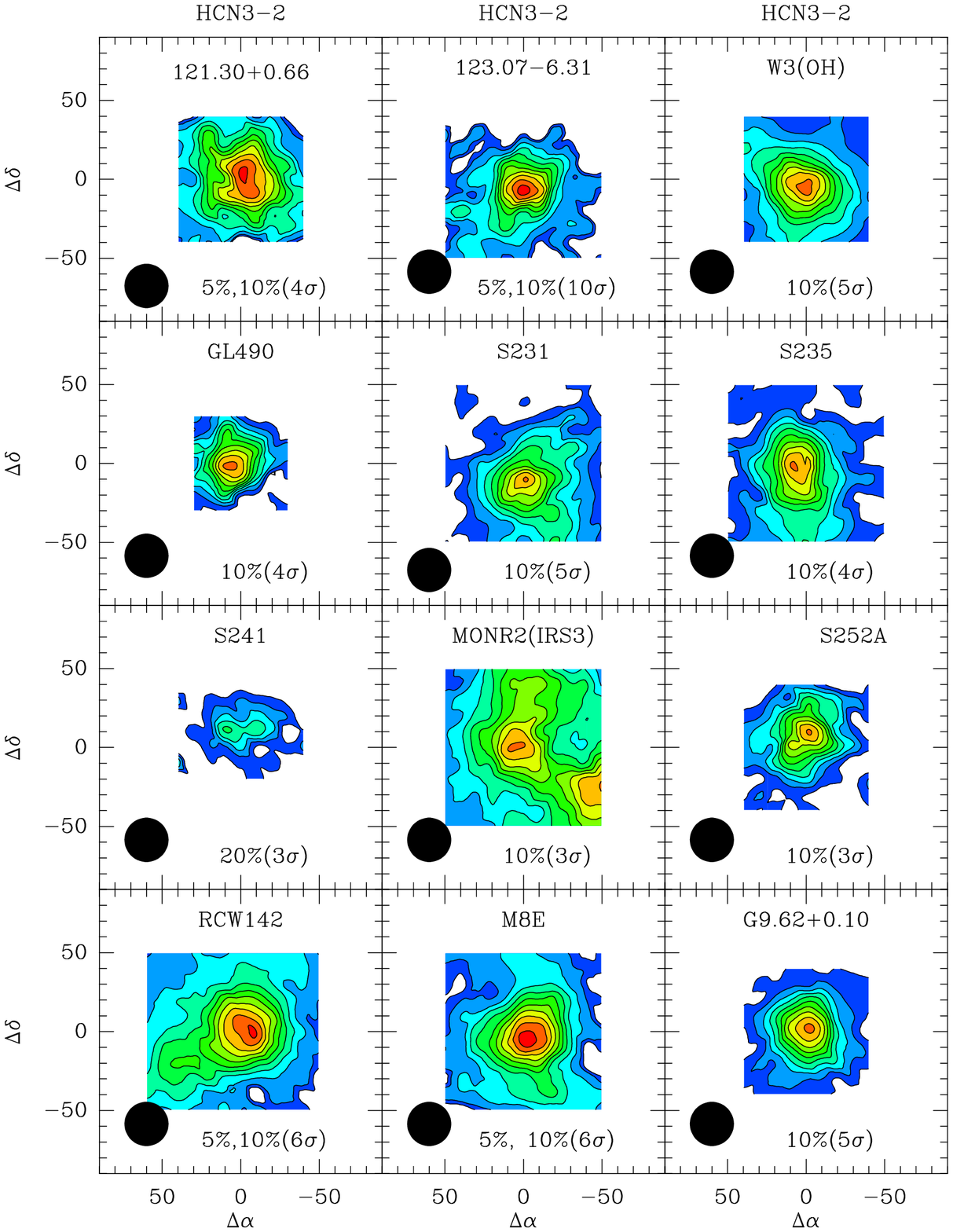}
\caption{\label{hcn321}HCN 3-2 contour maps of massive clumps. The lowest
contour level and
increasing step of contours are indicated in the plot.
The beam size is shown at the lower left of each map.}
\end{figure}

\begin{figure}[hbt!]
\epsscale{0.90}
\plotone{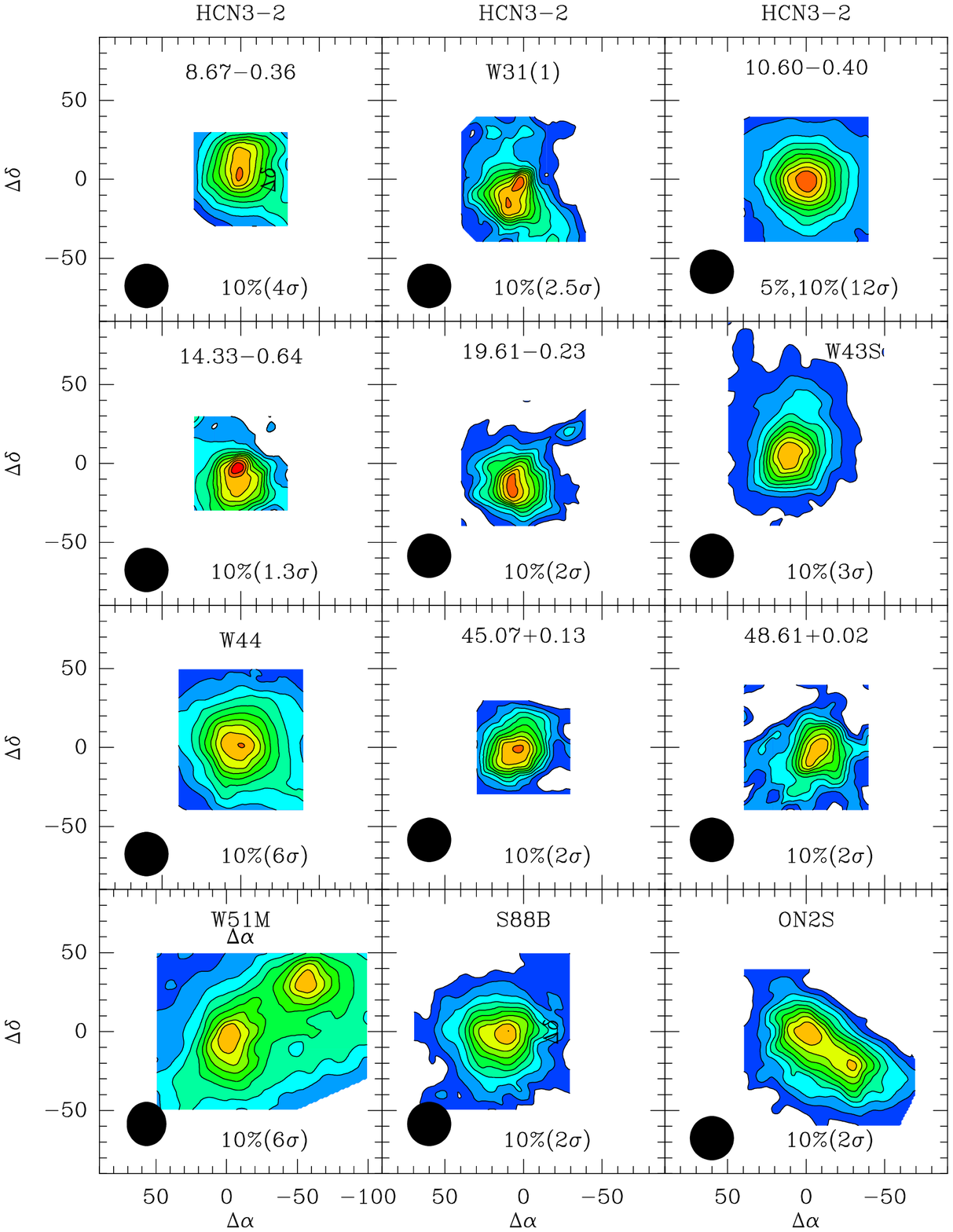}
\caption{\label{hcn322}HCN 3-2 contour maps continue... }
\end{figure}

\begin{figure}[hbt!]
\epsscale{0.90}
\plotone{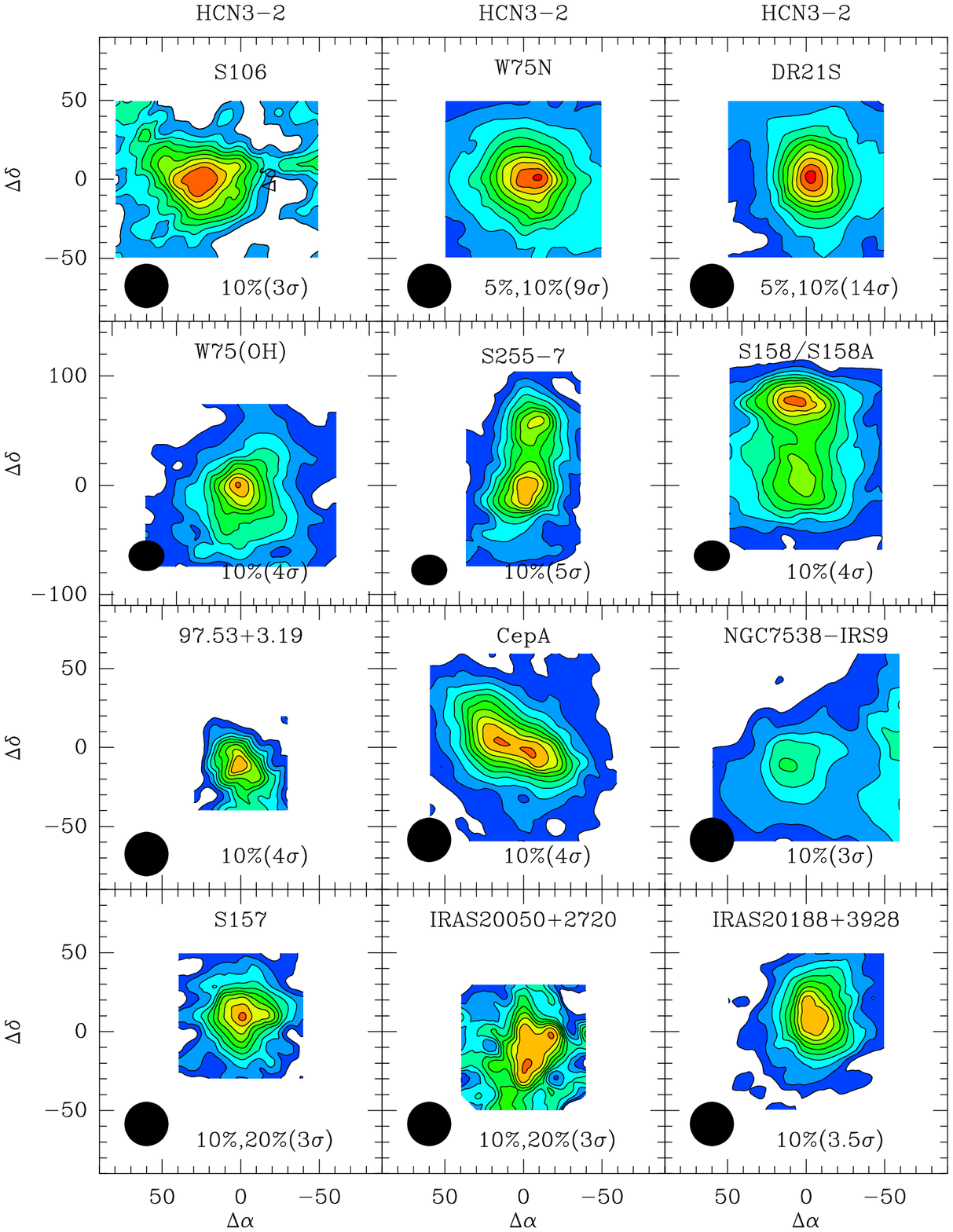}
\caption{\label{hcn323}HCN 3-2 contour maps continue... }
\end{figure}

\begin{figure}[hbt!]
\epsscale{0.90}
\plotone{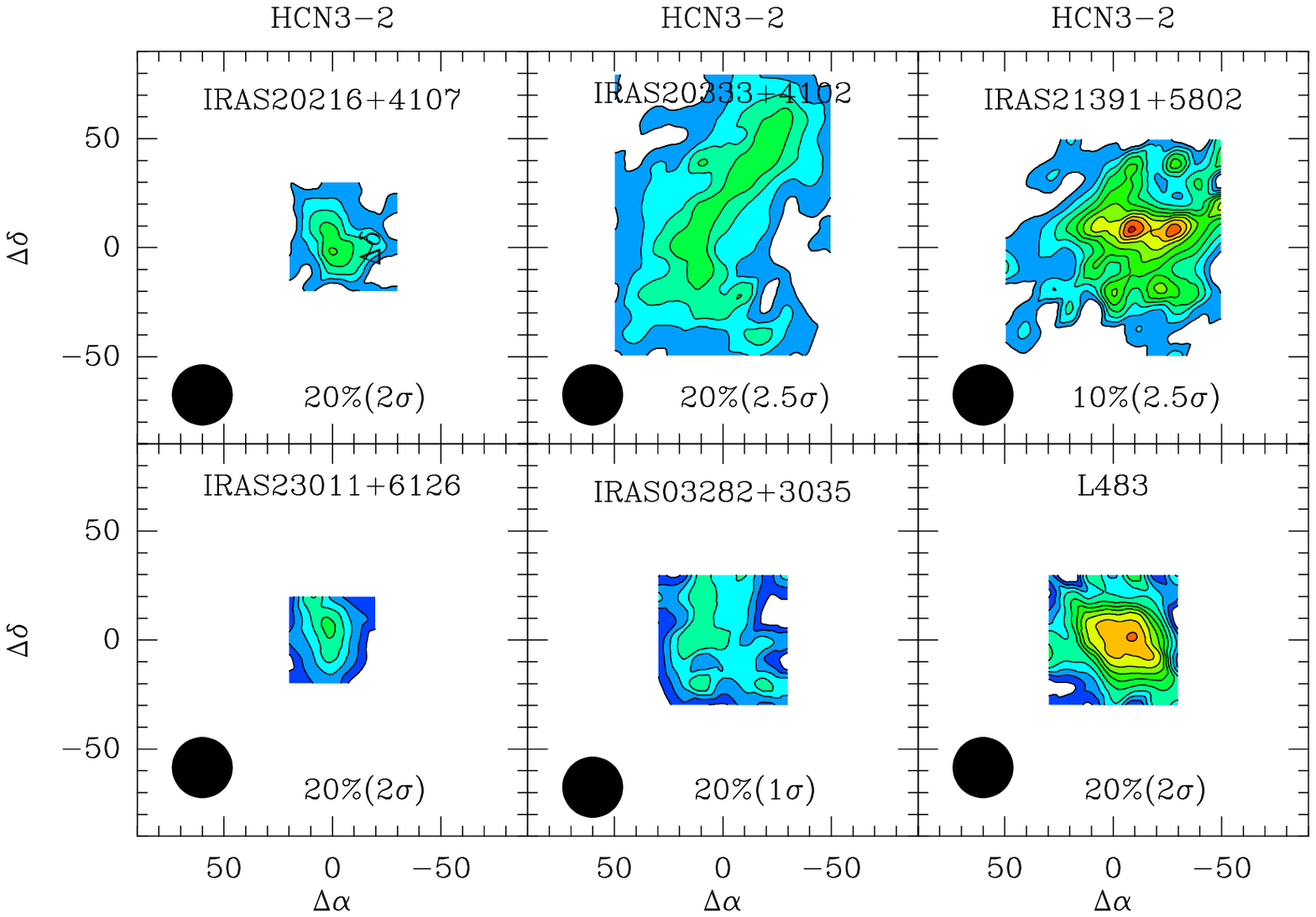}
\caption{\label{hcn324}HCN 3-2 contour maps continue... }
\end{figure}

\begin{figure}[hbt!]
\epsscale{0.85}
\plotone{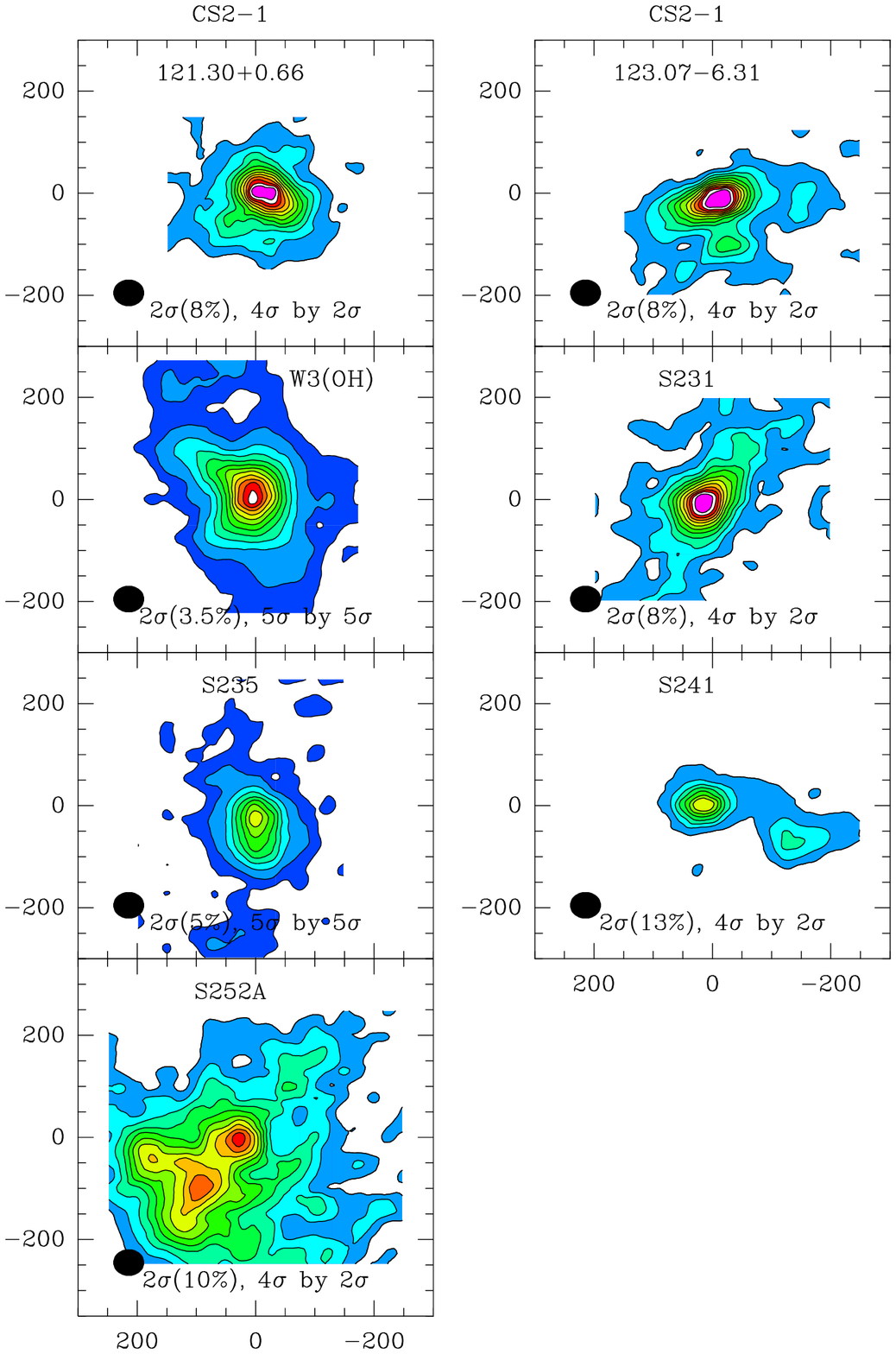}
\caption{\label{cs211}
CS 2-1 contour maps of massive clumps. The lowest contour level and
increasing step of contours are indicated in the plot.
The beam size is shown at the lower left of each map.}
\end{figure}

\begin{figure}[hbt!]
\epsscale{0.90}
\plotone{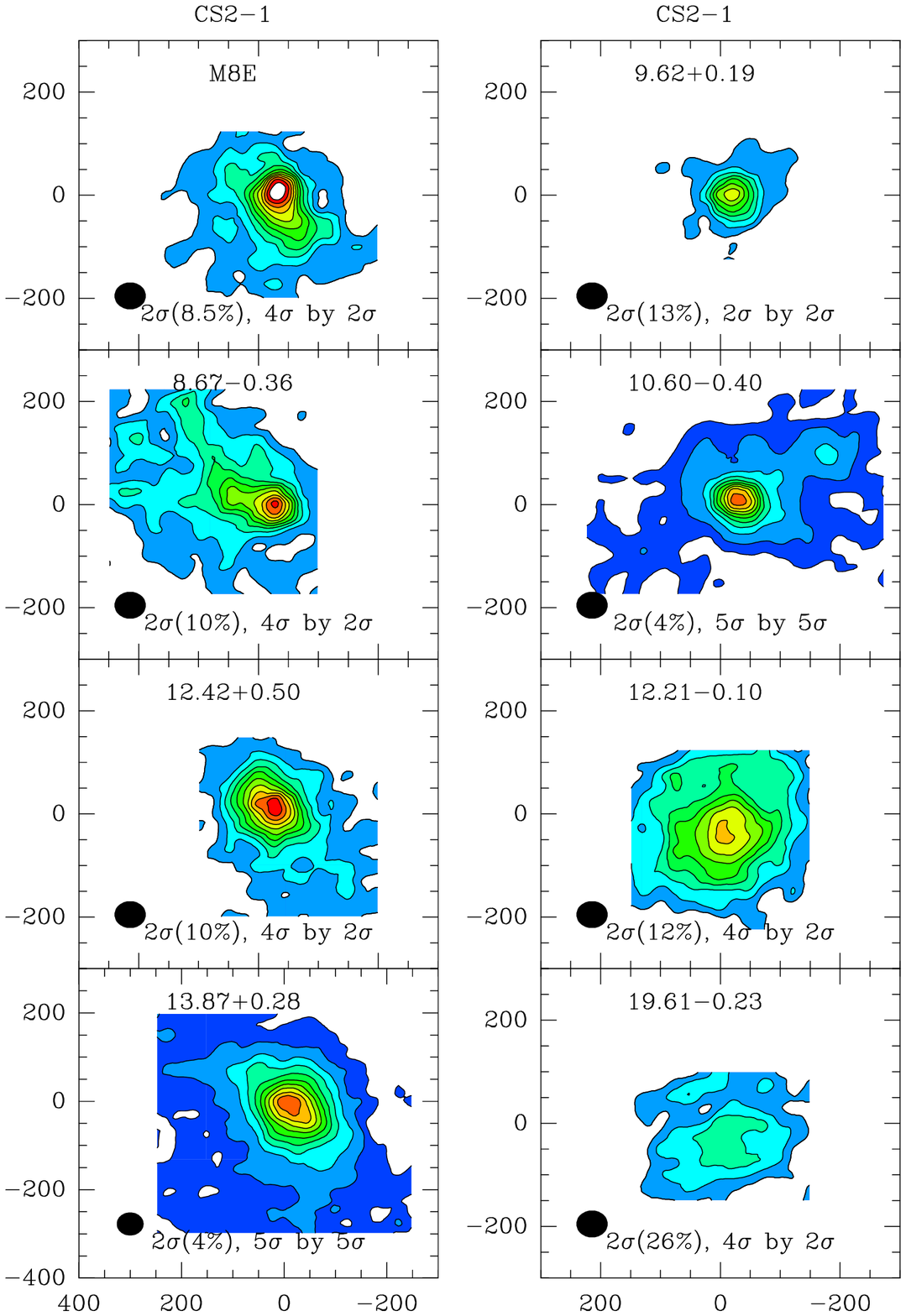}
\caption{\label{cs212}CS 2-1 contour maps continue... }
\end{figure}

\begin{figure}[hbt!]
\epsscale{0.90}
\plotone{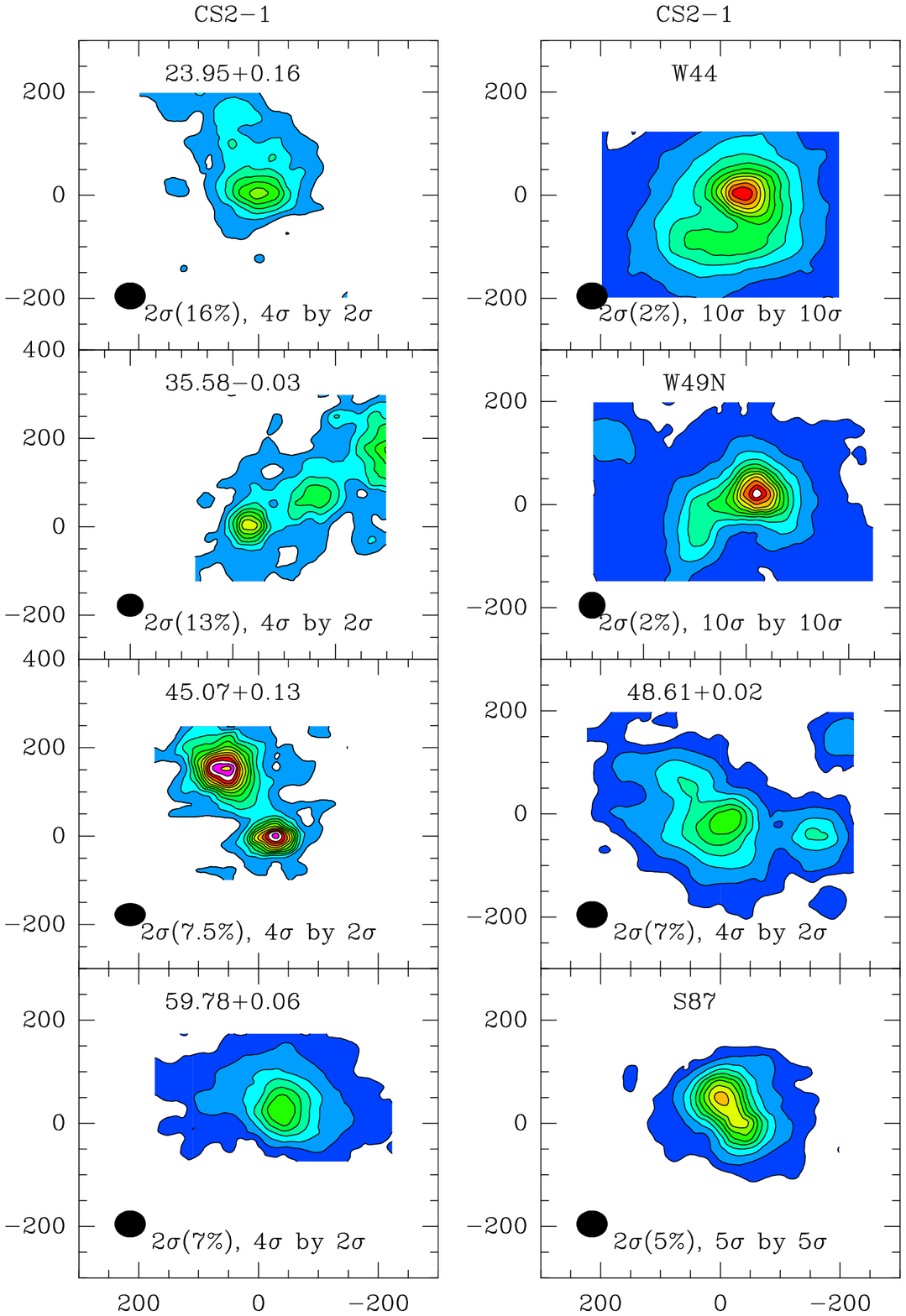}
\caption{\label{cs213}CS 2-1 contour maps continue... }
\end{figure}

\begin{figure}[hbt!]
\epsscale{0.90}
\plotone{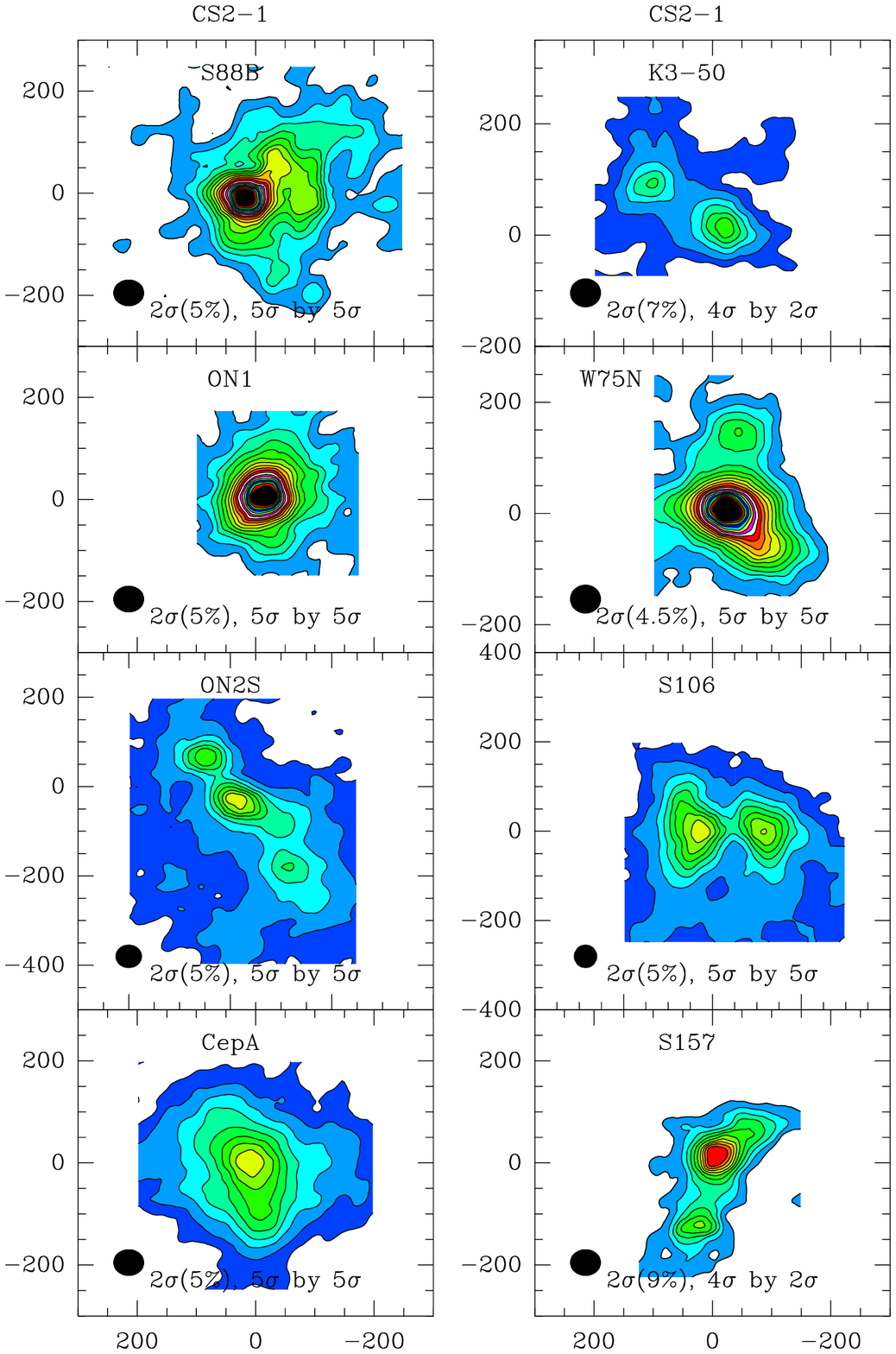}
\caption{\label{cs214}CS 2-1 contour maps continue... }
\end{figure}

\begin{figure}[hbt!]
\epsscale{0.90}
\plotone{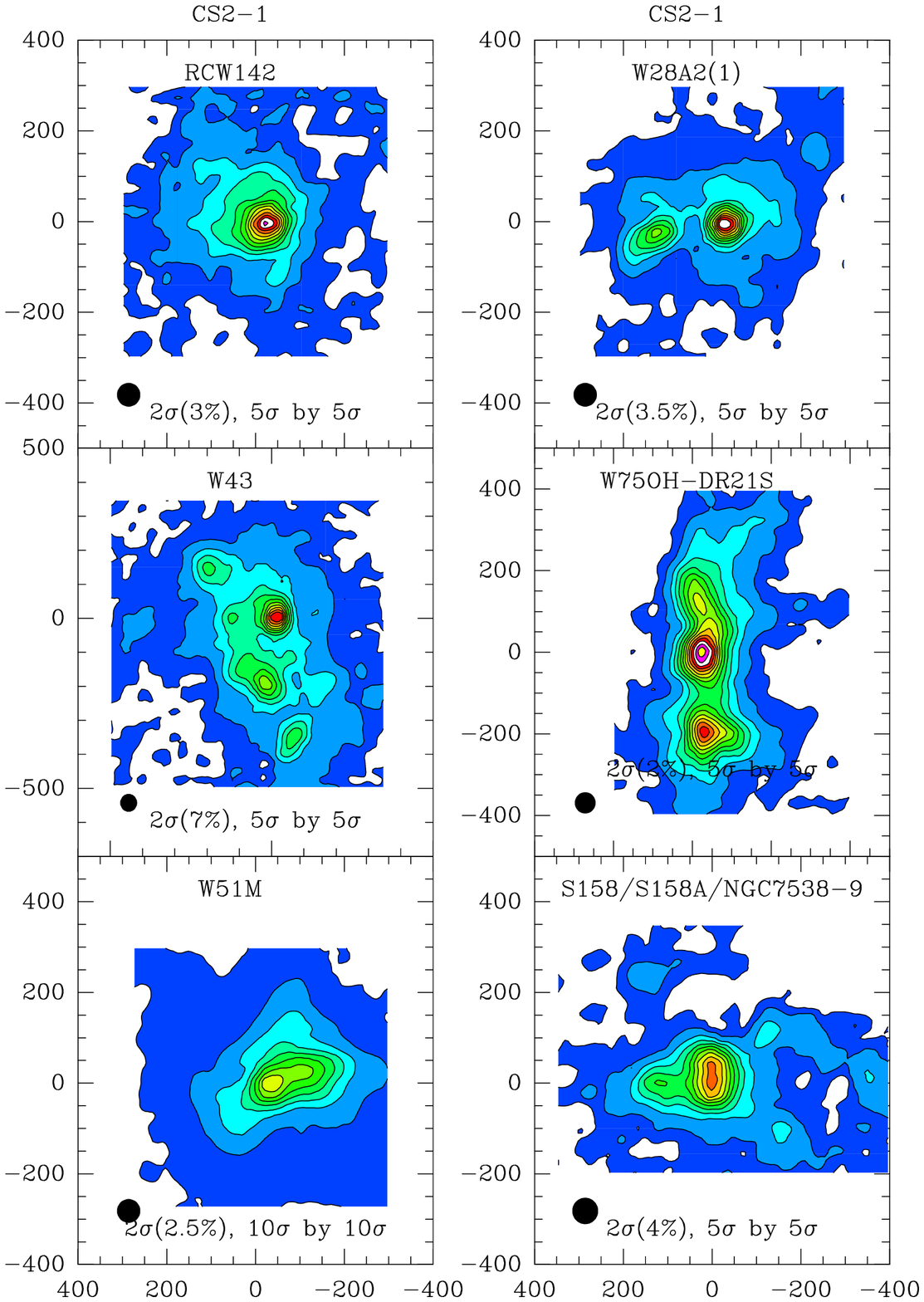}
\caption{\label{cs215}CS 2-1 contour maps continue... }
\end{figure}

\begin{figure}[hbt!]
\epsscale{0.90}
\plotone{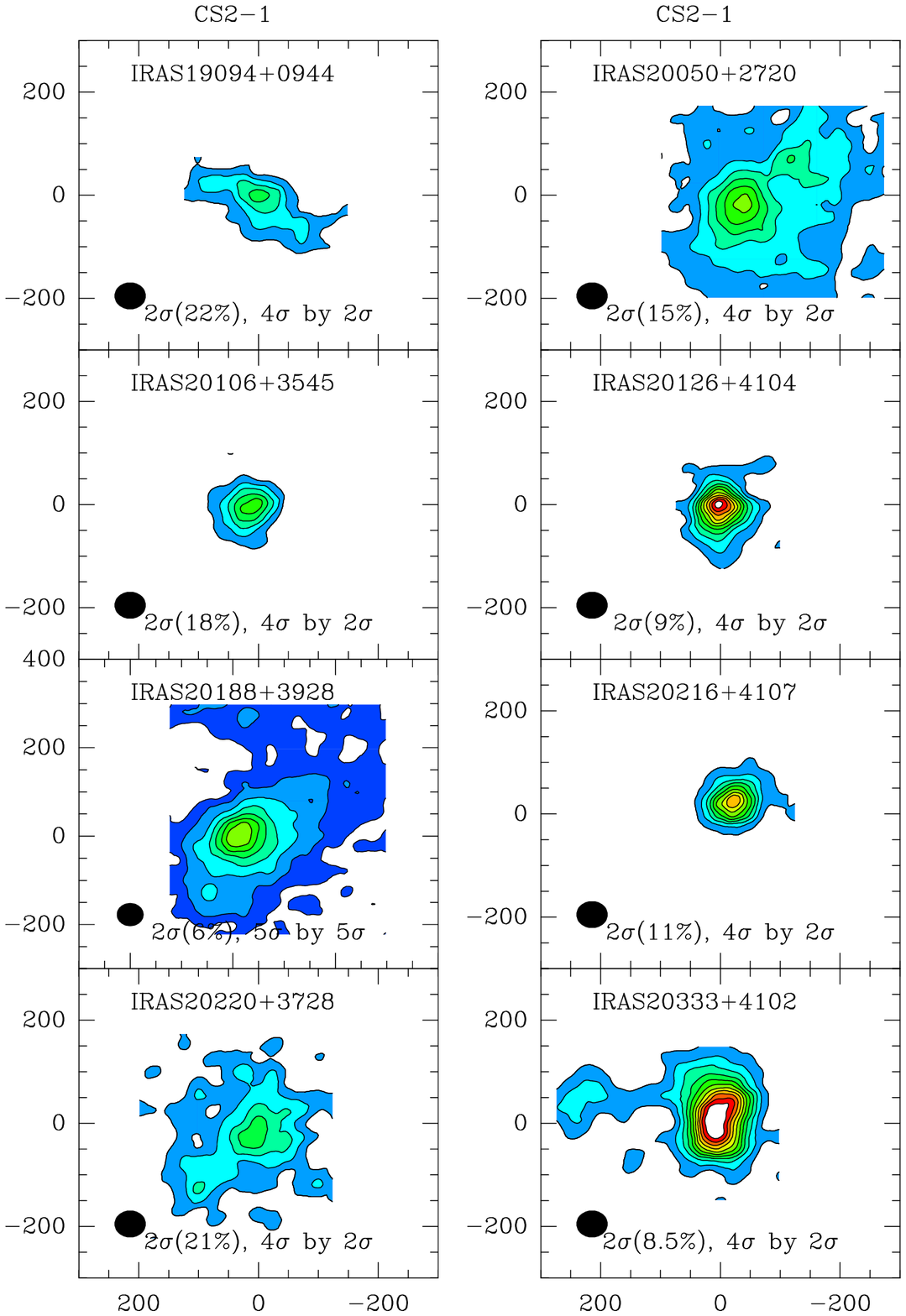}
\caption{\label{cs216}CS 2-1 contour maps continue... }
\end{figure}

\begin{figure}[hbt!]
\epsscale{0.90}
\plotone{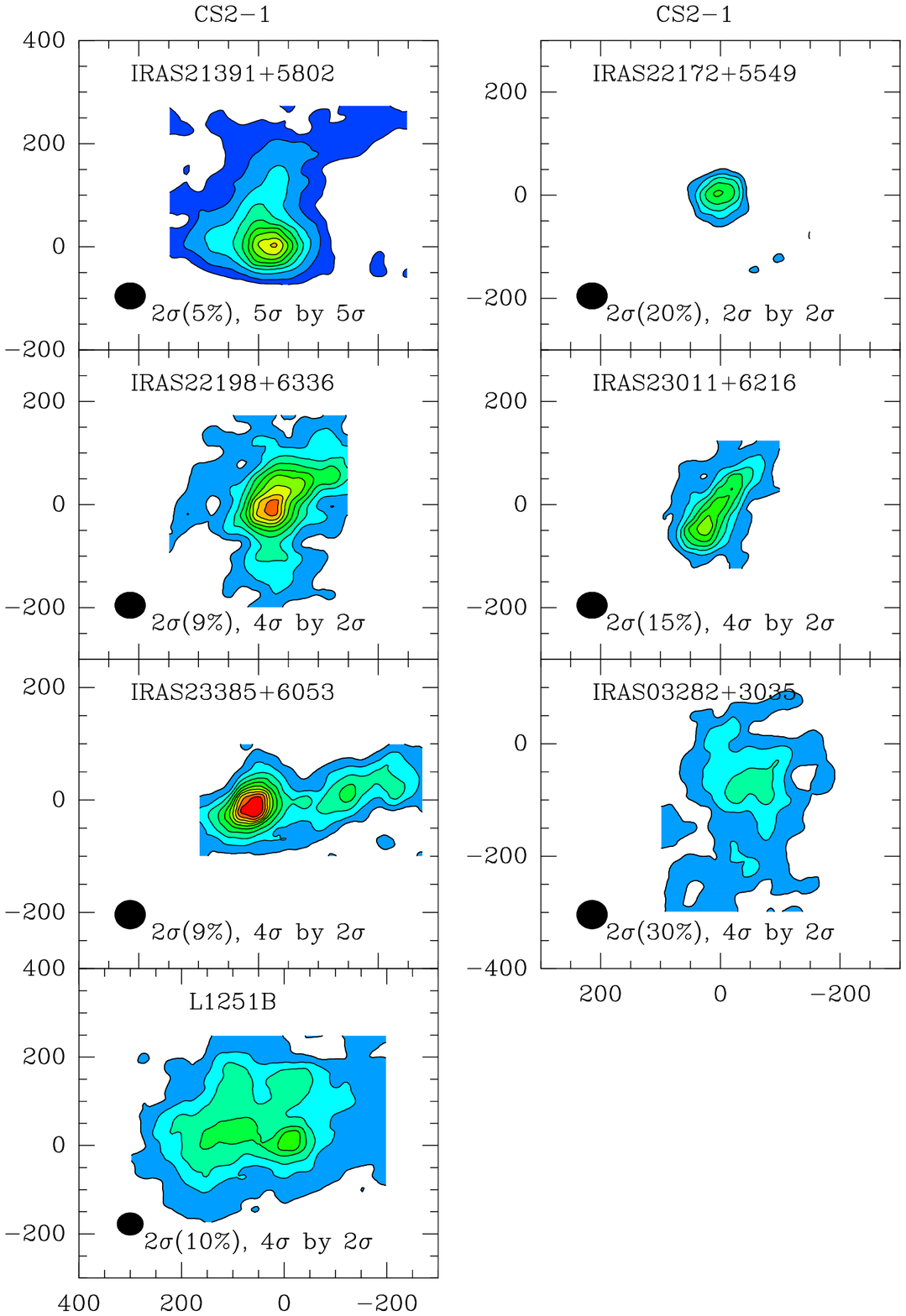}
\caption{\label{cs217}CS 2-1 contour maps continue... }
\end{figure}

\begin{figure}[hbt!]
\epsscale{0.90}
\plotone{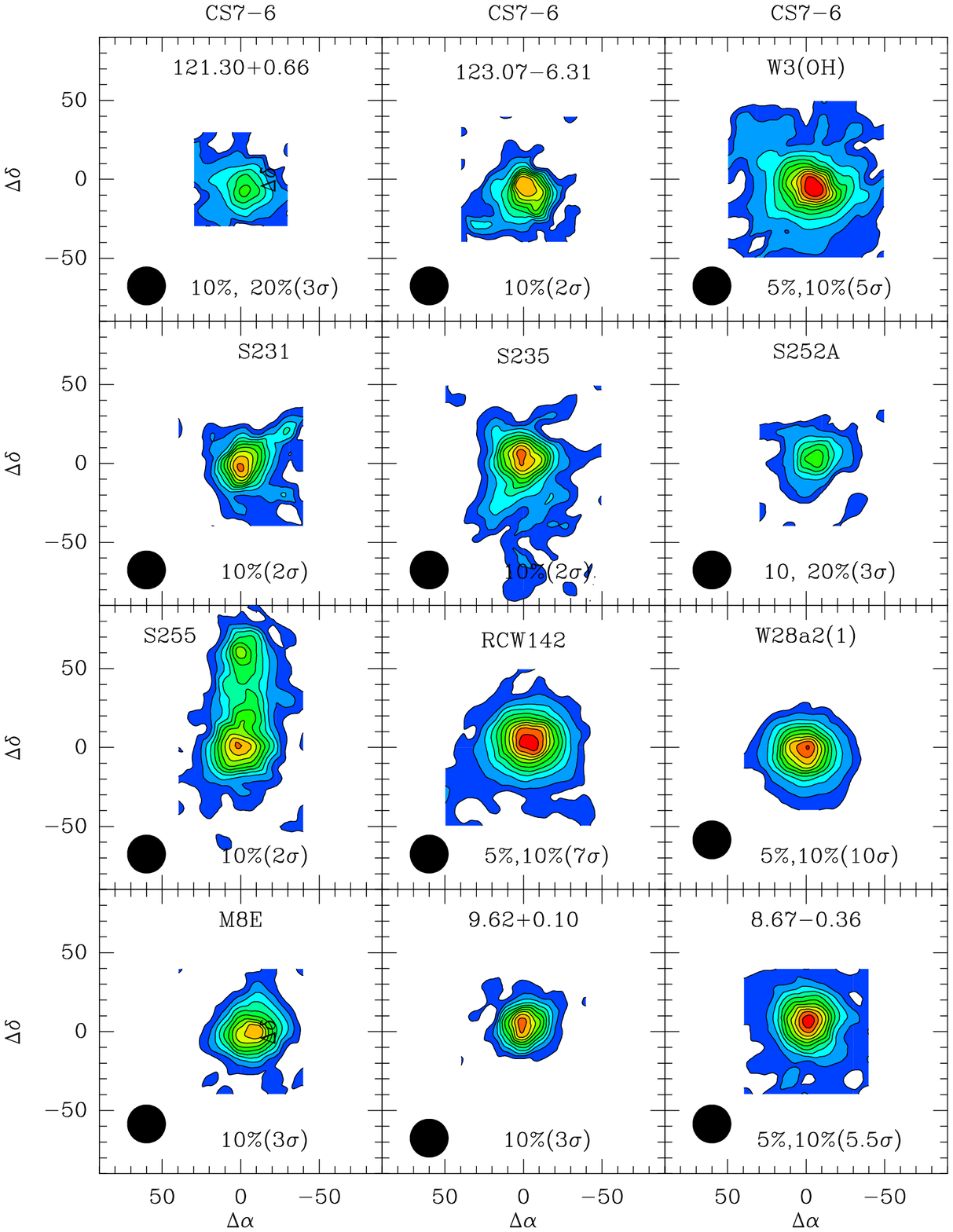}
\caption{\label{cs761}CS 7-6 contour maps of massive clumps. The lowest contour
level and
increasing step of contours are indicated in the plot.
The beam size is shown at the lower left of each map. }
\end{figure}

\begin{figure}[hbt!]
\epsscale{0.90}
\plotone{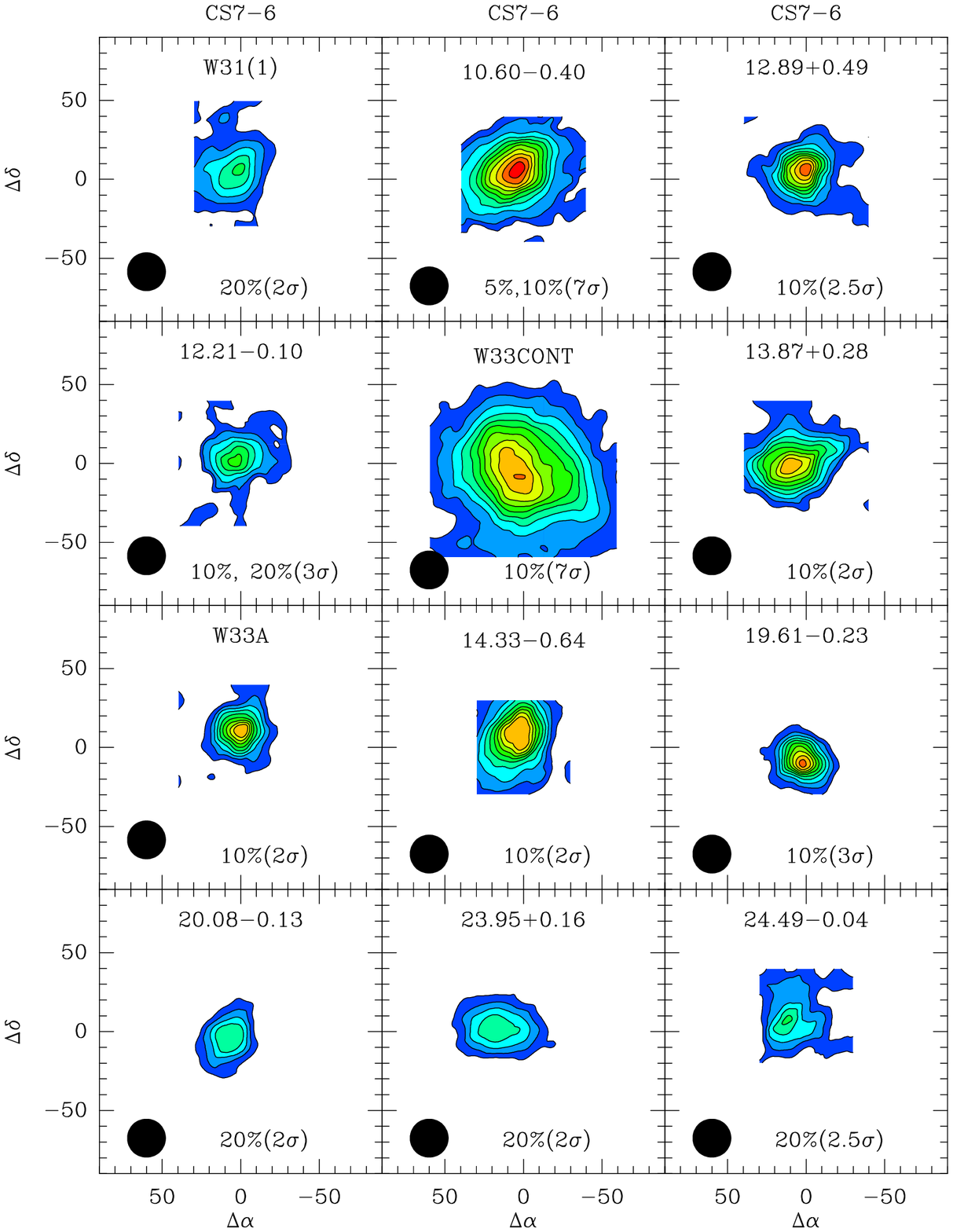}
\caption{\label{cs762}CS 7-6 contour maps continue... }
\end{figure}

\begin{figure}[hbt!]
\epsscale{0.90}
\plotone{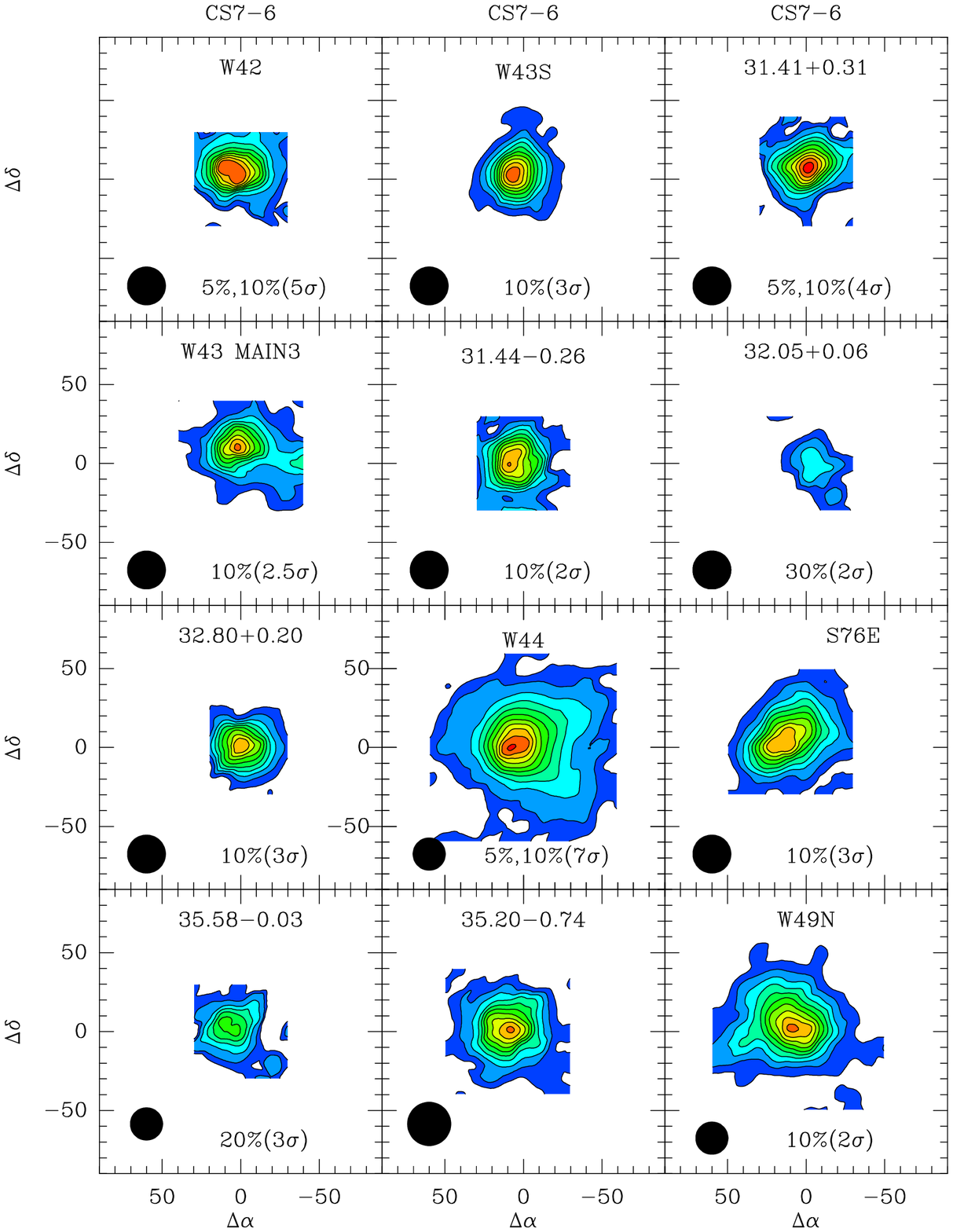}
\caption{\label{cs763}CS 7-6 contour maps continue... }
\end{figure}

\begin{figure}[hbt!]
\epsscale{0.90}
\plotone{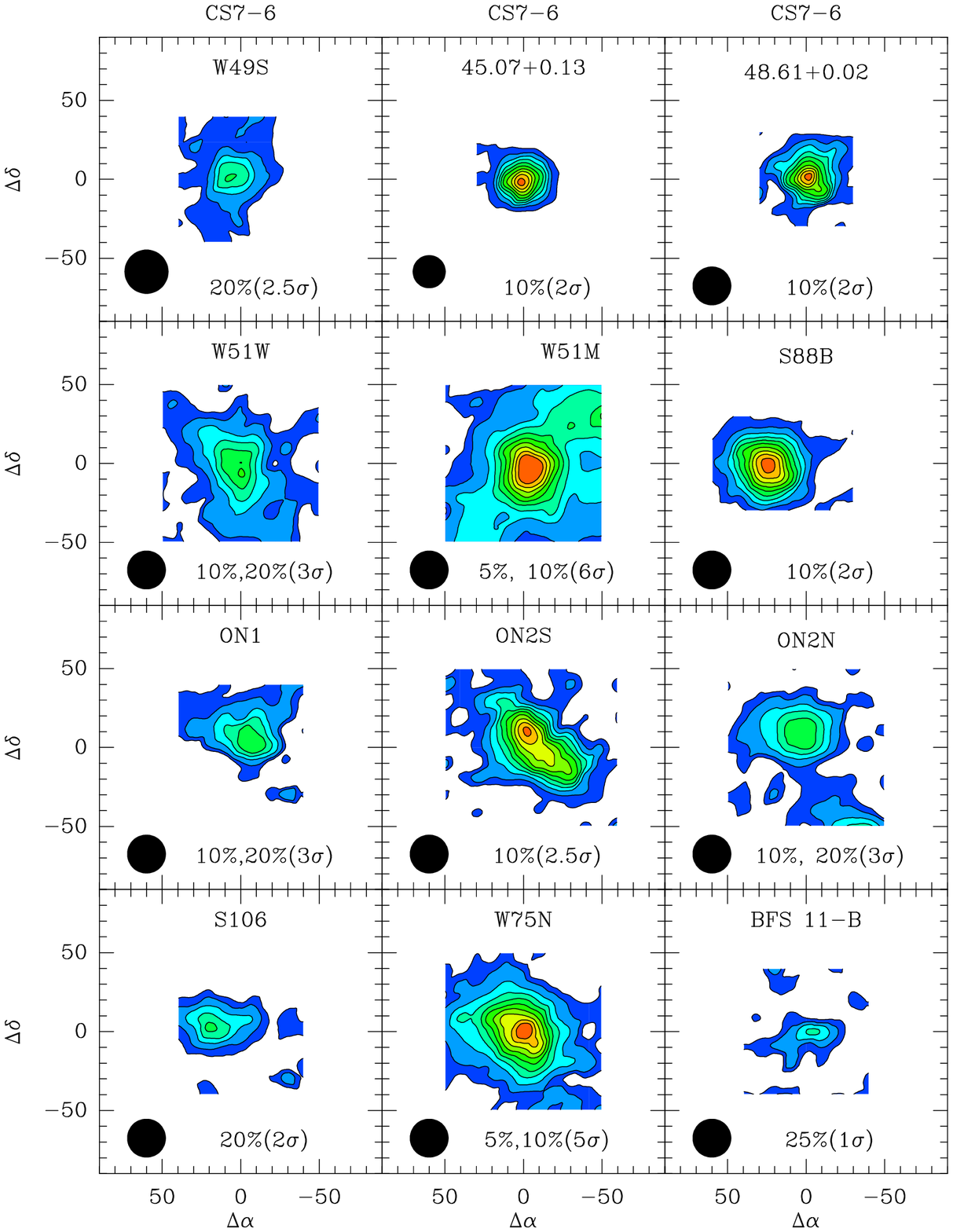}
\caption{\label{cs764}CS 7-6 contour maps continue... }
\end{figure}

\begin{figure}[hbt!]
\epsscale{0.90}
\plotone{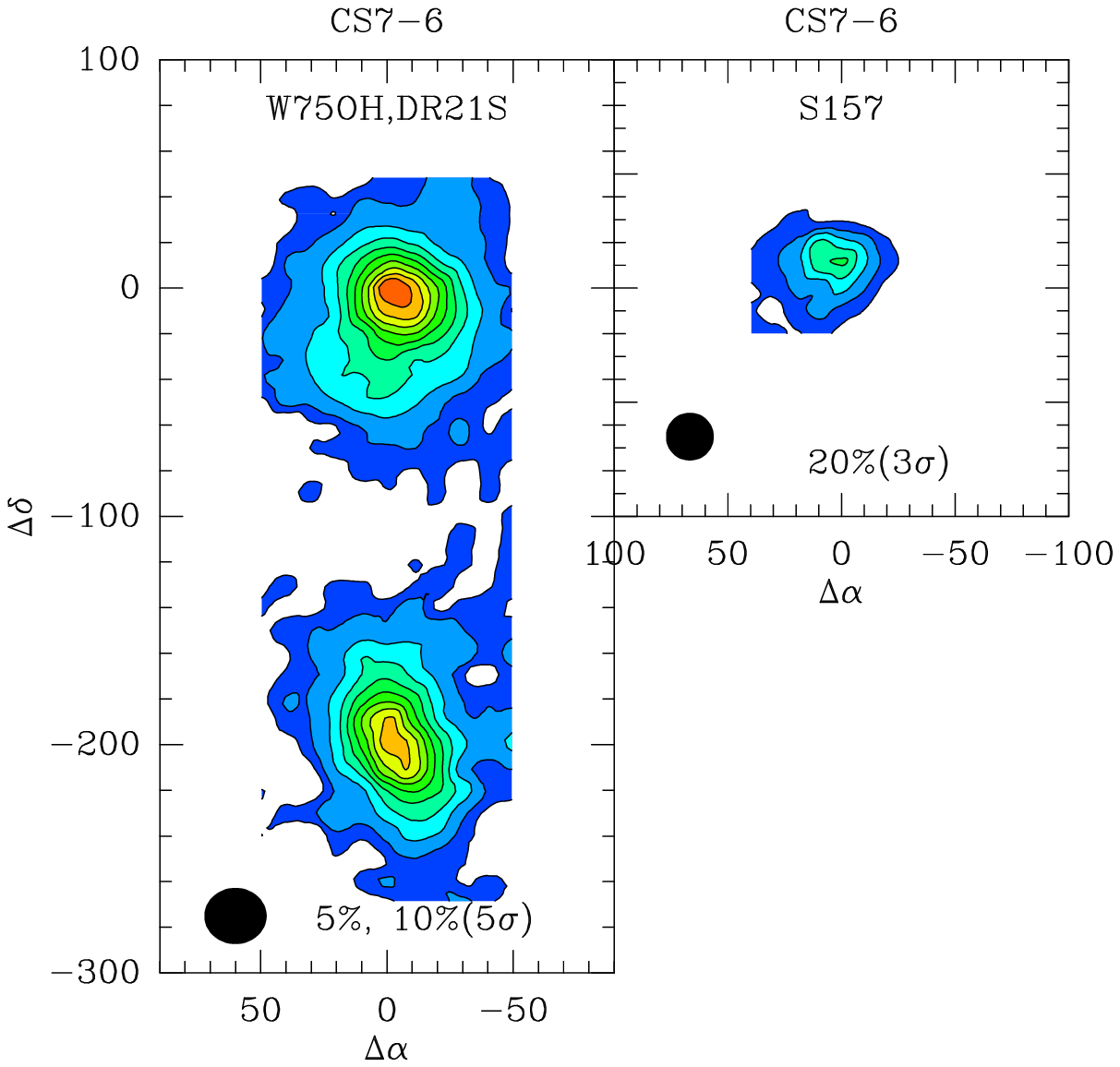}
\caption{\label{cs765}CS 7-6 contour maps continue... }
\end{figure}

Contour maps of HCN 1-0, HCN 3-2, CS 2-1, and CS 7-6 are presented in
Figures \ref{hcn1}-\ref{cs765}. These maps show a variety of morphologies,
with some multiple peaks. HCN 1-0 and CS 2-1 maps
cover a much larger area (typically greater than $7\arcmin \times  7\arcmin$)
than the higher J transition maps. A substantial fraction of
these low-J transition maps show multiple peaks. For HCN 1-0, 16 out of 53 maps
(30\%) show multiple peaks; For CS 2-1, 21 out of 53 maps show multiple
peaks (40\%). This ratio is lower for high-J HCN and CS tracers. Only
8 out of 42 HCN 3-2 maps (19\%) and 1 of 51 CS 7-6 (2\%) present multiple
peaks.
For CS 5-4 maps (Shirley et al. 2003), this ratio is 16\% (9 out of 55).
The majority of HCN 3-2 and CS 7-6 maps show a
single, well-peaked distribution of intensity, similar to earlier maps
of the CS 5-4 transition or the 350 \micron\ dust continuum emission (Shirley
at al. 2003, Mueller et al. 2002).

Considering only sources with maps of multiple lines of the same species,
there are 37 dense clumps mapped in both HCN 1-0 and HCN 3-2; 6 clumps that
are single clumps in HCN 1-0 maps break into multiple clumps in HCN 3-2.
Among the 38 clumps mapped by both CS 2-1 and CS 5-4, 4 single clumps
in CS 2-1 maps show multiple components in CS 5-4 maps.
This effect may be due to the higher resolution of the HCN 3-2 and
CS 5-4 observations. For the 4 clumps that have been resolved by CS 5-4 but
not CS 2-1 maps, they are either not observed by CS 7-6, or the other
components are beyond the CS 7-6 mapping area. Consequently, we do not know
if they are also separate peaks in the CS 7-6 maps.

\begin{figure}[hbt!]
\epsscale{0.80}
\plotone{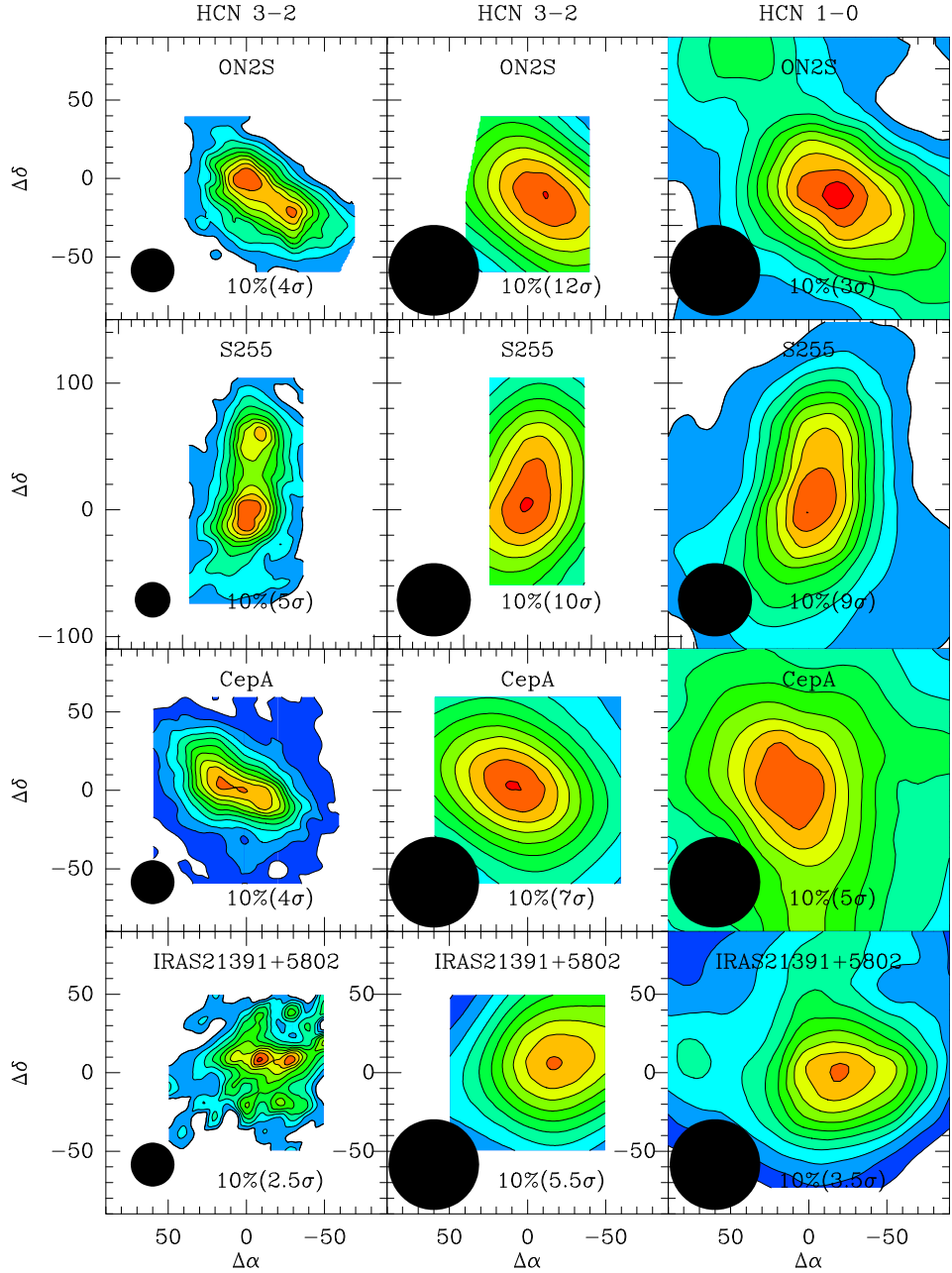}
\caption{\label{compare12}The plots to show the effect when we smooth the HCN 3-2 maps (left panels) of four clumps with more than one peaks, to the
resolution of HCN 1-0 data (middle panels), comparing to the real HCN 1-0 maps
(right panels).}
\end{figure}

The overall impression is that low excitation lines may show secondary peaks
that are not dense enough to show up in the lines requiring higher excitation,
while maps with higher resolution may find multiple peaks within the
low-resolution map. Because lines
of higher excitation in our study have better spatial resolution, those
multiple peaks may not be seen in the lower excitation tracers, just because
of instrumental limitations. To check this possibility, we have smoothed
the HCN 3-2 maps for four clumps with multiple peaks to the resolution
of the HCN 1-0 data and compare them to the HCN 1-0 data
(Figure \ref{compare12}).
The similarity of the smoothed HCN 3-2 maps to the HCN 1-0 maps supports
the suggestion that HCN 1-0 maps with higher resolution would also
resolve multiple peaks.

These few clumps aside, the general appearance of most maps at the
current resolution is not one of many cores within a clump, but of single
well-peaked distributions. Maps with higher resolution that still retain
sensitivity to extended structure are needed to study substructure in these
clumps.

\subsection{Clump Sizes}

\begin{figure}[hbt!]
\epsscale{0.70}
\rotatebox{270}{\plotone{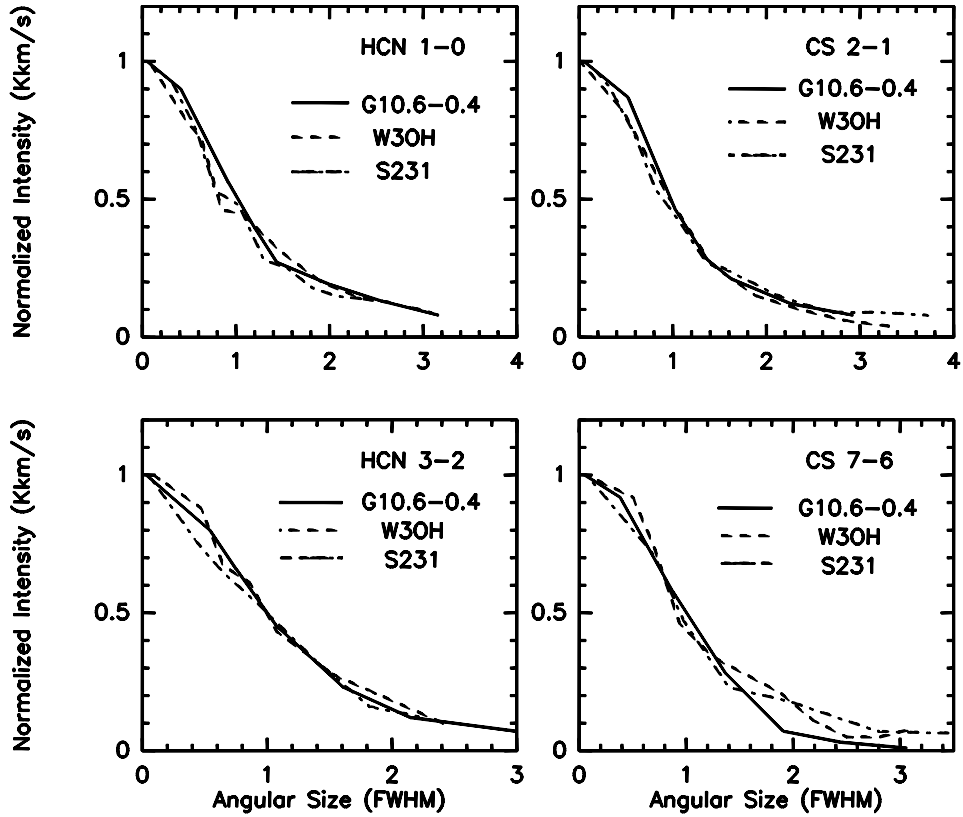}}
\caption{\label{rprofile}The one-dimensional radial profiles of some dense
clumps. The intensities have been normalized to the peak values, and the
angular radii (convolved with beam sizes) are in units of the FWHM
radius of each source.}
\end{figure}

The size of the clump is characterized by the angular diameter or
linear radius after beam deconvolution.
$\theta_{transition}$ is the angular diameter of a circle that has the
same area as the half peak intensity contour:
$$\theta_{transition}=2(\frac{A_{1/2}}{\pi}-\frac{\theta^{2}_{beam}}{4})^{1/2}, $$
and $R_{transition} = \theta_{transition} D /2,$
where $A_{1/2}$ is the area within the contour of half peak intensity,
$\theta_{beam}$ is the FWHM beam size, and $D$ is the distance to the source.
It is important to note that these sizes do not represent sharp boundaries.
To show this fact, we chose G10.6-0.4, W3(OH) and S231 as
examples, since these sources are round in the contour maps, thus better
approximated by a 1D radial profile. In Figure \ref{rprofile}, the emission
generally declines smoothly to the noise level, when plotted against the
angular source size, in units of the FWHM angular extent.

\begin{figure}[hbt!]
\epsscale{0.70}
\rotatebox{270}{\plotone{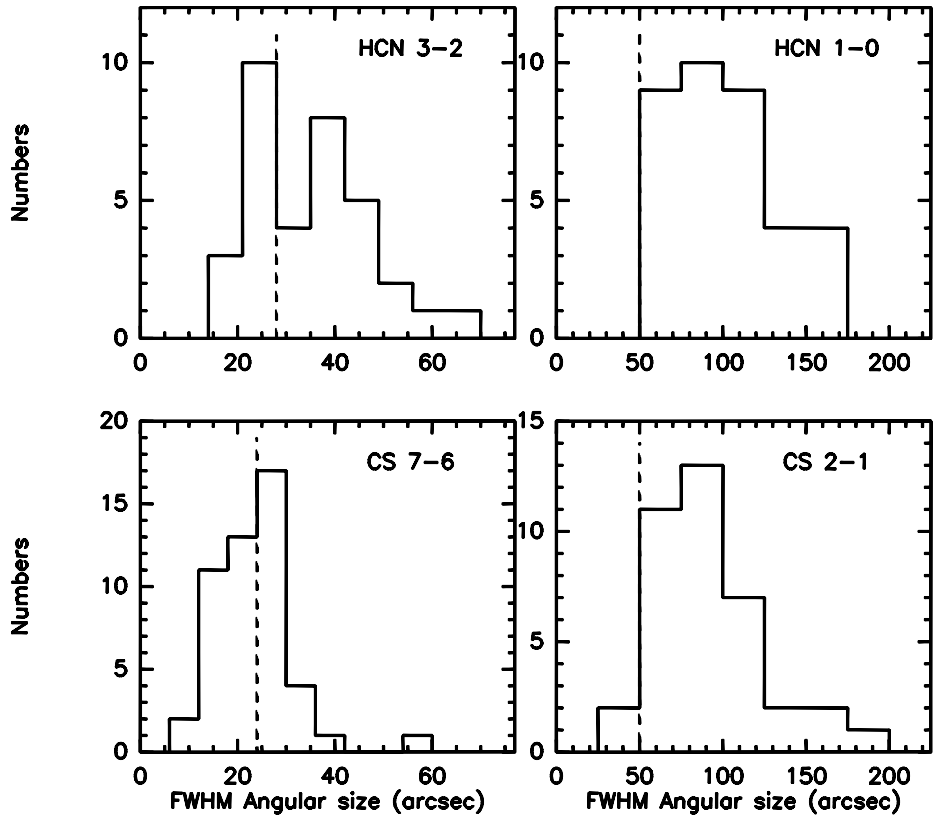}}
\caption{\label{anguhis}Histograms of the deconvolved angular size of massive
dense clumps in different transitions. The vertical dashed line in each panel
indicates
the beam size of each transitions. }
\end{figure}

\begin{figure}[hbt!]
\epsscale{0.70}
\rotatebox{270}{\plotone{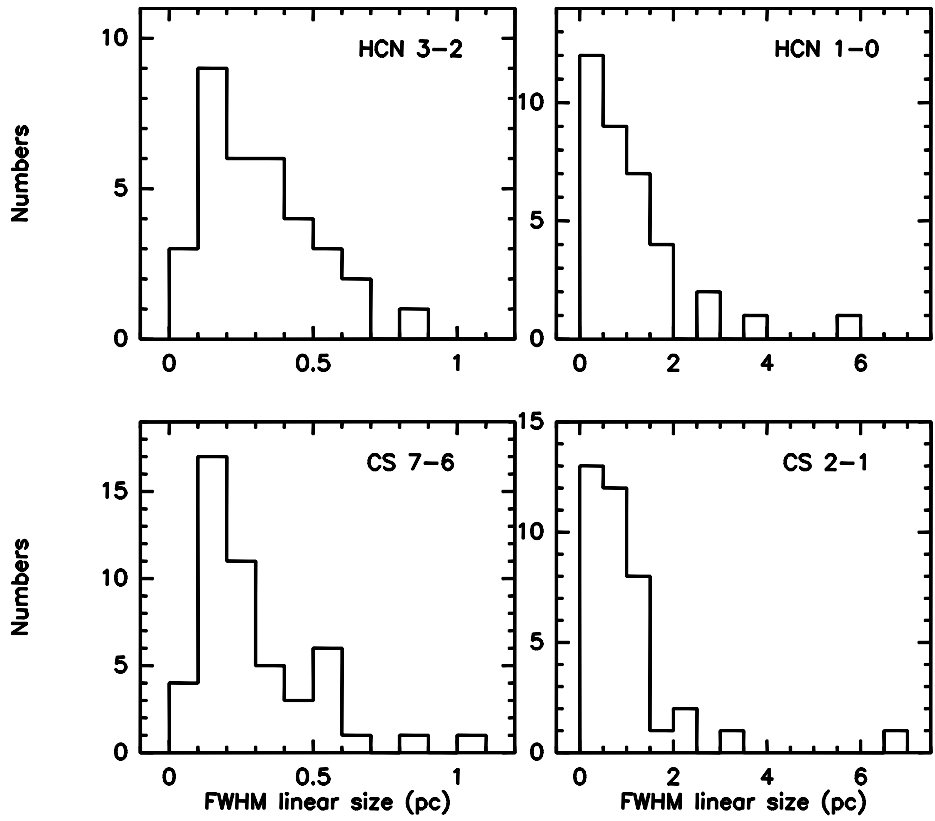}}
\caption{\label{linhis}Histograms of the deconvolved linear size of massive
dense clumps in different transitions. }
\end{figure}

The calculated sizes of clumps in each transition are
listed in Table \ref{sizetable}, with the mean and median values given at the
end of the Table.  The statistics on median and mean value include only the
massive sources, but not those sources below the horizontal line in the tables,
which are mostly not very massive and were added to study the lower luminosity
end of the star formation law.
The histogram of the deconvolved FWHM sizes
(angular size and linear size) of each transition are presented in
Figures \ref{anguhis} to \ref{linhis}, in which again we only include
massive dense clumps above the horizontal line in the tables.
Figure \ref{anguhis} shows the
histogram of deconvolved angular FWHM size of these
dense clumps, with a dashed line indicating the beam size in that transition.
In HCN 3-2 maps, 29\% of dense clumps have their deconvolved FWHM sizes
smaller than the beam size. For CS 7-6 maps, this ratio increases to 55\%,
including one source (G12.21-0.10) not being resolved at all. The beam size
effect can be nearly ignored for lower-J transitions; almost all the FWHM sizes
of HCN 1-0 and CS 2-1 are well above their beam sizes.

As Figure \ref{linhis} makes particularly clear, the size distribution
is skewed to low values before it is cut off by our resolution limits.
Note that the plots of the lower excitation lines use a larger scale because
they cover a substantially larger range of sizes.

\subsection{Luminosities}

For later comparison to work on other galaxies, we compute the line
luminosities and the infrared luminosities in the same way as is
done for extragalactic studies.
The line luminosity of each core, assuming a gaussian
brightness distribution for the source and a gaussian beam, is (we use HCN 1-0
for example)
\begin{equation}
L'_{HCN1-0}=23.5\times 10^{-6}\times D^{2}\times (\frac{\pi\times\theta_{s}^{2}}
{4ln 2})\times(\frac{\theta_{s}^{2}+\theta_{beam}^{2}}{\theta_{s}^{2}})
\times \int T_{R}dv  K \kms.
\end{equation}
Here D is the distance in  kpc, and $\theta_{s}$ and $\theta_{beam}$
are the angular size of the half-peak HCN 1-0 contour and the beam in
arcsecond.  This method is analogous to that of Gao \& Solomon (2004b),
but adapted to Galactic clumps.

The total infrared luminosity (8-1000 $\mu$m) was calculated
based on the 4 IRAS bands (Sanders and Mirabel 1996), as was done for
the galaxy sample of Gao \& Solomon (2004a):
\begin{equation}
L_{IR}  = 0.56\times D^{2}\times (13.48\times f_{12}+5.16\times f_{25}+
2.58\times f_{60}+f_{100}),
\end{equation}
where $f_x$ is the flux in band $x$ from the four IRAS bands in units
of Jy, D is in kpc, and L$_{IR}$ is in L$_{\odot}$. Some authors used the
bolometric luminosity (L$_{bol}$), which is calculated
from the SED of the source (e.g. Mueller et al. 2002).
Our statistics show that L$_{bol}$ calculated in Mueller et al. (2002)
is very close to the L$_{IR}$ calculated in this paper for the same
clumps. The ratio of L$_{IR}$ to L$_{bol}$ is 1.28 $\pm$0.90, and the median
is 1.09. The relatively large dispersion for this ratio is strongly affected
by two sources (S252A and CepA) whose ratio is much larger than others.
Excluding
these two sources, the mean ratio becomes 1.10$\pm$0.25, with the
median 1.08.
The mean and median values of \lir\ are ($4.7\pm1.2$)\ee5 \lsun\ and
1.06\ee5 \lsun, with values ranging from 708 \lsun\ to 5\ee6 \lsun,
considering only the sources in the original sample.

\begin{figure}[hbt!]
\epsscale{0.90}
\plotone{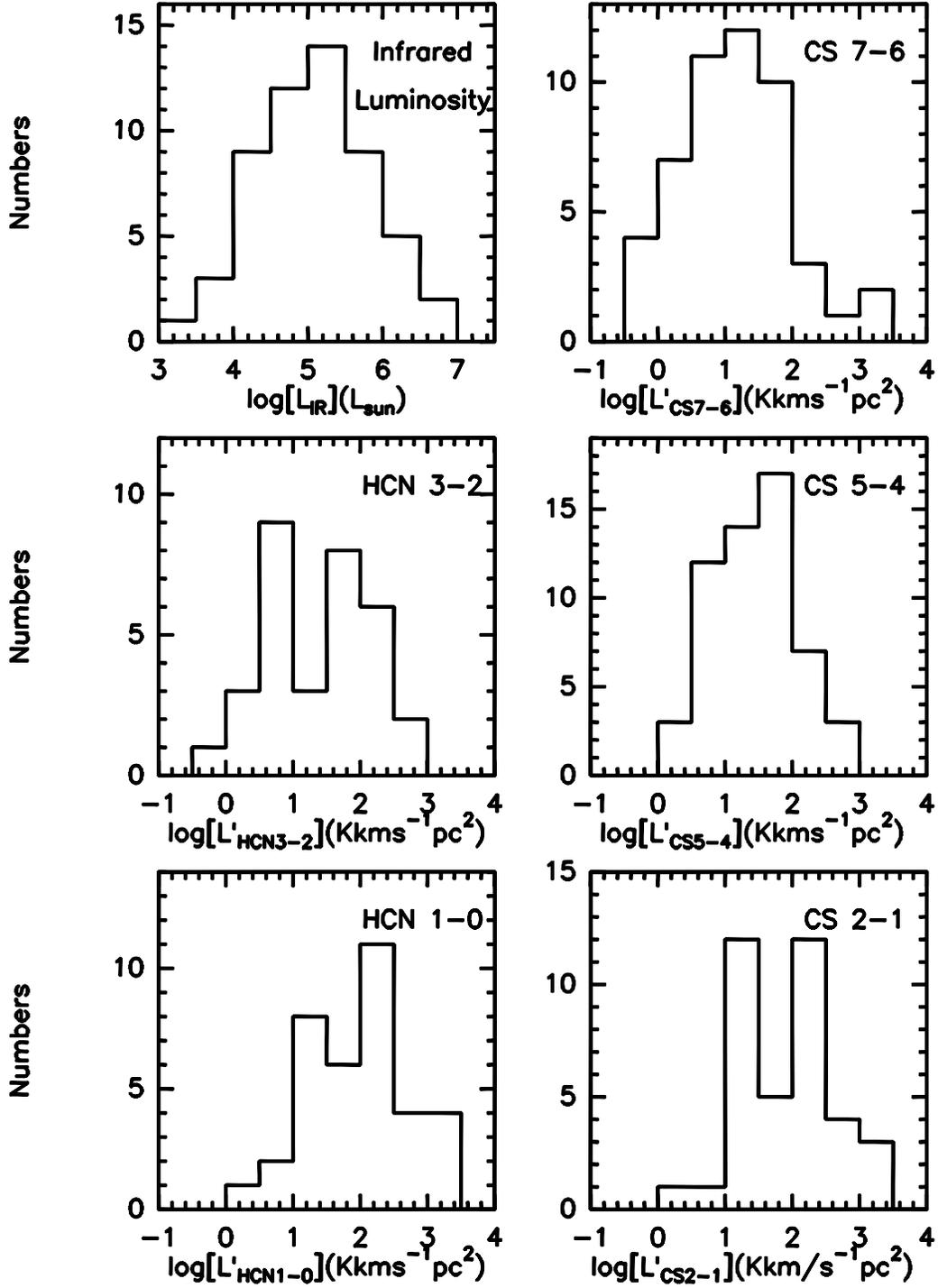}
\caption{\label{lumhis}Histograms of the infrared and
line luminosity of massive dense clumps
in different transitions. }
\end{figure}

The infrared and line luminosities are tabulated in Table \ref{lumtable}, with
mean and median values for the massive, dense clumps at the bottom.
The histograms of luminosity are shown in Figure \ref{lumhis}.
While the histograms appear to be well-peaked, we are systematically
biased against low luminosity sources. Together with the fact that
the histograms plot the logarithm of the luminosity, the highest
luminosity sources are clearly rare. For example,
the mean value of luminosity is typically 3 times the median.

\section{Discussion}

\subsection{Trends in Size with Tracer}

Although all these lines trace ``dense" gas, there are differences.
Higher J transitions of a given species trace denser parts of the core.
Crudely speaking, for a given
J level, HCN transitions have higher critical densities than do the
corresponding transitions in CS. The dependence on temperature also differs,
with the upper level of the CS 7-6 transition lying 66 K above ground, compared
to only 26 K above ground for the HCN 3-2 transition.

In Table \ref{ratiotable}, we list the properties of the transitions and the statistics of the
size ratios and luminosity ratios of maps derived from different tracers.
Table \ref{property} shows that maps of low-J transitions have larger
deconvolved sizes than those of higher J transitions.
For instance, the linear size of the HCN 3-2
map is 0.28 of the size of the HCN 1-0 maps, and the ratio is the same
for CS 7-6 versus CS 2-1.

This result would not be found in a ``raisin-pudding" model of clumps:
a roughly uniform pudding (the clump) with multiple dense cores sprinkled
randomly throughout. That model would predict similar sizes for maps of
transitions sensitive to different density regimes. Such a pattern was
seen in early studies of multiple transitions in a small number
of sources (Snell et al. 1984, Mundy et al. 1986, 1987).
The current results have higher spatial
resolution and more complete maps in the high-excitation lines than
the early results and do show that higher excitation lines are more centrally
condensed.
If the clumps are centrally condensed, with either a smooth increase
in density with decreasing radius, or an increased fraction of the volume
filled by dense ``raisins", the differences in size would be predicted as
the lines from higher-J transitions would be strongly enhanced in the
denser regions. Detailed models of individual dense clumps do support density
gradients with power laws providing good fits ($n(r) \propto r^{-p}$ with
$\mean{p} = 1.8$) (Mueller et al. 2002).

\subsection{An Inverse Size-Linewidth Relation}

The linewidth at the peak of the dense clumps for different tracers shows
an interesting trend. As presented in Table \ref{fwhm}, the mean and median
FWHM linewidth of the dense clumps increase for higher J transitions, from
about 4.8 \kms\ for CS 2-1, to about 5.9 \kms\ for CS 7-6.
If we look at a optically thin tracer, like C$^{34}$S 5-4, the mean FWHM is
4.7 \kms, which is closer to that of the CS 2-1 than higher J CS transitions.

Studies of molecular clouds have established a positive correlation of
cloud size, as mapped with a single tracer, and linewidth of that tracer
(Larson 1981, Myers 1985), which can be interpreted in terms of the virial
theorem. The linewidth-size relation within individual cores or clumps
is much less well-established (e.g., Evans 1999). Here we refer to the
useful distinction introduced by Goodman et al. (1998) between 4 different
kinds of relationships. A Goodman Type-4 relation plots linewidths of a
single tracer measured over regions of increasing radius; such plots show
little correlation (Goodman et al. 1998, Barranco \& Goodman 1998).
Goodman Type-1 relations plot sizes measured in multiple tracers for
multiple cores or clumps. Such plots generally show positive correlations
when using tracers of modest density applied to low-mass and intermediate-mass
cores or clumps; for example Caselli \& Myers (1995) found $\Delta v \propto
r^{0.53\pm0.7}$ with a correlation coefficient of 0.81 using \ammonia,
CS 2-1, and \cooo\ 1-0 maps of 8 starless cores.

In contrast, studies of the cores in this sample with Goodman Type-2
relations (single tracer, multicore) have shown no correlation between
linewidth and size (Plume et al. 1997 for CS 7-6) or a weak (r = 0.43)
correlation (Shirley et al. 2003 for C$^{34}$S 5-4). In both cases, the
linewidths were 4 to 5 times larger than would be predicted from the usual
linewidth-size relations developed from studies of clouds and low mass
cores. With maps of multiple tracers toward many cores, we could construct
Goodman Type 3 relations (multitracer, single core) for each core. Instead,
we try a modified version (Type 5?) in which we use the mean value over
many cores for each transition.

\begin{figure}[hbt!]
\epsscale{0.50}
\rotatebox{270}{\plotone{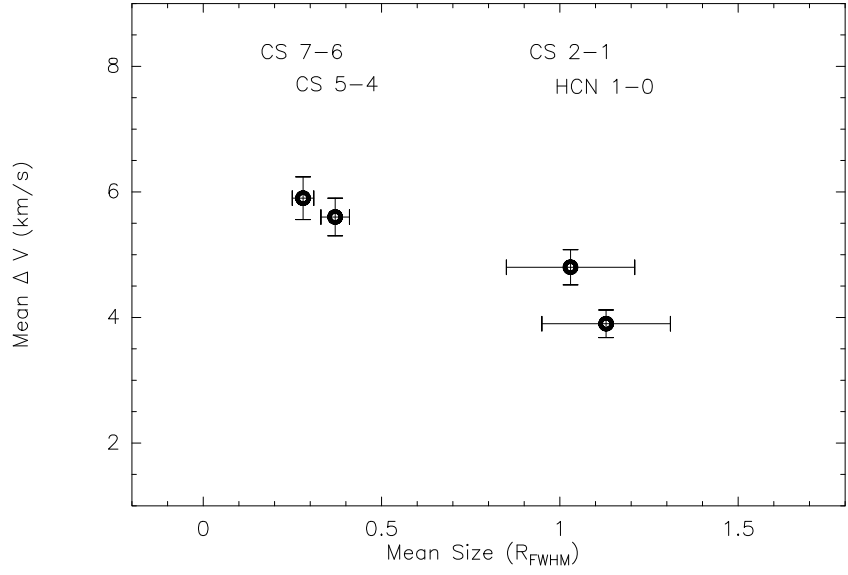}}
\caption{\label{sizedv}The inverse $\Delta V$-R$_{FWHM}$ relation between
different tracers for massive dense clumps. }
\end{figure}

Figure \ref{sizedv} shows the mean linewidth
versus the mean size of the dense clumps for CS 2-1, CS 5-4, CS 7-6 and HCN 1-0
transitions. HCN 3-2 lines have serious self-absorption, therefore they
are excluded from this figure.
HCN 1-0 has 3 hyperfine components; we took the mean linewidth of the
strongest component (F=2-1) for the plot.
The mean linewidth of the tracer has an {\it inverse} correlation with
the mean size of the dense clumps.  For example, while the CS 2-1 sizes are
on average 4 times larger than those of CS 7-6,
the mean of the linewidth ratio of CS 7-6 to CS 2-1 is 1.18$\pm$0.34 for
the massive dense clumps; the median ratio is 1.16.
The CS 2-1 emission traces a larger, quieter area than higher J transitions,
while CS 5-4 and CS 7-6 trace more compact and apparently more turbulent
regions. Alternatively, the larger linewidths toward the center could
in part reflect higher inflow motions.
A similar trend was seen in observations and models of G10.6 (Keto 1990).
It is also possible that the turbulence (here refers to all unresolved
gas motions, including inflow, outflow, complex fluid motions...) has
radial structure, which is higher close to the forming protostars
and therefore is being picked up better by the higher-J, dense gas
tracers.
The usual size-linewidth relations certainly do not apply to these regions
where massive stars are forming.

\subsection{What Physical Conditions Do the Lines ``Trace"?}

While we speak of a transition as ``tracing" gas of a certain density, this
formulation is too simplistic; the emission from a transition increases
smoothly and strongly over a wide range of densities and then levels out
just above the critical density. [This description applies only to lines
with frequencies high enough that radiative transitions stimulated by
cosmic background radiation or continuum emission from dust are unimportant
(Evans 1989).]
Nonetheless, it is common to speak of
a transition as tracing its critical density. If anything, the opposite
is true, as transitions fail to provide useful probes of density
{\it above} their critical density.  A more sensible fiducial
density is the effective density (Evans 1999), which is the density needed to
produce a line with a radiation temperature of 1 K for typical conditions.
For these comparisons, we use the critical and effective densities calculated
for a kinetic temperature of 100 K (Evans 1999).
Both these measures are given in Table \ref{property}.

The critical densities (n$_{crit}$) of the HCN 1-0 and CS 2-1 lines differ
by an order of
magnitude, but their effective densities (n$_{eff}$), are very similar.
The similarity of their mean sizes indicates that n$_{eff}$ better reflects
the excitation requirements. For HCN 3-2, CS 5-4, and CS 7-6, their sizes are
also
similar while the critical and effective densities make opposite predictions;
CS 7-6 has a much higher n$_{eff}$ than the other two, but HCN 3-2 has the
highest n$_{crit}$. The excitation energy of the upper level (E$_{up}$) is
considerably higher for the CS 7-6 transition. If we consider only the sources
that have been mapped by all three high-J transitions CS 7-6, CS 5-4, and HCN
3-2 (column 5 and 7 in Table \ref{property}),  statistics show that
the mean size and luminosity of the dense clumps
increase with decreasing $E_{up}$ of the transitions, as well as
decreasing n$_{eff}$, while the correlation with n$_{crit}$ is weaker.
The tracers that need higher excitation energy or density are confined
to smaller regions in the dense clumps, consistent with centrally
condensed structures. In addition, n$_{eff}$ better reflects the
excitation requirement than n$_{crit}$.

Empirically, we find that the HCN 1-0 and CS 2-1 lines form a matched set,
while HCN 3-2 and CS 5-4 lines make another pair; each pair has very similar
sizes and fairly similar luminosities in the mean. The CS 7-6 lines are close
to the properties of HCN 3-2 and CS 5-4 pair, but still more compact in size.

\subsection{Virial Mass, Surface Density, Volume Density, and Relations with
\lir}

The virial mass of the dense clumps with power-law density distributions
($n(r) \propto r^{-p}$) is calculated by
\begin{equation}
M_{Vir}(R)=\frac{5R\Delta v^{2}}{8a_{1}a_{2}Gln2}
 \approx 209\frac{(R/1 pc)(\Delta v/1 \kms)^{2}}{a_{1}a_{2}} M_{\odot}
\end{equation}
\begin{equation}
a_{1}=\frac{1-p/3}{1-2p/5}, \ p < 2.5.
\end{equation}
where $a_{1}$ is the correction for a power-law density distribution and
$a_{2}$ is the correction for a nonspherical shape (Bertoldi \& McKee 1992).
For aspect ratios less than 2, $a_{2}\sim 1$ and can be ignored for our
sample. To calculate $a_{1}$ we take the average p value ($1.77$) from Mueller
et al. (2002) for all massive dense clumps. A detailed discussion of
calculating the virial mass of dense clumps in this sample can be found in
Shirley et al. (2003).

The virial masses calculated from each transition
($M_{Vir}(R_{transition})$), for
HCN 1-0, 3-2, CS 7-6, and 2-1, are tabulated in Table \ref{virialtable}.
Rather than using the linewidth of the tracer, we used
the linewidth of C$^{34}$S 5-4 for all these tracers
because it is more reliably optically thin
than is the 3-2 line of H$^{13}$CN.
For lower mass sources (those below
the horizontal line in Table \ref{virialtable}), their C$^{34}$S 5-4 lines are
too weak to detect or to give a Gaussian fit except for IRAS20188+3928.
The FWHM linewidth of CS 2-1 is closest to that of C$^{34}$S 5-4 in
our sample, with a ratio of linewidth for CS 2-1 to  C$^{34}$S 5-4 of
1.2$\pm$0.1 and a median of 1.1. So, for these lower mass clumps, we used the
CS 2-1 linewidth to calculate their virial mass, and assigned an additional
20\% uncertainty to their linewidth uncertainties. While the linewidths vary
among transitions by about 25\%, some of this variation must be caused by
optical depth differences, and the differences in any case,
are small. The sense of these calculations is then an estimate of
the mass within the FWHM contour of each tracer.  The mean ratio of
$M_{vir}$ for CS 2-1 to CS7-6 is 4.0, with a median of 3.7;
the mean ratio for HCN 1-0 and 3-2 is 2.9, with a median of 2.8.
One would expect these ratios
to reflect the size ratios, which are 4.1 and 3.0 (the mean ratio),
respectively.

\begin{figure}[hbt!]
\epsscale{0.70}
\rotatebox{270}{\plotone{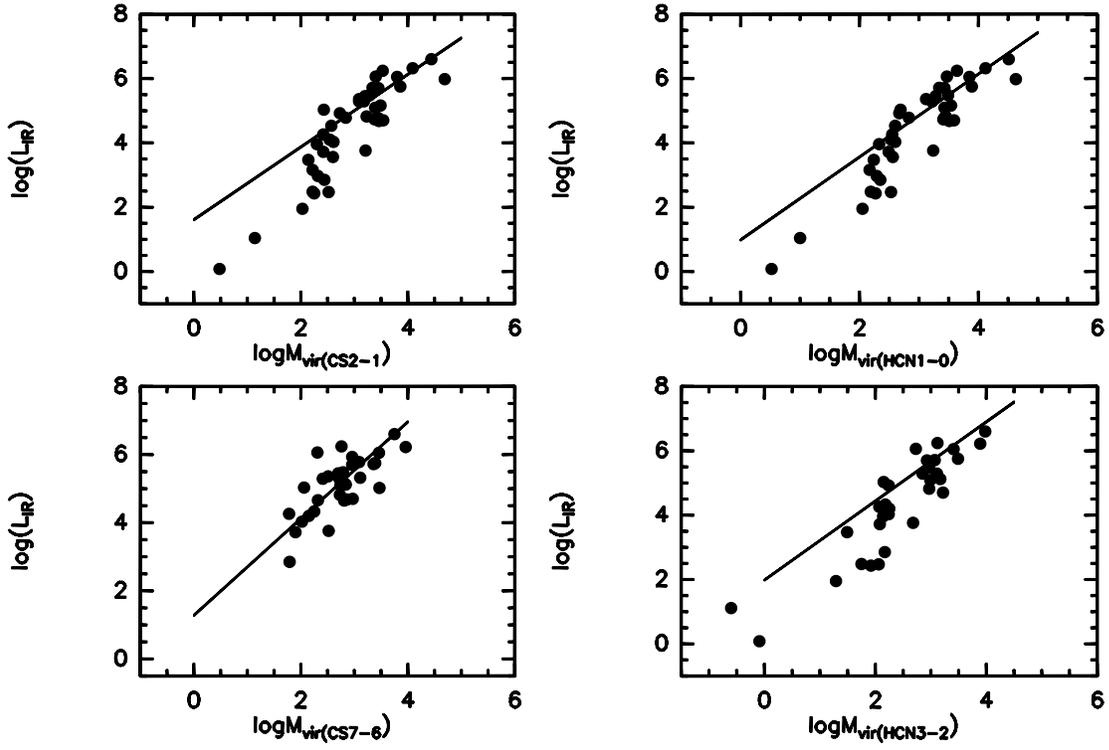}}
\caption{\label{mvirlir4} Correlations between the infrared luminosity
and the virial mass measured by various dense gas tracers. The fitted lines
show the linear least squares fit for clumps with $\lir > \eten{4.5}$ \lsun.}
\end{figure}

The virial mass enclosed by the FWHM contour of dense gas tracers
represents the mass, most of which is dense gas, that is available to
clustered star formation.
Therefore, one should expect a positive correlation between
the star formation rate, as indicated by the infrared luminosity,
and the virial mass.
One caveat is that the star formation rate can be underestimated by
L$_{IR}$ until the initial mass function is well sampled by the
star formation process (Krumholz \& Thompson 2007).
We plot \mvir\ versus  L$_{IR}$ for four transitions
in figure \ref{mvirlir4}.
It is obvious that the infrared luminosity is positively correlated with
the virial mass for these clumps.
We used linear fits based on least squares minimization and alternatively
on robust estimation, in which the absolute value of the deviation
is minimized. Comparing these two fits usually provides a more realistic
appraisal of the true uncertainties.

As noted by Wu et al. (2005),
there is a cut-off around L$\sim 10^{4.5}L_{\odot}$ in the
L$_{IR}$-$L'_{HCN1-0}$ correlation.
If we fit only the data for massive dense clumps with
$\lir > \eten{4.5}$ \lsun, the fitting results are as follows
(r is the correlation coefficient):

HCN 1-0:
$$Least\ Squares: log(L_{IR})=1.29(\pm0.08)\times {\rm
log}(M_{Vir}(R_{HCN1-0}))+0.98(\pm0.28); r=0.69 $$
$$Robust\ Fit: log(L_{IR})=0.70\times {\rm log}(M_{Vir}(R_{HCN1-0}))+3.03 $$
HCN 3-2:
$$Least\ Squares: log(L_{IR})=1.23(\pm0.09)\times {\rm
log}(M_{Vir}(R_{HCN3-2}))+1.98(\pm0.24); r=0.61 $$
$$Robust\ Fit: log(L_{IR})=0.79\times {\rm log}(M_{Vir}(R_{HCN3-2}))+3.16 $$
CS 2-1:
$$Least\ Squares: log(L_{IR})=1.13(\pm0.06)\times {\rm
log}(M_{Vir}(R_{CS2-1}))+1.61(\pm0.19); r=0.67 $$
$$Robust\ Fit: log(L_{IR})=0.98\times {\rm log}(M_{Vir}(R_{CS2-1}))+2.24 $$
CS 7-6:
$$Least\ Squares: log(L_{IR})=1.42(\pm0.07)\times {\rm
log}(M_{Vir}(R_{CS7-6}))+1.28(\pm0.23);r=0.49 $$
$$Robust\ Fit: log(L_{IR})=0.62\times {\rm log}(M_{Vir}(R_{CS2-1}))+3.76 $$

There are substantial variations between fitting method employed and among the
line used. If we
average the two fitting methods for each line and then average over all lines,
the resulting slope is $1.02\pm0.03$. Thus, L$_{IR}$ is roughly linearly
correlated with the virial mass for dense clumps with
$\lir > \eten{4.5}$ \lsun.

Below this threshold, the slope is steeper.
We also provide linear fits and robust fits to the
sources with $\lir < \eten{4.5}$ \lsun\ (the CS 7-6 maps have too few sources
below this luminosity to make a reasonable fitting):

HCN 1-0:
$$Least\ Squares: log(L_{IR})=2.02(\pm0.13)\times {\rm
log}(M_{Vir}(R_{HCN1-0}))-1.43(\pm0.30); r=0.87 $$
$$Robust\ Fit: log(L_{IR})=1.71\times {\rm log}(M_{Vir}(R_{HCN1-0}))-0.81 $$
HCN 3-2:
$$Least\ Squares: log(L_{IR})=2.30(\pm0.25)\times {\rm
log}(M_{Vir}(R_{HCN3-2}))-0.98(\pm0.53); r=0.83 $$
$$Robust\ Fit: log(L_{IR})=0.97\times {\rm log}(M_{Vir}(R_{HCN3-2}))+1.69 $$
CS 2-1:
$$Least\ Squares: log(L_{IR})=2.46(\pm0.23)\times {\rm
log}(M_{Vir}(R_{CS2-1}))-2.44(\pm0.56); r=0.87 $$
$$Robust\ Fit: log(L_{IR})=1.59\times {\rm log}(M_{Vir}(R_{CS2-1}))-0.68 $$

When only clumps with L$_{IR} < 10^{4.5}$ are included in the fit,
the L$_{IR}$ increases steeply with the  mass of
dense gas with a power law ranging from $1.5$ to $2.5$.
Using the same averaging method yields a mean slope of $1.84\pm0.20$.

The surface density of a molecular clump should have an important impact on the
time scale and mass accretion rate of star formation in the dense
clump (McKee \& Tan 2002, 2003). Recent work by Krumholz \& McKee (2008)
proposed that only clouds with column densities of at least 1 g cm$^{-2}$
can form massive stars. We can test this idea on these dense clumps with
different dense gas tracers. We calculate the surface density within the FWHM
size of each transition, following the method used by Shirley et al. (2003)
for CS 5-4:
\begin{equation}
\Sigma_{transition}=\frac{M_{vir}(R_{transition})}{\pi R^{2}_{transition}}
\approx 0.665\frac{(M_{vir}/ 1.0 \times 10^{4} M_{\odot})}{(R_{transition}/ 1
pc)^{2}}
\end{equation}

The calculated surface densities are listed in Table \ref{virialtable}.
The surface densities
are lower than 1 g cm$^{-2}$ within the FWHM size of CS 2-1 and HCN 1-0
maps. For massive dense clumps, the
mean is 0.29 g cm$^{-2}$ for HCN 1-0, with a median 0.28 g cm$^{-2}$,
and 0.33 g cm$^{-2}$ for CS 2-1, with a median 0.31 g cm$^{-2}$.
Almost no source has surface density greater than 1 g cm$^{-2}$, when
measured by HCN 1-0 or CS 2-1.
Higher J transitions trace more compact, denser regions, resulting a larger
surface densities. The mean surface density within the FWHM contour of HCN 3-2
is 0.78$\pm$0.55 g cm$^{-2}$, with a median of 0.68 g cm$^{-2}$. This result
is similar to the result from CS 5-4 maps (the mean is 0.82 g cm$^{-2}$,
with a median of 0.60 g cm$^{-2}$ (Shirley et al. 2003).
CS 7-6 traces the most compact part of the
clumps in this survey, and its surface density is closest to the theoretical
threshold of 1 g cm$^{-2}$. The mean surface density within CS 7-6 clumps is
1.09 g cm$^{-2}$, with a median of 0.89 g cm$^{-2}$. The percentage of
sources that have surface density larger than 1 g cm$^{-2}$ is 26\%, 31\%,
and 38\% for clumps traced by CS 5-4, HCN 3-2, and CS 7-6, respectively.

As one might have expected, tracers with higher $n_{eff}$ trace higher
surface densities. Do the surface densities correlate with other
properties?  One might expect that
the higher the surface density, the larger the virial mass of the clumps
within the same FWHM area.
According to  Krumholz \& McKee's (2008) model,
a surface density of 1 g cm$^{-2}$ allows the gravity of the core
to overcome the radiation pressure of the new born stars, in order to
keep the accretion going and form massive stars inside.
If the surface density of a dense clump is high,
one may also expect that the star formation rate indicator L$_{IR}$ and the
indicator of star formation efficiency (L$_{IR}$/M$_{Vir}$) are also high.

\begin{figure}[hbt!]
\epsscale{0.70}
\rotatebox{270}{\plotone{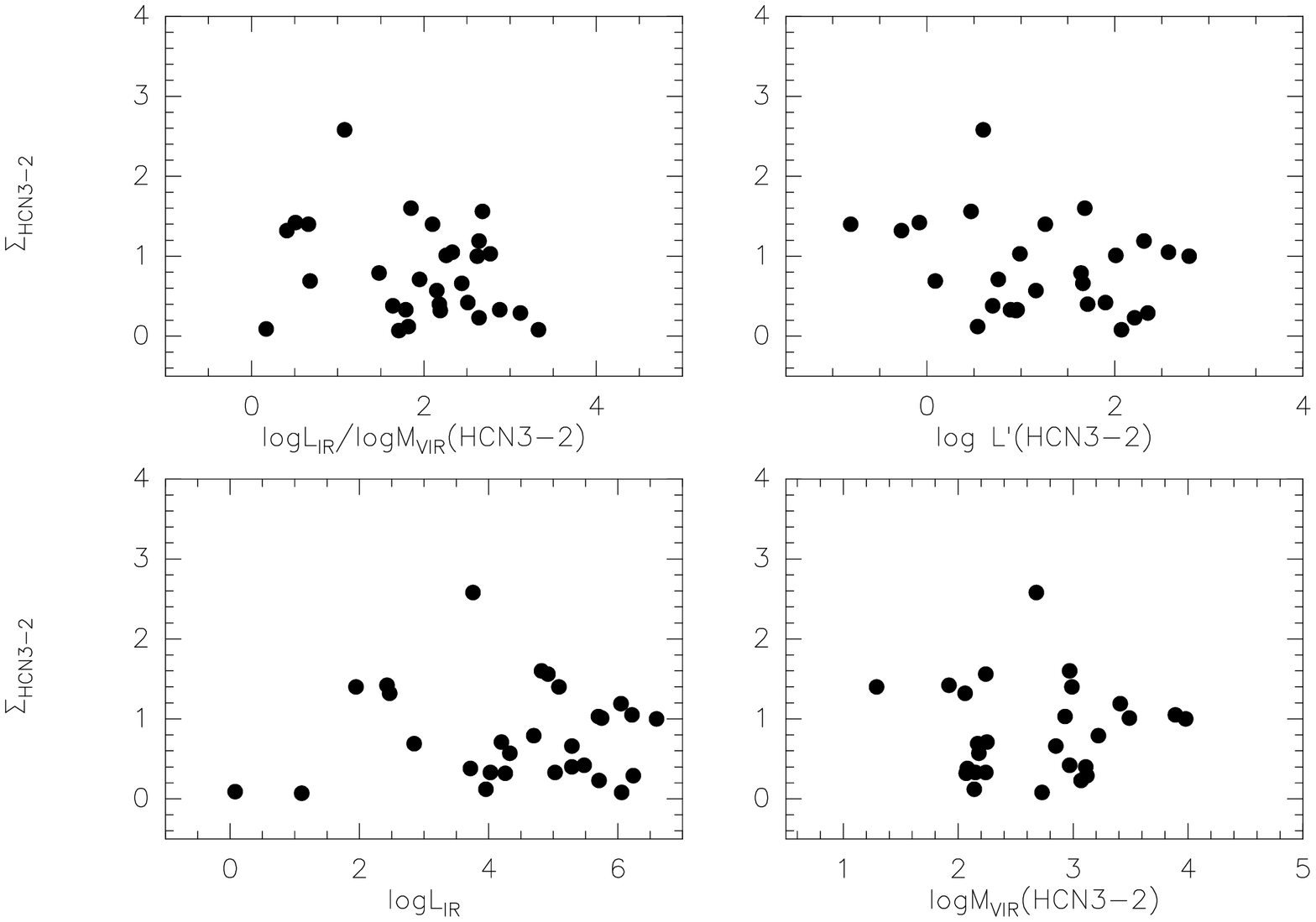}}
\caption{\label{lhcn32sigma4} Plots of the surface density vs. the
ratio of infrared luminosity to virial mass, infrared luminosity, line
luminosity, and virial mass derived from HCN 3-2 maps. }
\end{figure}

\begin{figure}[hbt!]
\epsscale{0.70}
\rotatebox{270}{\plotone{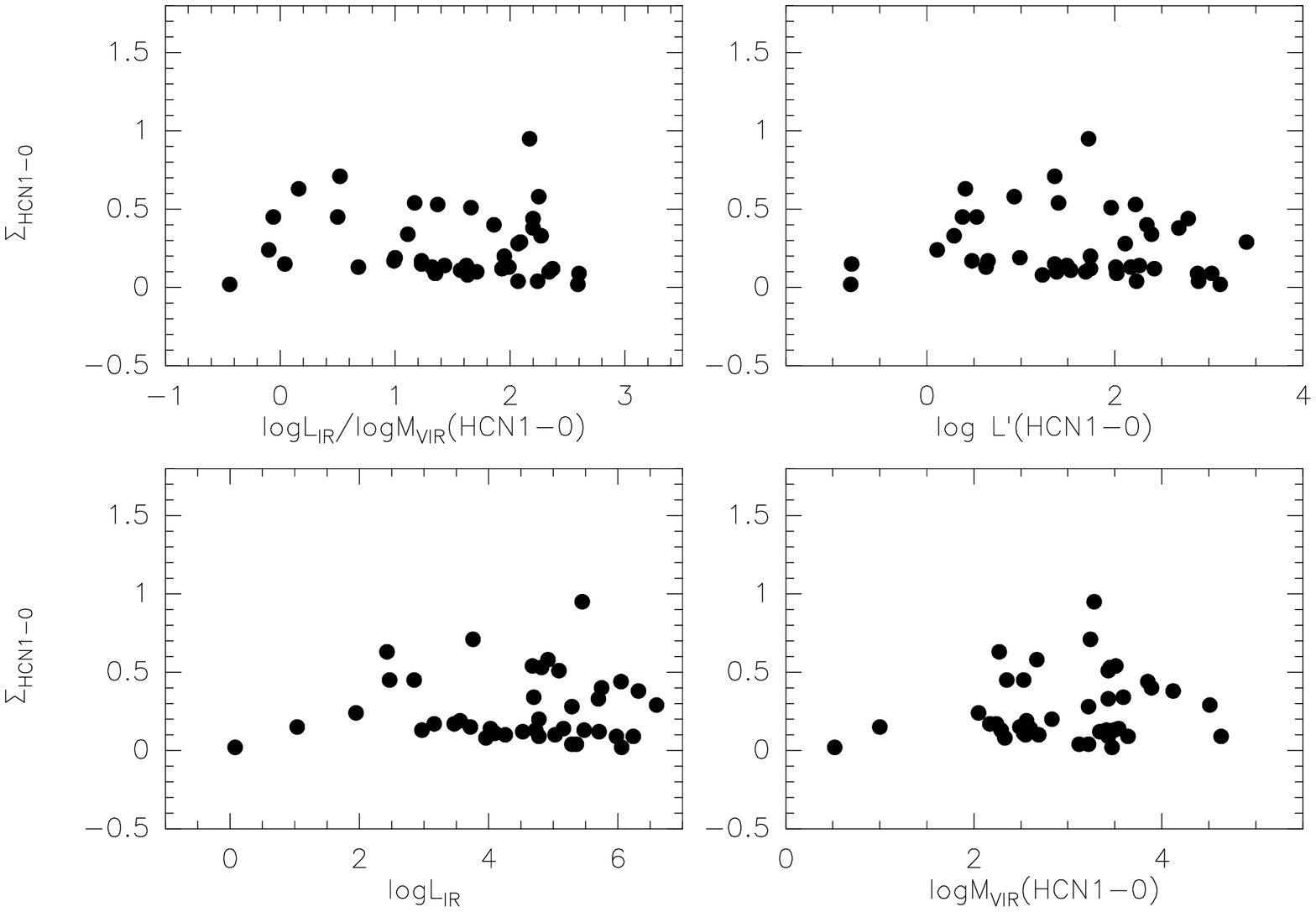}}
\caption{\label{lhcn10sigma4} Plots of the surface density vs. the
ratio of infrared luminosity to virial mass, infrared luminosity, line
luminosity, and virial mass derived from HCN 1-0 maps. }
\end{figure}

\begin{figure}[hbt!]
\epsscale{0.70}
\rotatebox{270}{\plotone{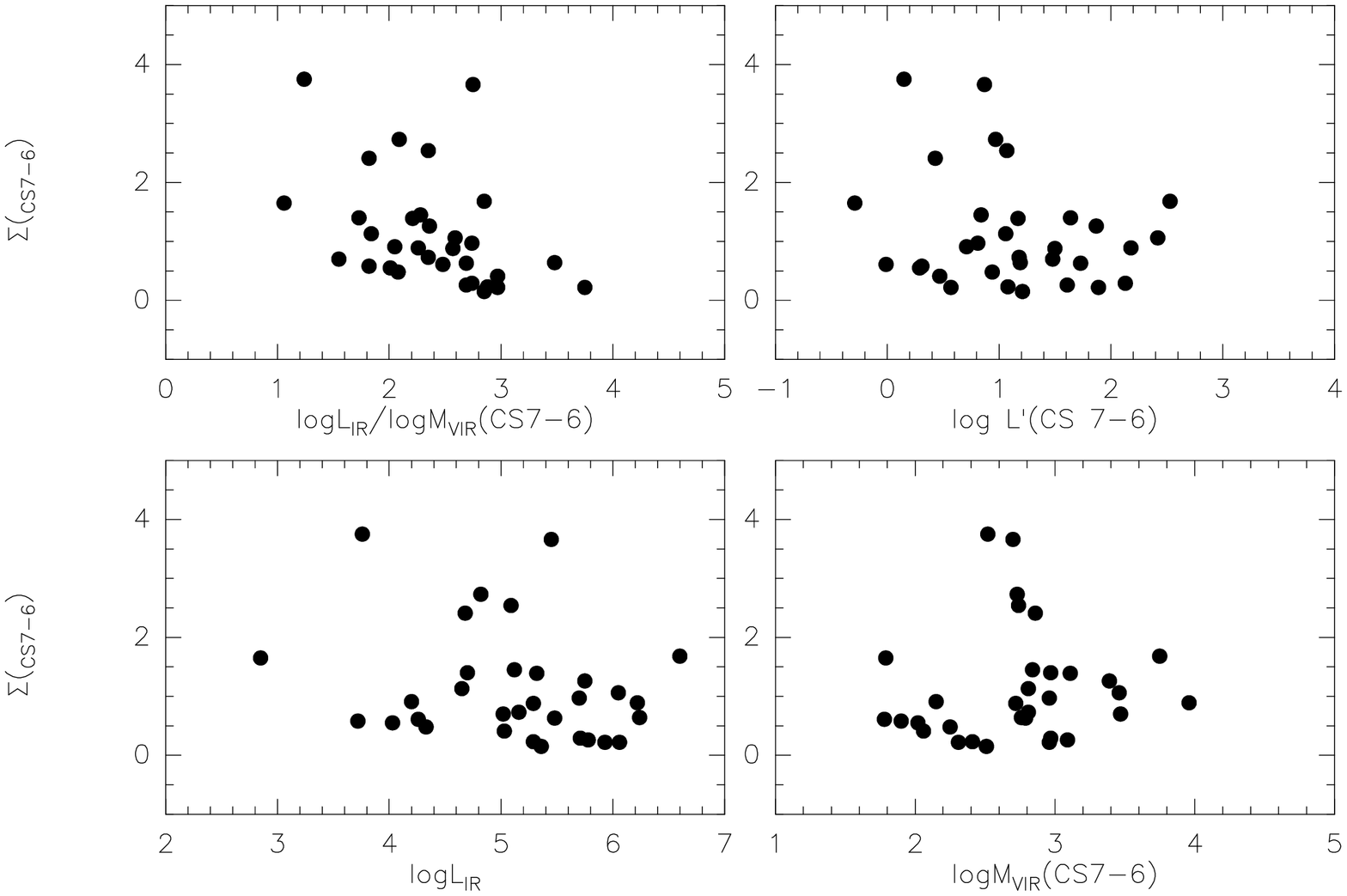}}
\caption{\label{lcs76sigma4} Plots of the surface density vs. the
ratio of infrared luminosity to virial mass, infrared luminosity, line
luminosity, and virial mass derived from CS 7-6 maps. }
\end{figure}

\begin{figure}[hbt!]
\epsscale{0.70}
\rotatebox{270}{\plotone{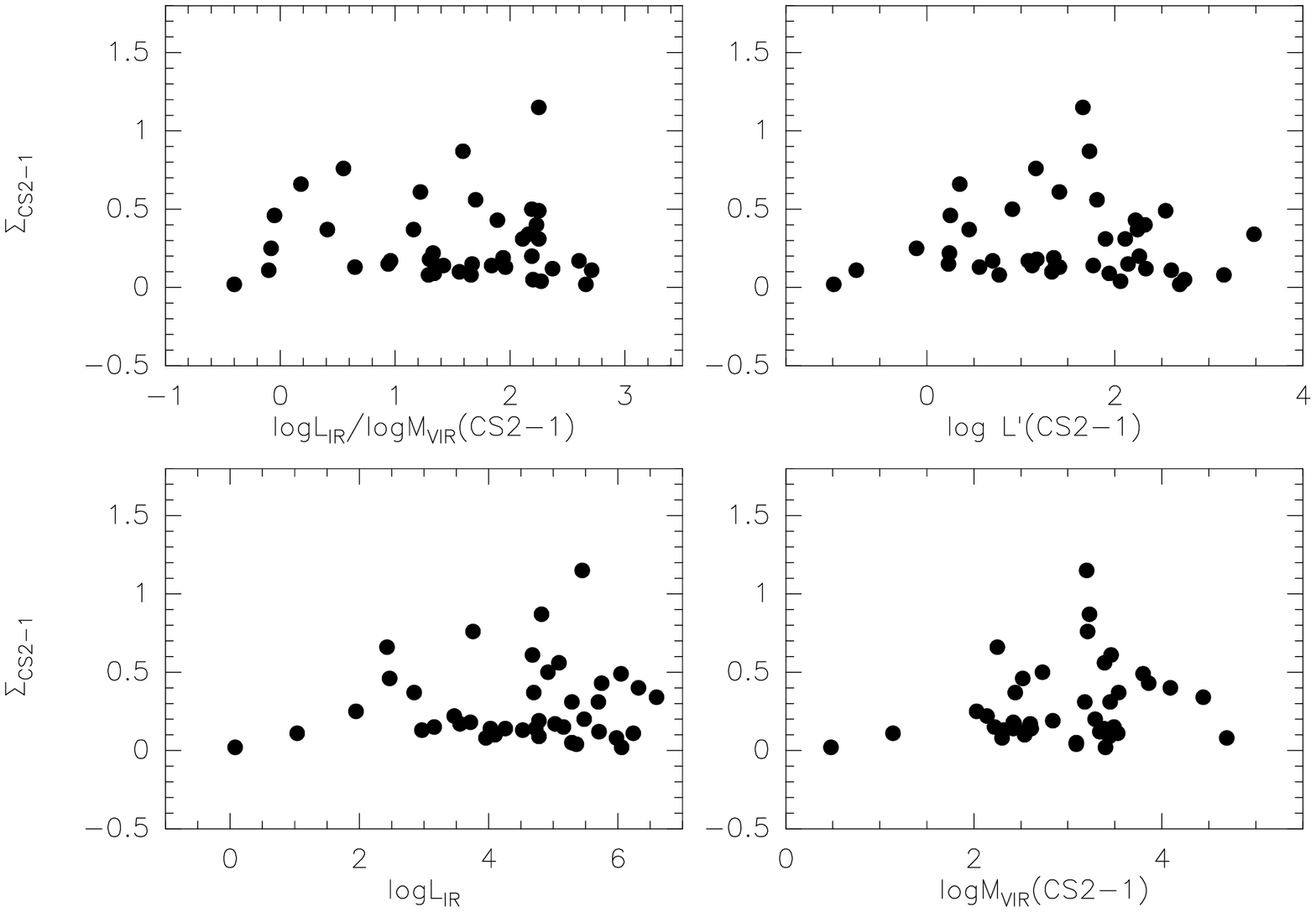}}
\caption{\label{lcs21sigma4} Plots of the surface density vs. the
ratio of infrared luminosity to virial mass, infrared luminosity, line
luminosity, and virial mass derived from CS 2-1 maps. }
\end{figure}

We check for any correlations between the surface density and
$M_{Vir}$, $L_{IR}$, $L_{IR}/M_{Vir}$, and line luminosity L$'$
for the four dense gas tracers (HCN1-0, HCN 3-2, CS 2-1, CS 7-6)
in Figure \ref{lhcn32sigma4} to \ref{lcs21sigma4}.
There are no obvious correlations found between the
surface density and these parameters for dense clumps, except for the
very lower end of the luminosity for a few low-mass cores, whose surface
density is lower than massive clumps. In particular, $L_{IR}$ ranges
over four orders of magnitude within a total variation of $\Sigma_{HCN3-2}$
of about a factor of 10 for dense clumps. We see no evidence for a sharp
threshold in surface density for massive dense clumps.

Of course, the threshold is probably not exact, and our values have
substantial uncertainty, so generally speaking, these massive clumps are in the
right range to form massive stars. With density gradients,
most of the dense clumps in our sample will reach the critical value of
1 g cm$^{-2}$ at some radius.  The amount
of mass above the threshold might help to determine the mass of the
stars that can form. The CS 7-6 line comes closest to being a tracer of
surface density of 1 g cm$^{-2}$. The mass inside the CS 7-6 contour is
on average about 0.25 of the mass inside the CS 2-1 contour and has
a mean value of 1300 \msun. Thus, these clumps on average have plenty of
mass at high enough surface densities to form clusters of massive stars.

The volume density of a cloud is often taken to control the star
formation rate of the cloud. We calculated the mean volume density
($n \equiv n(\hh) + n(He) + ... = \rho/\mu m_H$, with $\mu = 2.37$
(Kauffmann et al. 2008) within
the FWHM size of the clumps for each transition, according the the following
equation:
\begin{equation}
\nbar_{transition}=\frac{M_{vir}(R_{transition})}{\frac{4}{3}\pi \mu m_H R^{3}_{transition}}
\approx 4.08\frac{(M_{vir}/ M_{\odot})}{(R_{transition}/ 1
pc)^{3}}
\end{equation}
The derived volume densities for each transition are tabulated in Table
\ref{virialtable}.

The logarithmic averages over all clumps of the mean volume density
\mean{\rm{log} \nbar} are 4.1$\pm$0.1, 4.3$\pm$0.1, 5.1$\pm$0.1, and
5.4$\pm$0.1 for HCN 1-0, CS 2-1, HCN 3-2, and CS 7-6 respectively.
The uncertainties are the standard deviation of the mean.
In Plume et al. (1997), the volume density of a similar sample of clumps
 were derived by fitting multiple CS and C$^{34}$S spectral lines
towards the center position with an LVG model for the excitation.
The logarithmic average over the sample was $\mean{\rm{log}n_{LVG}}
=  5.9\pm0.2$.

The excitation densities (\nlvg) derived from the LVG model (Plume
et al. 1997) are larger than
mean densities (\nbar) derived from the mass and size for two reasons.
The value of \nbar\ is an average over all the gas within the FWHM contour
traced by a certain tracer. Even the CS 7-6 maps cover more than one to a few
beam sizes. The density derived from single point spectrum with an LVG model
was calculated from one beam measurement at the peak. Since the clumps are
centrally condensed, \nbar\ will be less than $n$ from the LVG model.
In addition, the excitation requirements of the hardest-to-excite transition
in the LVG fit push the fit to the higher densities in the region.
The fact that the values of \nbar\ from the highest excitation lines
are closest to the values from the LVG models is a reflection of this point.
These points were discussed in detail by Plume et al. (1997).

\begin{figure}[hbt!]
\epsscale{0.70}
\rotatebox{270}{\plotone{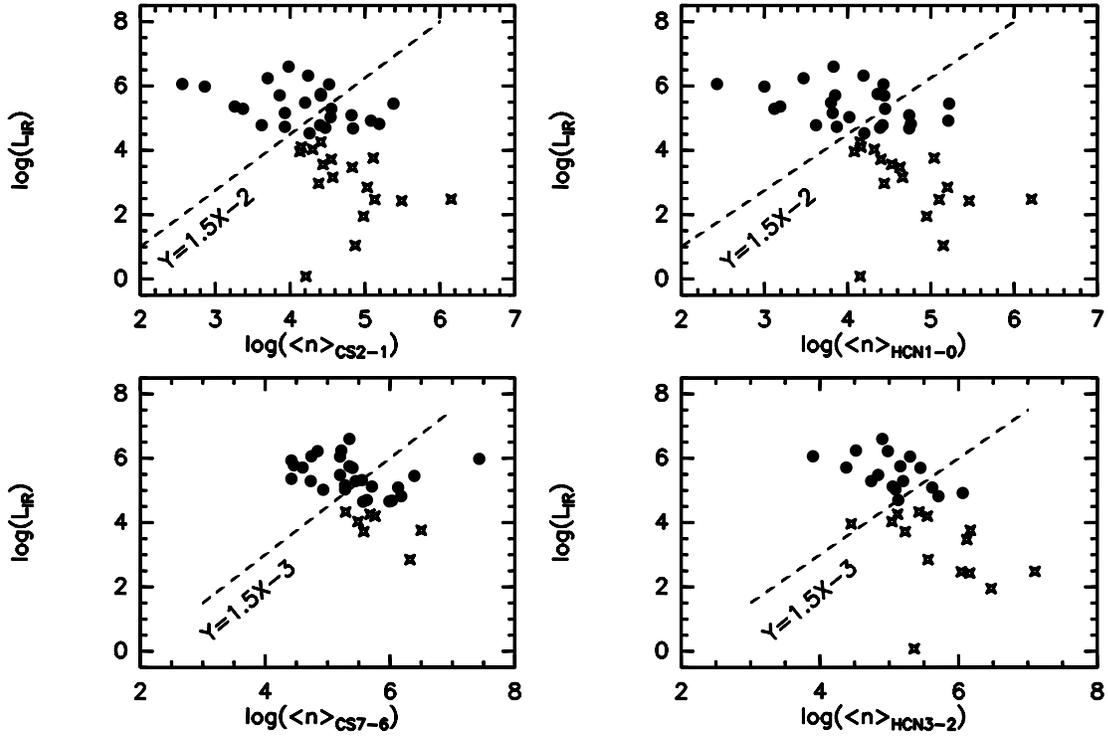}}
\caption{\label{lirnv4} Correlation between infrared luminosity and the
mean volume density \mean{n} (\nbar\ in the text) within the FWHM contour of
CS 2-1, CS 7-6, HCN 1-0, and HCN 3-2 maps. The dashed line has the slope
predicted by Krumholz \& McKee (2008) and Narayanan et al. (2008) and
an arbitrary offset. The clumps with L$_{IR}$ above and below the threshold luminosity 10$^{4.5}$L$_{\odot}$ are plot with solid circles and hollow stars, respectively.}
\end{figure}

If the star formation rate is proportional to the amount of gas and
inversely proportional to the free-fall time, a simple prediction is that
$\sfr \propto \nbar/\tff \propto \nbar^{1.5}$. This idea has been used
[e.g., Krumholz et al. (2007), Narayanan et al. (2008)]
to explain the basis of the observed Kennicutt-Schmidt relations for
star formation in galaxies.
Is the star formation rate proportional to the mean density to the
power of 1.5, as theorists have suggested? We plot the mean volume density
of each tracer vs. infrared luminosity in Figure \ref{lirnv4}.
Linear least square fits with uncertainty give the following results:
$$Least\ Squares: log(L_{IR})=-3.42(\pm0.28)\times {\rm log}(\nbar(R_{HCN1-0}))+
20.11(\pm1.26), r=-0.55 $$
$$Robust\ Fit: log(L_{IR})=-1.02 \times {\rm log}(\nbar(R_{HCN1-0}))+ 8.78 $$
$$Least\ Squares: log(L_{IR})=-5.28(\pm0.73)\times {\rm log}(\nbar(R_{HCN3-2}))+
33.35(\pm3.99), r=-0.57 $$
$$Robust\ Fit: log(L_{IR})=-1.2\times {\rm log}(\nbar(R_{HCN3-2}))+ 10.96 $$
$$Least\ Squares: log(L_{IR})=-3.46(\pm0.25)\times {\rm log}(\nbar(R_{CS2-1}))
+20.46(\pm1.14), r=-0.49 $$
$$Robust\ Fit: log(L_{IR})=-0.90\times {\rm log}(\nbar(R_{CS2-1}))+8.35 $$
$$Least\ Squares: log(L_{IR})=-1.18(\pm0.51)\times {\rm log}(\nbar(R_{CS7-6}))+
11.78(\pm0.30), r=-0.37 $$
$$Robust\ Fit: log(L_{IR})=-0.70\times {\rm log}(\nbar(R_{CS7-6}))+8.90 $$

\begin{figure}[hbt!]
\epsscale{0.70}
\rotatebox{270}{\plotone{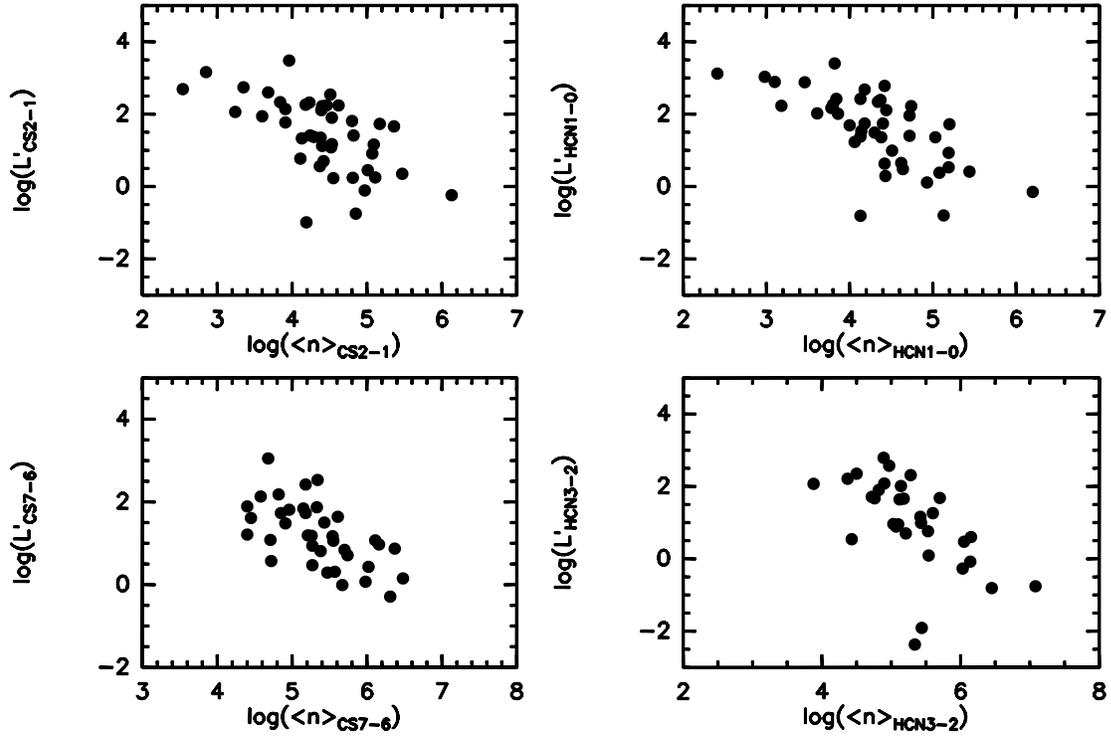}}
\caption{\label{lmol_nv4} Correlation between line luminosity and the
mean volume density \mean{n} (\nbar\ in the text) within the FWHM contour of
CS 2-1, CS 7-6, HCN 1-0, and HCN 3-2 maps.}
\end{figure}

Far from the prediction that  \nbar\  and L$_{IR}$ are positively correlated,
in general, L$_{IR}$ decreases with increasing \nbar. (There may be a hint
of positive correlation for the few low-luminosity cores with adequate data.)
Interestingly, all the fits from robust estimation have slopes near $-1$
with mean value over all tracers of $-0.96\pm 0.21$.
Since most massive clumps are large, they have lower {\it mean} densities;
this fact probably explains the negative correlation if luminosity is just
proportional to mass above a threshold density. A similar effect is
seen in the plot of L$'_{mol}$ versus \nbar\ (Figure \ref{lmol_nv4}).
This result is also plausible because L$'_{mol}$ is almost
linearly correlated with L$_{IR}$ (see \S 5.2).

\begin{figure}[hbt!]
\epsscale{0.50}
\rotatebox{270}{\plotone{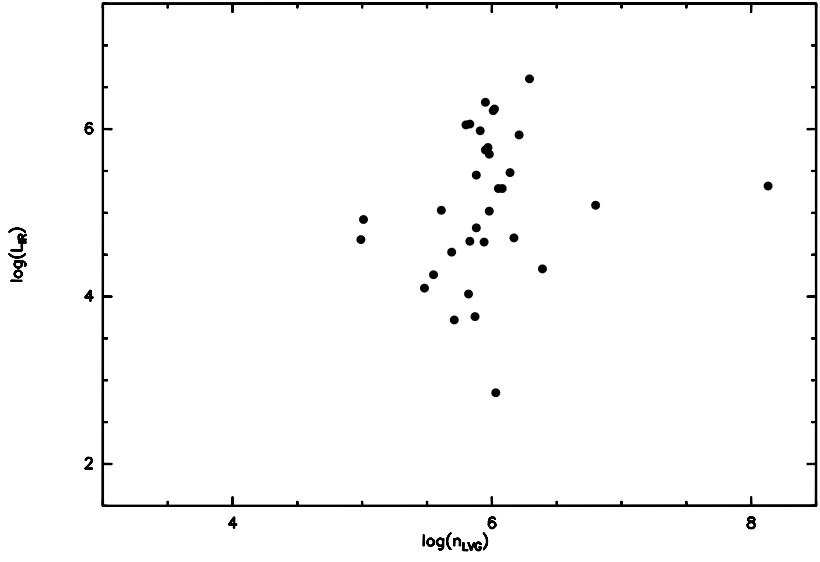}}
\caption{\label{lirnv97} L$_{IR}$ vesus volume density derived from LVG model (Plume et al. (1997). }
\end{figure}

We can also plot L$_{IR}$ versus \nlvg, which traces
the denser part of a clump with a density gradient.
The plot of L$_{IR}$ vs. \nlvg\ (Figure \ref{lirnv97}), reveals a very
weak positive correlation with L$_{IR}$. The linear least square fit with
uncertainty gives
$$ log(L_{IR})=15.7(\pm10.1)\times {\rm log}(n(LVG))-87.9(\pm119.3), r=0.18 $$

We see no sign of $L_{IR}  \propto n^{1.5} $ for these massive
dense clumps, regardless of how density is defined.

\begin{figure}[hbt!]
\epsscale{0.70}
\rotatebox{270}{\plotone{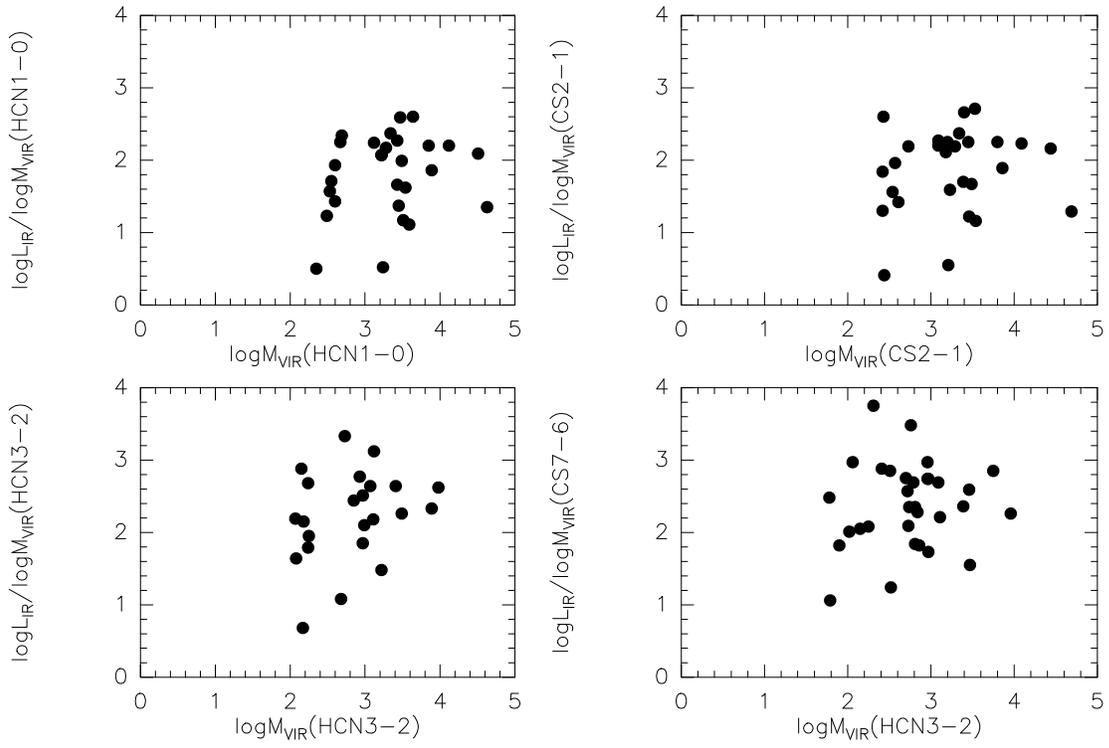}}
\caption{\label{lirdmvir}The star formation efficiency indicated by the
infrared luminosity per unit of dense gas mass vs. virial mass.  }
\end{figure}

We also ploted the correlation between the star formation 
efficiency (indicated by the infrared luminosity devided by rivial mass) and 
the virial mass for different tracers (Figure \ref{lirdmvir}). It shows no 
trend between them

Comparing the correlations between $L_{IR}$ and virial mass, surface density,
and volume density for dense clumps, we found that only the virial mass has a
strong correlation with $L_{IR}$; the correlations between $L_{IR}$ and
surface density or mean volume density are weak. It suggests that the most
relevant parameter for the star formation rate in a dense clump is how much
gas above a threshold density it contains. The surface density or mean
density within the clump are not the controlling factors.

\subsection{Do Line Luminosity Ratios Correlate with Other Properties?}

\begin{figure}[hbt!]
\epsscale{0.70}
\rotatebox{270}{\plotone{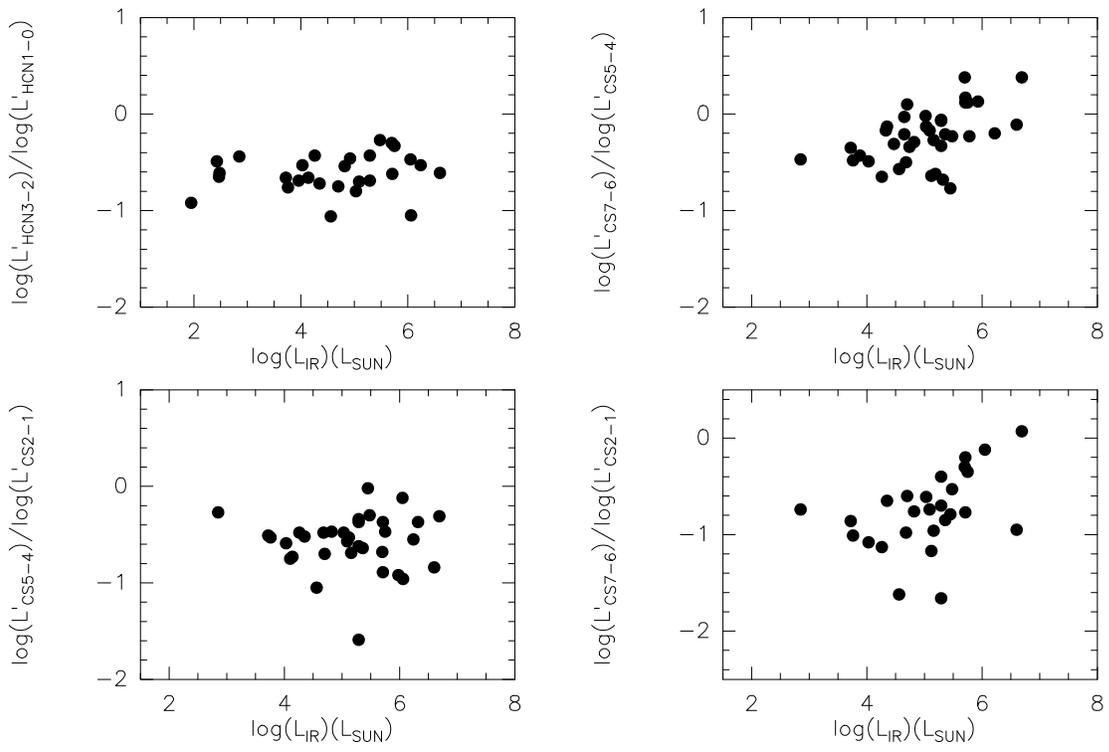}}
\caption{\label{lirlratio}Luminosity ratios between different J transitions
versus infrared
luminosity. }
\end{figure}

Table \ref{ratiotable} gives the mean, median, and standard deviation of
the ratios between the various lines.
The luminosity ratio of high J to low J transitions could reflect the
ratio of very dense gas to less dense gas in a clump.  They could also
depend on the infrared luminosity because a higher luminosity will result
in a greater fraction of the gas being warm enough to emit in the higher
J lines.  We have tested this latter possibility by plotting
line luminosity ratios versus infrared luminosity L$_{IR}$.
Figure \ref{lirlratio} indicates that only the ratio $\frac{L'_{CS 7-6}}{L'_{CS
5-4}}$ shows a weak correlation with L$_{IR}$.  It can be fitted by
a linear correlation:
$${\rm log}(\frac{L'_{CS7-6}}{L'_{CS5-4}})=0.18(\pm0.05)\times log(L_{IR})-1.11(\pm0.26).$$
The correlation coefficient r is 0.49.  Furthermore, the spread
in values of the ratio is modest for most ratios (Table \ref{ratiotable}).
The linear fit for other ratios versus $L_{IR}$ show weak or no correlations:
$${\rm log}(\frac{L'_{CS7-6}}{L'_{CS2-1}})=0.09(\pm0.10)\times log(L_{IR})-1.28(\pm0.53), r=0.17 $$
$${\rm log}(\frac{L'_{CS5-4}}{L'_{CS2-1}})=-0.01(\pm0.06)\times log(L_{IR})-0.54(\pm0.30), r=-0.03 $$
$${\rm log}(\frac{L'_{HCN3-2}}{L'_{HCN1-0}})=0.084(\pm0.03)\times log(L_{IR})-0.10(\pm0.14), r=0.46 $$
Thus, with the possible
exception of the CS 7-6 line, we find little dependence of the line
ratios on luminosity, suggesting that temperature differences are not
important for most ratios. In particular, the ratio of $\frac{L'_{HCN
1-0}}{L'_{HCN 3-2}}$ is nearly constant for these clumps, with much less
scatter than other ratios. These ratios are primarily sensitive to the
density structure for these massive clumps, and the uniformity of the
values reflects the fact that most dense clumps have power law density
distributions (Mueller et al. 2002, Shirley et al. 2003).

\begin{figure}[hbt!]
\epsscale{0.70}
\rotatebox{270}{\plotone{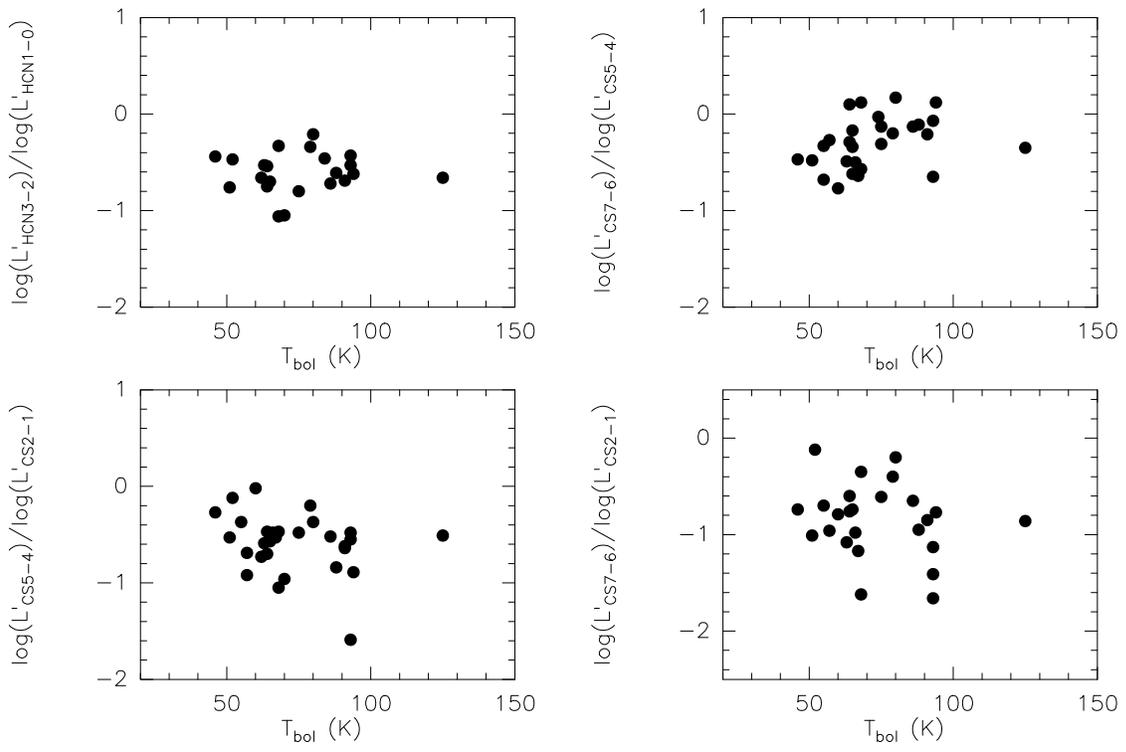}}
\caption
{\label{tblratio}Luminosity ratios between different J transitions versus
bolometric
temperature.}
\end{figure}

We can also correlate the line ratios with the
bolometric temperature (T$_{bol}$), which is the temperature of a blackbody
with the same mean frequency as the actual SED.
In low mass cores, \tbol\ is related to the evolutionary status of a core
(Chen et al. 1995), with lower \tbol\ corresponding to earlier stages.
In these sources, it might also reflect the relative amounts of gas
at higher and lower temperatures. The plots in Figure \ref{tblratio} show
little correlation between line ratios and
\tbol.  The linear least square fits to these correlations give:
$${\rm log}(\frac{L'_{CS7-6}}{L'_{CS5-4}})=0.005(\pm0.003)\times  \tbol\ -0.67(\pm0.22), r=0.32 $$
$${\rm log}(\frac{L'_{CS7-6}}{L'_{CS2-1}})=-0.005(\pm0.004)\times  \tbol\ -0.50(\pm0.32). r=-0.22 $$.
$${\rm log}(\frac{L'_{CS5-4}}{L'_{CS2-1}})=-0.005(\pm0.003)\times  \tbol\ -0.25(\pm0.24). r=-0.27 $$.
$${\rm log}(\frac{L'_{HCN3-2}}{L'_{HCN1-0}})=0.0006(\pm0.002)\times \tbol\ -0.65(\pm0.20). r=0.05 $$
This result suggests that the fraction of gas in various density
regimes does not depend strongly on evolutionary stage, at least for
those stages probed by our sample.

\begin{figure}[hbt!]
\epsscale{0.70}
\rotatebox{270}{\plotone{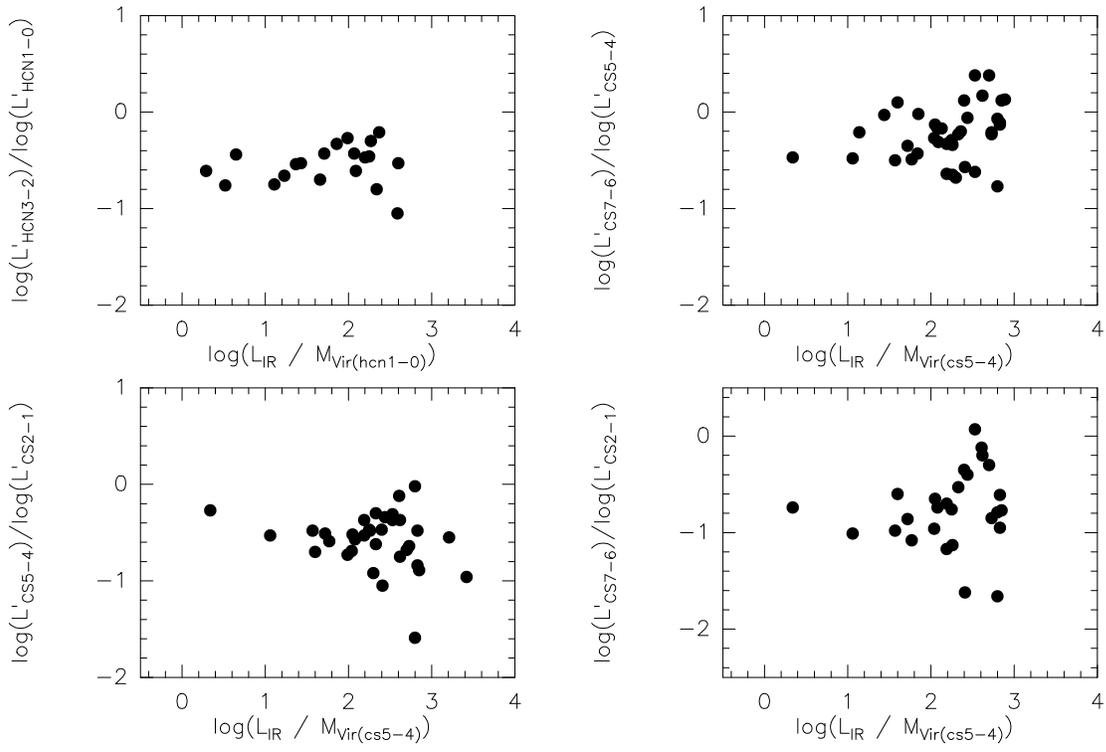}}
\caption
{\label{ldmhcn}Luminosity ratios between different J transitions versus star
formation efficiency. }
\end{figure}

Do the line ratios correlate with $L_{IR}$/M$_{vir}$, which is
a star formation efficiency indicator, if \lir\ is tracing star formation
rate? The plots in Figure \ref{ldmhcn} show no clear trends. These results
suggest that all of these tracers are probing gas that is dense enough
to be involved in massive star formation. The lower J lines simply trace a
larger region of somewhat lower density and thus have larger map sizes
and luminosities.

\begin{figure}[hbt!]
\epsscale{0.75}
\rotatebox{270}{\plotone{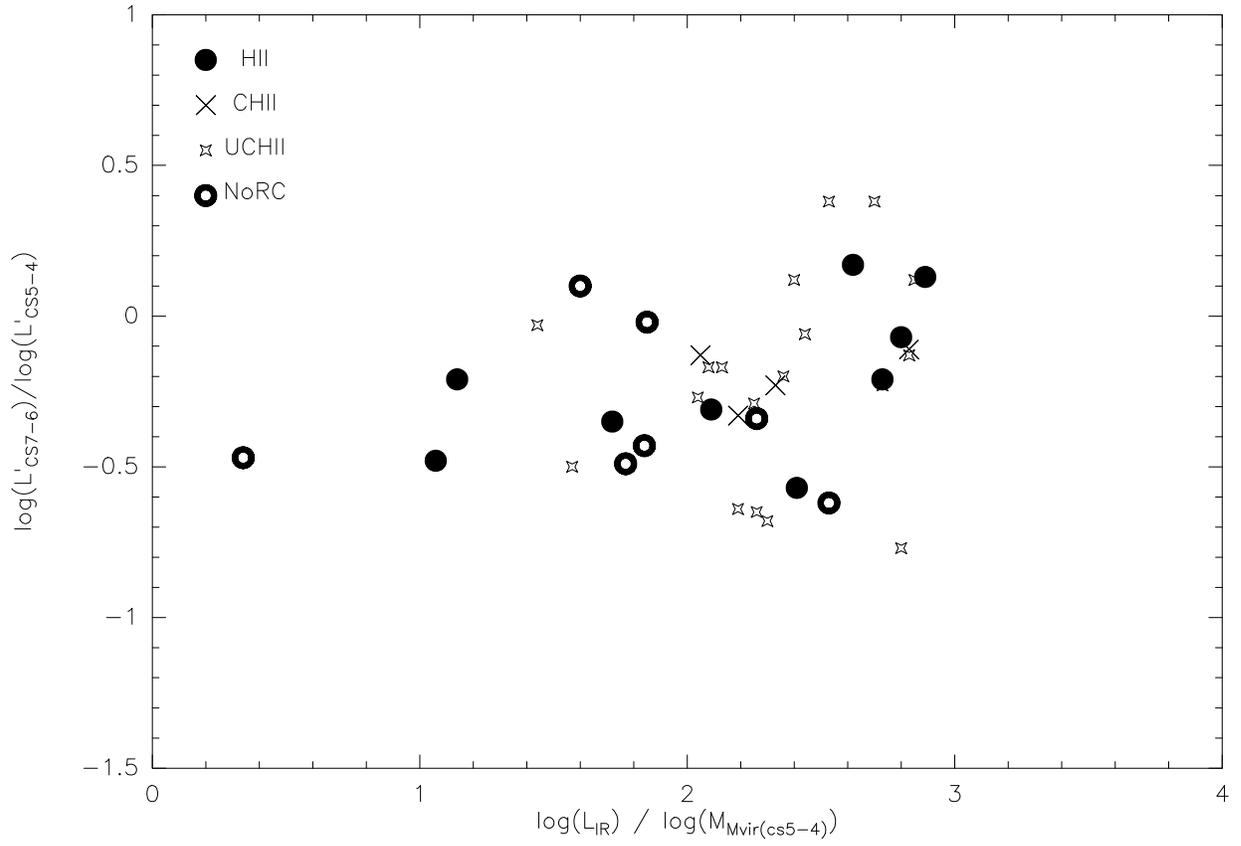}}
\caption
{\label{lirdmvl7654}The distribution of clumps at different evolutionary states
in the line luminosity ratio versus star formation efficiency plot. }
\end{figure}

Another approach to an evolutionary sequence for massive star-forming
regions is based on the evolution of HII regions.  In this picture,
a clump would evolve from having no radio continuum
(NoRC), to an Ultra-compact HII region stage (UCHII), to a
Compact HII region stage, and finally to a complete HII region stage.
In the paper of Shirley et al. (2003), an extensive literature search using
the SIMBAD database was performed to find HII regions associated with
these dense clumps. Using the taxonomy of Kurtz (2002), HII regions are
classified as UCHII if the diameter of 2 cm radio continuum emission
is $\leq$0.1 pc, CHII is the diameter is $\leq$0.5 pc, and an extended HII
region if the diameter is greater than 0.5 pc or clearly associated with a
classical HII region. The classification of the evolutionary states of these
dense clumps refers to Table 1 of Shirley et al. (2003).
If we classify the massive clumps in the sample according to this scheme, do we
see any trends?
We plot the line ratio of $\frac{L'(CS7-6)}{L'(CS5-4)}$ (the
ratio of two tracers that form a well matched sample in our survey, so we have
enough points to make the statistics) versus star formation
efficiency in Figure \ref{lirdmvl7654} with different symbols for each
HII region status.
No obvious differences emerge in the plot.
The mean, median, and standard deviation of this ratio for each evolutionary
stage are tabulated in Table \ref{ratio7654}. We see no obvious trend for the
ratio of $\frac{L'(CS7-6)}{L'(CS5-4)}$ along the evolutionary sequence, except
that the mean and median ratio is lower for sources with no radio
continuum emission than other three categories, indicating either that
the dense gas ratio is lower or the tempertures are lower in this stage.

\subsection{Dynamical Properties}\label{dynamics}

The dynamical state of molecular clumps is an important input into
star formation theories. As has been discussed many times, the linewidths
of massive clumps are much wider than expected for thermal broadening, and
some form of turbulence is clearly important.
For example, the characteristic temperature in these dense clumps is
less than about 100K. At this temperature, the sound speed is 0.7 \kms.
The mean FWHM linewidth from an optically thin tracer C$^{34}$S 5-4 for these
massive sources is 4.7 \kms. This gives a Mach number of 6.7. Thermal
broadening can only account for a small fraction of this broad linewidth.
The turbulence is clearly supersonic and seems to increase on smaller scales,
as discussed above.

Evidence for infalling gas can be seen in the form of line profiles
skewed to the blue in optically thick lines compared to optically thin lines.
Evidence for a significant excess of blue skewed profiles over red-skewed
profiles was seen in low-mass Class 0 (Gregersen et al. 1997) and Class I
sources (Gregersen et al. 2000). An excess was also found in a sample
of low-mass starless cores (Gregersen \& Evans 2000).
A recent study of somewhat more massive
starless cores in Orion found no excess of blue-skewed profiles
(Velusamy et al.  2008). While Velusamy et al. do not give masses of
their cores, they are drawn from the sample of Li et al. (2007), which
has a maximum mass of 46 \msun. These studies all refer to what are
now called cores, plausible precursors of single stars or small groups.
The massive clumps in this study are much more massive and are likely
to be the formation sites of clusters. In such sources, a blue-skewed
profile would not reflect infall onto a forming star, but {\it inflow}
toward the central parts of the clump.

The turbulence in these massive clumps makes detection of inflow motions
difficult, but some
sources have shown line profiles similar to those that are interpreted
as evidence of inflow in lower mass cores. In particular, we found
an excess of ``blue" profiles over ``red" profiles in initial observations
of the HCN 3-2 line
for this survey  (Wu \& Evans 2003), and a similar result was found using
\hcop\ by Fuller, Williams, \& Sridharan (2005).

Since we have observed more sources with the HCN 3-2 line and an optically
thin line (48) since the paper (Wu \& Evans 2003) on the statistics of
blue profiles of 28 massive clumps, we can reassess this situation.
The definitions of ``blue" and ``red" profiles have been discussed by
Wu \& Evans (2003) and references therein. Here we note only that inflow
{\bf may} produce an asymmetric line profile in an optically thick line,
skewing the peak toward the lower velocities relative to the peak of
an optically thin line. This offset, measured in units of the width of
an optically thin line, is defined by

\begin{equation}
\delta v = \frac{v_{thick} - v_{thin}}{\delta v_{thin}}.
\end{equation}

An absolute value of $\delta v \geq 0.25$ is considered significant.
In addition, the line profile may be double peaked
with the blue peak stronger than the red peak. We refer to this signature
as the blue peak. Since some profiles show the opposite asymmetries, one
can characterize a sample by the excess, defined as

\begin{equation}
E = \frac{N_{blue}-N_{red}}{N_{total}}
\end{equation}

We find that H$^{13}$CN 3-2 is already optically thick
in some sources that show absorption features, while C$^{34}$S 5-4 may be a
better optically thin tracer for calculating $\delta v$.
The mean ratio of the FWHM of
H$^{13}$CN $3-2$ to C$^{34}$S 5-4  for our sample is 1.27$\pm$0.32. So we use
C$^{34}$S 5-4 as the optically thin line to calculate $\delta v$ in this paper,
and we
use H$^{13}$CN $3-2$ only when the C$^{34}$S 5-4 line is not available.
The $\delta v$ values are listed in Table 4. Using this measure,
we identified 21 blue and 15 red profiles out of 48 sources; this leads to
$E=0.13$ compared to $E = 0.29$ in Wu \& Evans (2003).
Using the blue peak method, we identified
16 blue and 9 red profiles and $E=0.15$, compared to 0.29 in Wu \& Evans
(2003).
Thus, the inflow signature is not as common in the full sample
as it was in the initial results.
However, there are still more blue asymmetries than red, suggesting that
mass inflow is occurring in at least some of the sample.

For comparison, Fuller et al. (2005) found values of $E$ of 0.15 for the
\hcop\ 1-0 line and 0.19 for the \form\ \jkkjkk{2}{1}{2}{1}{1}{1}\
line in their
study of 112 objects. They found no significant excess in the 3-2 or 4-3
lines of \hcop.  There are only 3 sources in both the sample of Fuller et al.
(2005) and our sample: S231, G59.78+0.06 and IRAS20106+3545 (called WFS2, 79
and 93 in Fuller's sample). In both samples, S231 was classified as a red
profile and G59.78+0.06 as a blue profile (infall candidate).
We don't have optically thin line data for IRAS20106, but its HCN 3-2 line
show no asymmetry, which agrees with the observations by Fuller et al. (2005).

We also tabulate in Table \ref{evoltable} evidence for outflows based on the
outflow
catalog of Wu et al. (2005). Noted that this outflow catalog doesn't cover
all the sources in our sample. It is a collective catalog that provides all
the reported outflows by 2004. Therefore the following statistics of outflows
are based on the current available outflow reports, but not complete surveys.
Usually outflow and infall signatures are not present in the same source, but
there are obvious exceptions.
Of the 46 sources whose $\delta v$ are available, outflow has been
reported in 16. Of these, 8 have red asymmetry with $\delta v > 0.20$, 6 have
blue asymmetry with  $\delta v < -0.20$, in which outflows and infall
may occur at the same time. If we look the rate of the appearance of the outflow or infall
signature as the signature of ``dynamical activity'', the rate of the sources
that have infall or outflow to the total for each evolutionary stages are
55\% for regions without HII regions, 66\% for UCHII regions, 64\% for CHII
regions, and 33\% for fully formed HII regions. These numbers suggest that
the UCHII/CHII stages are more dynamically active than later stages.

\section{Star Formation Laws: Connecting to Extragalactic Star Formation}

A linear $L_{IR}$-$L'_{HCN 1-0}$ correlation has been reported (Gao \& Solomon
2004a) for nearby and
distant starburst galaxies. A similar correlation has been found for Galactic
clumps when studying CS 5-4
and dust emission maps (Shirley 2003, Mueller et al. 2002). However, the study
of the Kennicutt-Schmidt law with CO
and HI shows a  super-linear correlation (Kennicutt 1998, Kennicutt et al.
2007).  Why are these two correlations different?

Two distinct explanations have been offered. Gao \& Solomon (2004a) and
Wu et al. (2005) suggested that the line luminosity traces the mass of
dense gas; the linear relation between $L_{IR}$ and line luminosity then
reflects the fact that the star formation rate is linearly proportional to the
amount of gas above the threshold for star formation. The fact that
massive dense clumps in our Galaxy lie on the same relationship seen for
starburst galaxies indicates that starburst galaxies can be understood as
large collections of dense clumps, such as we observe in our Galaxy.
For a worked example of this idea, see Wu et al. (2009).

Krumholz \& Thompson (2007) and Narayanan et al. (2008) argue that the
underlying star formation law is superlinear and the linear relations
seen for dense gas tracers result from cancellation from a superlinear
dependence of star formation rate on density. As explained by Narayanan
et al. (2008), the following two relations,

$$ SFR \propto \rho^N $$

$$ \lmol \propto \rho^{\beta}, $$

can be combined to give

$$ SFR \propto \lmol^{\alpha},$$

with $\alpha = N/\beta$.
In this picture, the linearity for dense gas tracers ($\alpha \sim 1$)
results from the coincidence that $N$ and $\beta$ have similar values.
They predict that values of $\alpha$ will decline as higher excitation
lines are used because $\beta$ is larger for subthermal lines.
While Narayanan et al. (2008) are somewhat vague about the definition of
$\rho$, it is generally discussed in terms of mean density along the line
of sight, hence proportional to surface density.

The current surveys allow us to examine the correlations for
dense clumps in our Galaxy for five different tracers of dense gas.
For tracers with sufficient data on other galaxies, we can compare the
Galactic data on individual clumps to that for other galaxies.

\subsection{What do \lmol\ Measurements Trace?}

\begin{figure}[hbt!]
\epsscale{0.85}
\plotone{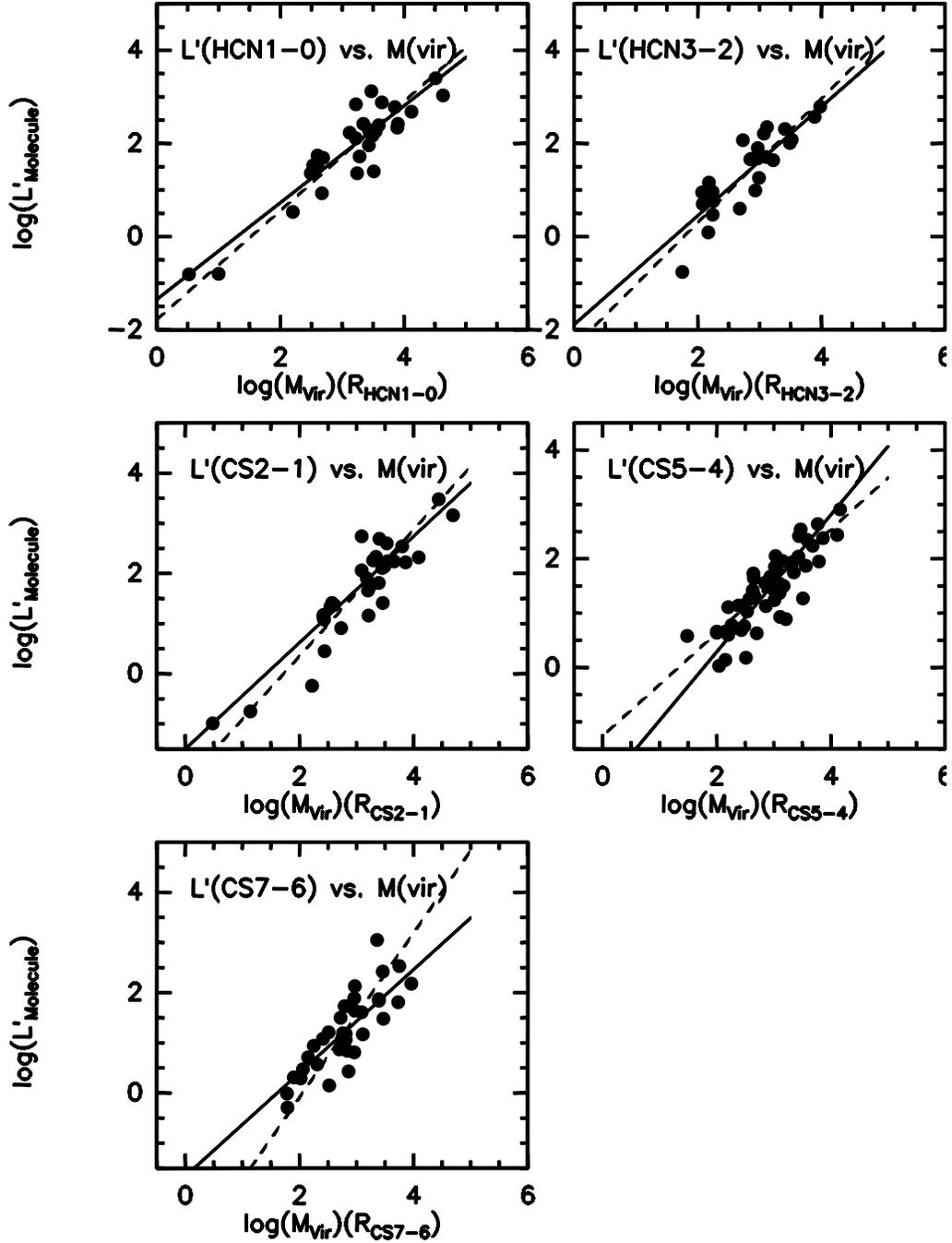}
\caption{\label{lmolmvir5} Correlation between the line luminosity of different
transitions and the virial mass within FWHM of contour maps. The dashed-line
shows the linear least squares fit with uncertainties; the solid line indicates
the robust fit.  Data of CS 5-4 comes from Shirley et al. (2003). }
!\label{f3}
\end{figure}

A first question is whether $L'_{HCN1-0}$ traces the mass of dense gas,
as suggested by Gao \& Solomon (2004a).
We plot the $L'_{HCN1-0}$-$M_{VIR}(R_{HCN1-0})$
correlation in figure \ref{lmolmvir5}. For comparison, we also plot the
$L_{molecule}$ vs. $M_{VIR}(R_{molecule})$ for the other four dense gas
tracers in figure \ref{lmolmvir5}. For all the panels, we show
fits based on least squares minimization (dashed line) and robust estimation
(solid line).
Here are the results from the fits:

HCN 1-0:
$$Robust\ Fit: log(L'_{HCN1-0})=1.04\times {\rm log}(M_{Vir}(R_{HCN1-0}))-1.35 $$
$$Least\ Squares: log(L'_{HCN1-0})=1.17(\pm0.05)\times {\rm
log}(M_{Vir}(R_{HCN1-0}))-1.78(\pm0.16); r=0.91 $$
HCN 3-2:
$$Robust\ Fit: log(L'_{HCN3-2})=1.17\times {\rm log}(M_{Vir}(R_{HCN3-2}))-1.89 $$
$$Least\ Squares: log(L'_{HCN3-2})=1.33(\pm0.06)\times {\rm
log}(M_{Vir}(R_{HCN3-2}))-2.36(\pm0.18); r=0.93 $$
CS 2-1:
$$Robust\ Fit: log(L'_{CS2-1})=1.06\times {\rm log}(M_{Vir}(R_{CS2-1}))-1.50 $$
$$Least\ Squares: log(L'_{CS2-1})=1.26(\pm0.05)\times {\rm
log}(M_{Vir}(R_{CS2-1}))-2.17(\pm0.16); r=0.91 $$
CS 5-4:
$$Robust\ Fit: log(L'_{CS5-4})=0.95\times {\rm log}(M_{Vir}(R_{CS5-4}))-1.26 $$
$$Least\ Squares: log(L'_{CS5-4})=1.26(\pm0.06)\times {\rm
log}(M_{Vir}(R_{CS5-4}))-2.23(\pm0.17); r=0.85 $$
CS7-6:
$$Robust\ Fit: log(L'_{CS7-6})=1.03\times {\rm log}(M_{Vir}(R_{CS2-1}))-1.66 $$
$$Least\ Squares: log(L'_{CS7-6})=1.64(\pm0.05)\times {\rm
log}(M_{Vir}(R_{CS7-6}))-3.37(\pm0.15); r=0.83 $$

The robust estimation fit consistently gives values near unity, indicating
linear relations between \lmol\ and \mvir. The least squares method yields
slight superlinear relations, with values around 1.25, except for CS 7-6.
More CS 7-6 maps on the lower luminosity may help to clarify
which fit is better. More importantly, these correlations are by far
the strongest we have plotted so far, confirming that
\lmol\ is an excellent tracer of \mvir\ for the dense gas.

In Table \ref{ratiomvdl} we give the statistics on the ratios of logarithms
of the virial mass to line luminosity for each tracer.
The scatter is about a factor of 3.
All these lines  provide good tracers for the mass of dense gas,
as required for the Gao \& Solomon interpretation. That statement is true
even though these
tracers are optically thick. The explanation is probably similar
to explanations given for the fact that CO traces the mass of lower density
material. Because CO is thermalized at densities above about 1000 \cmv\,
it is not sensitive to the dense gas that we are studying with these tracers.
Questions have been raised about the ability of HCN and CS to trace the
mass because they require high densities to excite (Krumholz \& Thompson 2007,
Narayanan et al. 2008). In our view, that is exactly their
virtue: they trace only the dense gas, which CO can fail to trace. The
density of the material they are tracing depends on their sensitivity to
density and temperature, as discussed above. What is clear is that they
trace the regions where clusters of stars, including massive stars, are
forming and are thus most relevant to estimating the mass of gas that is
immediately relevant for star formation.

The good correlation between \lmol\ and \mvir\ stands in sharp
contrast to the lack of correlation between surface density (a good
tracer for the mean density of the clump) and \lmol\ (Figures
\ref{lhcn32sigma4} to \ref{lcs21sigma4}). The spread in
\lmol\ for a given surface density is about three orders of magnitude.
We see no evidence in the data for the relation between \lmol\ and
mean density posited by Narayanan et al. (2008).

\subsection{Relations between \lir\ and \lmol}

The next question is whether the relations between \lir\ and \lmol\
seen in other galaxies (Gao \& Solomon 2004a) extend down to the scale
of individual clumps in our Galaxy. The initial results of our survey
of HCN 1-0 indicated that they did, except that there is a threshold,
below which there is not enough dense gas to sample the IMF.
Because \lfir\ is dominated by massive stars, it can trace the star formation
rate in the same way as galaxies, only with a reasonable sampling of the IMF.
Even then, it will be subject to large statistical fluctuations from
clump to clump, and it may underestimate the star formation rate because
it takes some time for massive stars to form (Krumholz \& Thompson 2007).
Therefore, we must expect substantial scatter in this relationship.

\begin{figure}[hbt!]
\epsscale{0.78}
\plotone{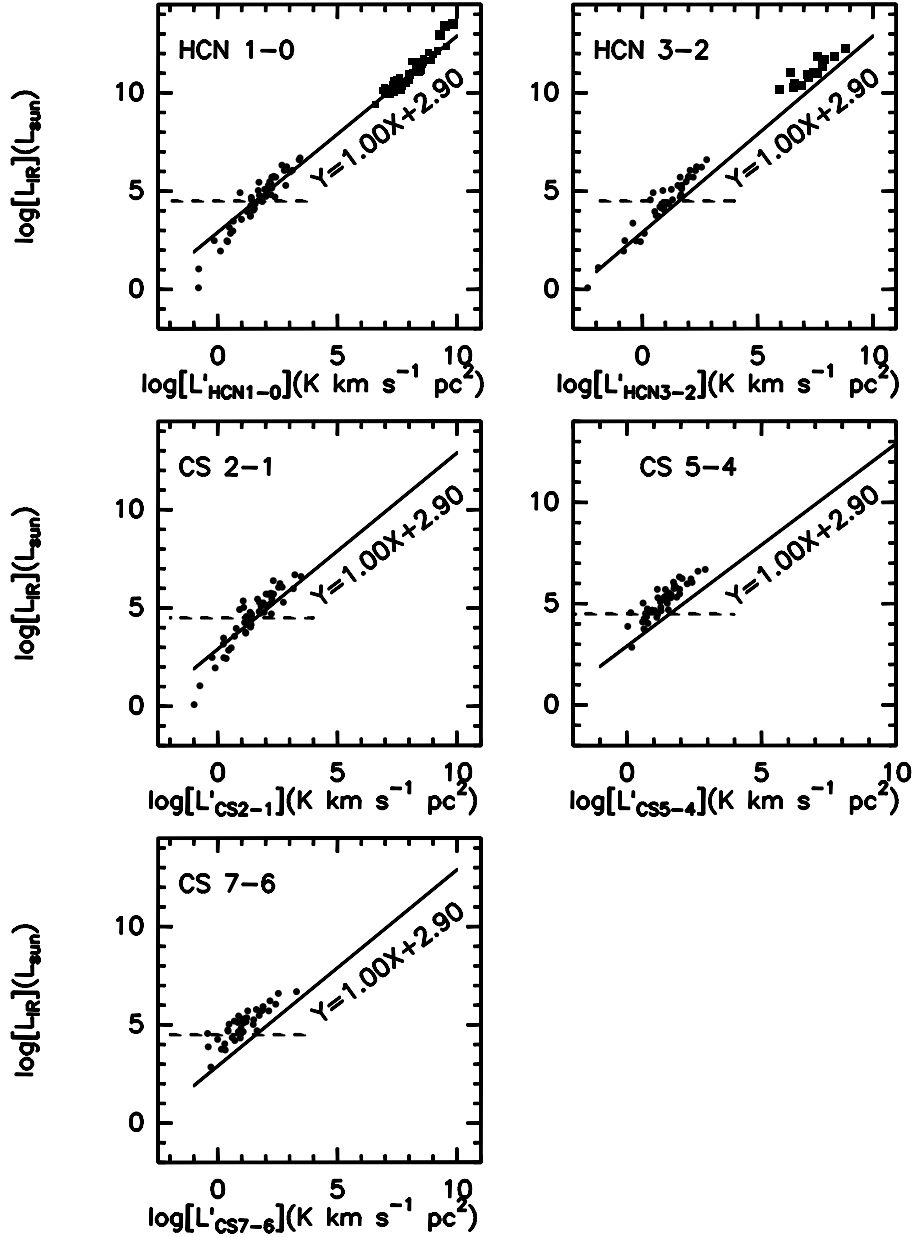}
\caption{\label{lirlmol5}L$_{IR}$-$L'_{mol}$ correlation for HCN 1-0, HCN 3-2,
CS 2-1, CS 5-4
and CS 7-6 Galactic clumps. Data of CS 5-4 are taken from Shirley et al. 2003.
The squares in the upper left panel are HCN 1-0
observations of galaxies(Gao \& Solomon 2004a) with 3 high-z detections (
Solomon et al. 2003, Vanden Bout et al. 2004, and Carilli et al. 2005). The
squares in the upper
right panel show the HCN 3-2 observations of galaxies (Bussmann et al. 2008).
The solid line in the first panel shows the best fit L$_{IR}$-$L'_{HCN1-0}$
correlation from galaxies,
and is shown in other plots to indicate the correlation shifts between HCN 1-0
and other tracers.
The dashed line shows where L$_{IR}$=10$^{4.5}$L$_{\odot}$ }
\end{figure}

\begin{figure}[hbt!]
\epsscale{0.78}
\plotone{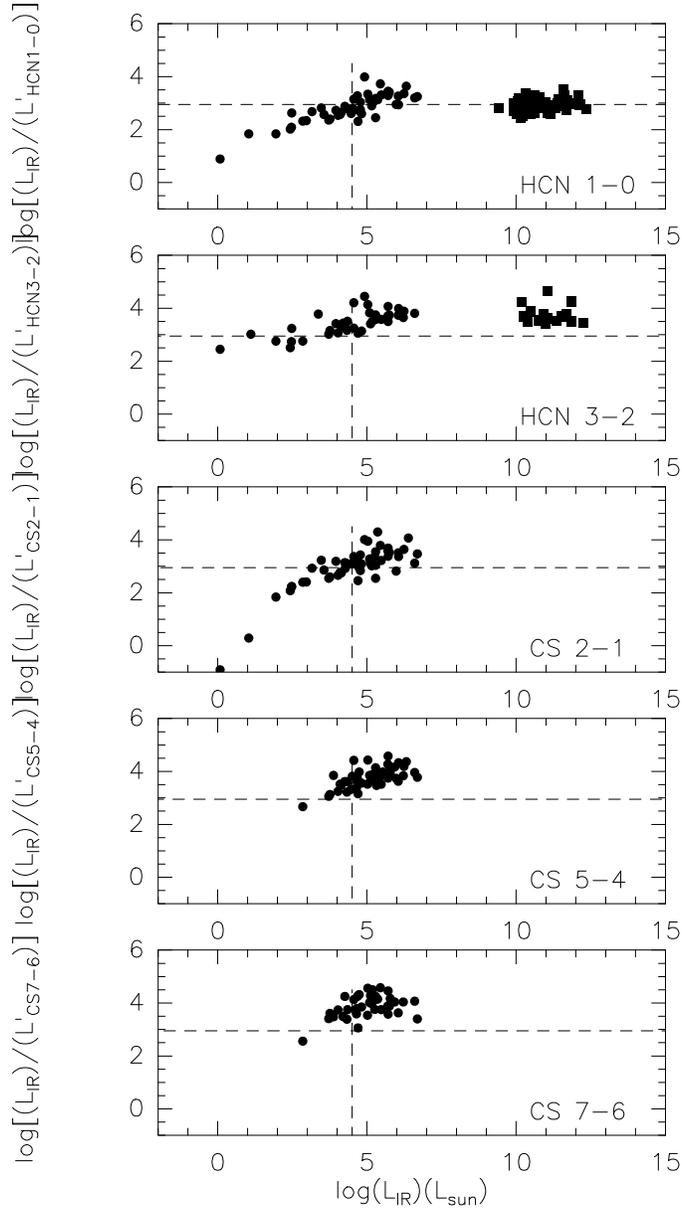}
\caption{\label{lirratio5}Correlations between the distant independent ratio
L$_{IR}$/$L'_{mol}$
vs. L$_{IR}$ for different tracers.
The squares in figures are HCN 1-0 observations of galaxies (Gao \& Solomon
2004a) in the
first panel, and HCN 3-2 observations of galaxies (Bussmann et al. 2008) in the
second panel.
The horizontal dashed line in the top plot indicates the averaged
L$_{IR}$/$L'_{HCN 1-0}$ ratio
for galaxies; the vertical dashed line in the top plot shows the cutoff at
L$_{IR}$=10$^{4.5}$L$_{SUN}$.
These two lines are also shown in other plots to indicate the relative shifts
of L$_{IR}$/$L'_{mol}$
between HCN 1-0 and other tracers.  }
\end{figure}

With these caveats in mind, we can ask two questions. Does \lfir\ correlate
with \lmol? Does the relation steepen for tracers with higher critical
density, as predicted by Narayanan et al. (2008)?
We now consider these questions for the four tracers in this survey, as
well as CS 5-4 mapping survey from Shirley et al. (2003).
In Figure \ref{lirlmol5} we plot \lir\ versus \lmol\ for HCN 1-0, HCN 3-2,
CS 2-1, CS 5-4 and CS 7-6. Figure \ref{lirratio5} shows the correlation
between the distance independent ratio
\lir/\lmol, an indicator of star formation efficiency, and \lir\
for different tracers.
Extragalactic data are lacking for the CS transitions, so we show the
extragalactic data only for the HCN transitions.
It is important to note that \lir\ is the same for each tracer, as it
is a property of the entire clump and is measured from emission in large
beams (e.g., IRAS) or by integrating over maps at submillimeter wavelengths.
This point has been missed by some analysts.

First we consider the plots using $L'_{HCN}(1-0)$ as the dense gas tracer..
The correlation between \lir\ and $L'_{HCN}(1-0)$ found in the
galaxies extends to the scale of dense clumps in our Galaxy, as originally
found by Wu et al. (2005).
There is a threshold \lmol, or equivalently, \lir, which is most easily
seen in Figure \ref{lirratio5}. For $\lir > \eten{4.5}$ \lsun,
all dense gas tracers show a nearly linear correlation with \lir.
Above the threshold,
a linear least squares fit with uncertainties in both axes gives
$$log(L_{IR})=1.07(\pm0.06)\times {\rm log}(L'_{HCN1-0})+2.98(\pm0.14); r=0.85$$
and a robust fit gives
$$(L_{IR})=1.03\times {\rm log}(L'_{HCN1-0})+3.12.$$
The fit to this relation in other galaxies is shown on Figure \ref{lirlmol5};
it is remarkably similar to that for massive, dense clumps in our Galaxy.

Data for other lines in galaxies are scarce.
The other panels on Figure \ref{lirlmol5} and \ref{lirratio5} show the
relations for the clumps in this study, along with the fit for the HCN
1-0 line in other galaxies. All show correlations with \lfir\ above the
threshold,
We give here (in order of increasing $n_{crit}$) the
linear least squares fit with uncertainties in both axes, and the robust fit
for other dense gas tracers for cores with $\lir > \eten{4.5}$ \lsun.

CS 2-1:
$$ Least\ Squares:log(L_{IR})=1.03(\pm0.05)\times {\rm log}(L'_{CS2-1})+3.25(\pm0.11); r=0.80 $$
$$ Robust\ Fit: log(L_{IR})=0.87\times {\rm log}(L'_{CS2-1})+3.56$$

CS 5-4:
$$Least\ Squares fit: log(L_{IR})=1.05(\pm0.05)\times {\rm log}(L'_{CS5-4})+3.77(\pm0.08); r=0.86 $$
$$Robust\ Fit:log(L_{IR})=0.86\times {\rm log}(L'_{CS5-4})+3.90$$

CS 7-6:
$$Least\ Squares fit: log(L_{IR})=0.81(\pm0.04)\times {\rm log}(L'_{CS7-6})+4.31(\pm0.06); r=0.81 $$
$$Robust\ Fit: log(L_{IR})=0.64\times {\rm log}(L'_{CS7-6})+4.58$$

HCN 3-2:
$$Least\ Squares fit: log(L_{IR})=0.88(\pm0.06)\times {\rm
log}(L'_{HCN3-2})+3.94(\pm0.11); r=0.82 $$
$$Robust\ Fit: log(L_{IR})=0.79\times {\rm log}(L'_{HCN3-2})+4.09$$

All the relations are close to linear. There is some tendency for the
tracers with higher $n_{crit}$ to have slopes less than unity, as
predicted by Narayanan et al. (2008). However they predict a slope of
about 0.7 (ranges from 0.63$\sim$0.73 from their Fig.7) for HCN 3-2
whereas we see $0.88\pm0.06$. They do not predict values
for CS transitions.

The ratio of \lir\ to \lmol\ depends on the tracer. The constant term in the
fit increases steadily as a function of excitation requirements.
As discussed in \S 4.1, the luminosity of the higher J tracers
is less for a given clump because of their higher requirements for both
density and temperature. It is then natural that the relation depends on
the tracer used. The dependence on critical density (or more generally
on excitation requirements) is primarily in this constant term (which
increases by an order of magnitude from CS 2-1 to CS 7-6) rather than in
the slope in the log-log plot.

\subsection{Comparing Explanations}

In order to explain the threshold and the linear correlation over a large
range of scales, we have proposed a model of clustered star formation
(Wu et al. 2005).  We suggested that there is
a basic unit of clustered formation. For $M_{dense}$ less than
the mass of this unit,
$L_{IR}/M_{dense}$ rises rapidly with $M_{dense}$, as higher mass stars can
form, and stellar luminosity rises rapidly with stellar mass.
For $M_{dense}$ greater than the mass of this unit, the IMF is reasonably
sampled and further increases in mass produce more units, but no further
change in $L_{IR}/M_{dense}$. If we suppose that larger scale cluster
formation is built up by adding more and more such units,
then the linear correlation between the total $L_{IR}$ and $M_{dense}$ is a
natural result, since $L'_{HCN}$ traces $M_{dense}$.
In that case, the only difference between star
formation on different scales and in different environments-- from
molecular clouds in our Galaxy to massive starbursts-- is just how
many such units they contain. Table \ref{statishm} and Table \ref{statisgt45}
contain the statistics of infrared luminosity, virial mass,
and L$_{IR}$/M$_{virial}$  of all the massive dense clumps,
and of the sample of massive
dense clumps with $\lfir > 10^{4.5}$ \lsun\ only, respectively.

The advantage of this model is that it can easily explain not only the linear
correlation, but also the luminosity threshold:  above a luminosity of about
\eten5\ \lsun, the IMF is reasonably well sampled, and the dependence
of stellar luminosity on stellar mass is less steep.
The luminosity threshold is set by the requirement to sample the IMF.
A threshold of $\lfir > 10^{4.5}$ \lsun\ is consistent with
a value ($\sim 10^{5}L_{\odot}$) that dominates the luminosity function of
the UCHII regions in the  Milky Way (Cassassus et al. 2000).
This model can explain the fact that starburst galaxies have a
very high star formation rate, but a similar star formation
efficiency with respect to the dense gas as individual massive dense
clumps in the Milky Way.
Extreme starbursts in galaxies happen when entire
molecular clouds become filled with dense gas, likely
due to the huge pressure in the extreme environments.

To summarize, the model proposed by Wu et al. (2005) starts with
what we know about well-studied regions of star formation in our Galaxy
and tries to extend that picture to other galaxies.

The alternative models assume that star formation laws derived from
extragalactic studies covering large scales can be extended to the
scales where stars actually form.  For example, Krumholz
\& Thompson (2007) assumed that a Schmidt law in the form
involving volume density describes star formation:
 $\dot \rho_{\star} \propto \rho_{g}^{1.5}$ over the full range of densities.
They further assumed that the distribution of densities in a molecular
cloud follows a lognormal probability distribution.
They argued that a molecular tracer essentially measures the mass of
gas with density equal to or greater than the critical density for that
transition. With a simple model of radiative transfer to correct the critical
density for trapping, they showed that these
assumptions led to a linear relation between star formation rate and
line luminosity of a transition when the mean density of the galaxy
was less than the critical density of the transition, as seen for HCN 1-0.
If, on the other hand, the mean density of the gas (\mean{n})
is greater than the
critical density of the tracer, then the $L_{IR}-L'_{tracer}$ correlation will
be super-linear, as is the case for CO. Narayanan et al. (2008) add to
this picture the idea that index is flattened by the increasing index
($\beta$) in the relation between \lmol\ and density.

The average density determined from CS excitation  of the massive clumps
in our sample is about $10^{5.9}$ \cmv\ (Plume et al. 1997), less than
the critical density of all the tracers in this study except for the
CS 2-1 line (Table \ref{property}), but greater than the effective density
(Table \ref{property})
and the density that was found to contribute most to the HCN 1-0 line
in the simulations of Krumholz and Thompson (2007). In fact, a density
derived from excitation analysis is biased towards the densest regions
and the mean density of the clumps in the sense of mass divided by
volume is generally less (e.g., Shirley et al. 2003).
As noted above, the relations we find do not support the suggestions
by Krumholz \& Thompson (2007) or Narayanan et al. (2008).

Another problem with the Krumholz-Thompson argument is that there is
no evidence in our Galaxy that the star formation density depends on
the local gas density in the way assumed for a Schmidt law. Rather,
there is a threshold density below which there is essentially no
star formation. Above that density, clumps and cores do not have lognormal
distributions, but instead have centrally condensed power law distributions
(Mueller et al. 2002, Shirley et al. 2003).

More detailed discussion about the origin of the $L_{IR}-L'_{tracer}$
correlation is beyond the scope of this paper. More discussion can be
found in recent reviews and papers (e.g.,
Shirley et al. 2008; Narayanan et al. 2008; Evans 2008; Baan et al. 2008; Inno et al. 2009; Juneau et al. 2009).

\subsection{Caveats: Different Beams and Different Distances}

The beam sizes used to map the lower excitation lines were
roughly twice as large as those used for the higher excitation lines
(Table 2). As shown by Liszt \& Linke (1975), derivation of densities
via excitation analysis with such data can be very misleading.
Densities from such an analysis with better-matched beams have been
obtained for this sample by Plume et al. (1997), and we have not
analyzed the current data in that way. Instead, we focused on the new feature
of these data, which are maps of multiple transitions. We treated each
transition separately and derived mean surface and volume densities
for each tracer (\S 4.4). We then discussed the difference between these
mean densities and those derived by Plume et al. (1997) from excitation
analysis.

\begin{figure}[hbt!]
\epsscale{0.70}
\rotatebox{270}{\plotone{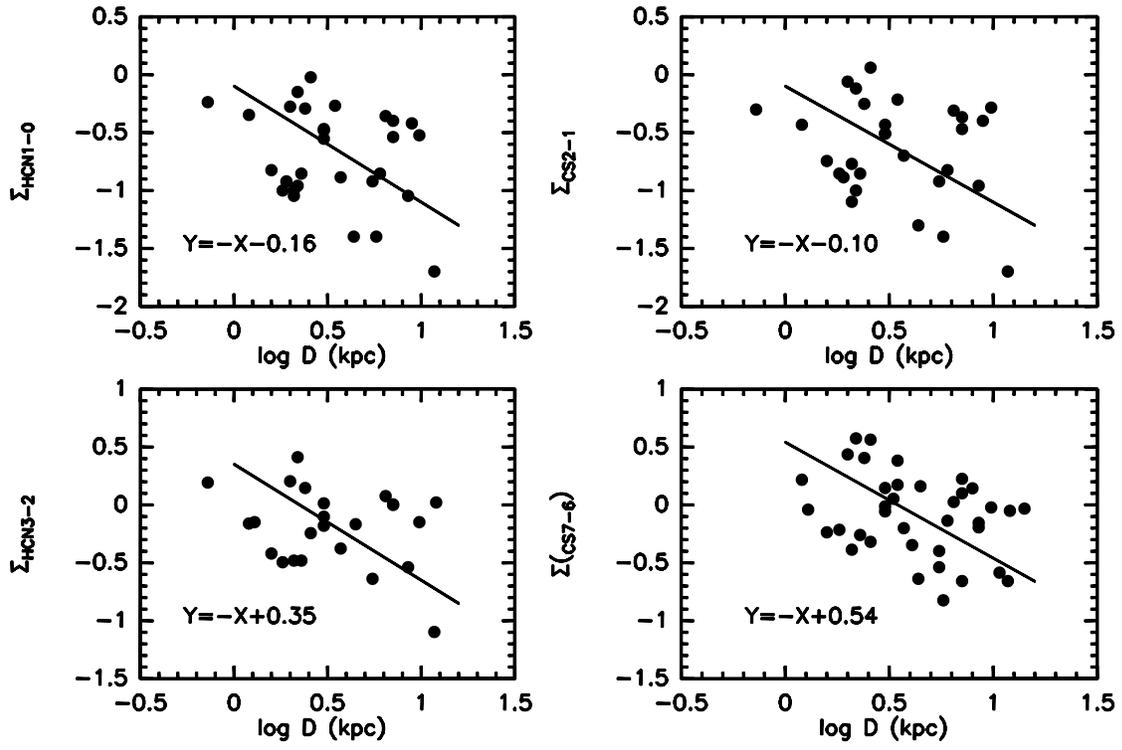}}
\caption{\label{logDS4} Log-log scale plots of the surface density vs.
distance for massive clumps for different transitions. The lines in the plots
have fixed slopes of $-1$ with different offsets.}
\end{figure}

\begin{figure}[hbt!]
\epsscale{0.70}
\rotatebox{270}{\plotone{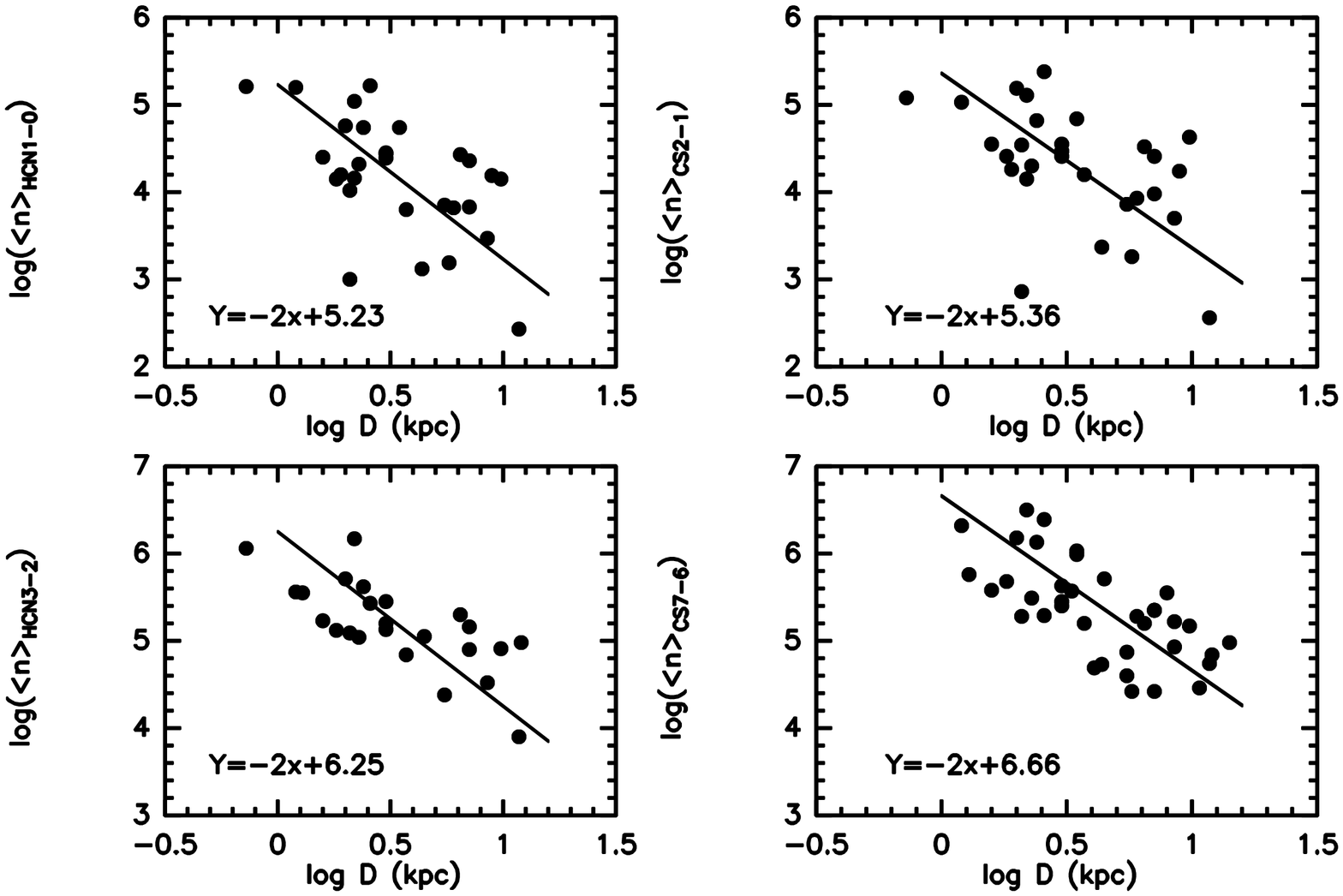}}
\caption{\label{logDN4} Log-log scale plots of the areraged volume density vs.
distance for massive clumps with different transitions. The lines in the plots
have fixed slopes of $-2$ with different offsets.}
\end{figure}

The sources in this study lie at a large range of distances (0.2 to 13.7 kpc).
This is an inevitable consequence of our attempt to capture the full
range of properties of massive dense clumps, as many of the most
extreme star-forming clumps are at large distances. The effective
linear resolution is much worse for the distant clumps, which could
bias the determinations of some quantities, such as surface and
volume densities, to lower values. Examination of equations 3, 5, and 6
would suggest that $\Sigma \propto d^{-1}$ and $\nbar \propto d^{-2}$
if there were no trends in mass with distance. To check the importance of
this effect, we plot in Figure \ref{logDS4} and Figure \ref{logDN4}
the logarithms of
surface and mean volume densities versus the logarithm of distance.
The lines in the plots have slopes fixed to the values suggested above and
only the offset is fitted. With substantial scatter, the expected relations
are seen. Note that these relations assume that there is no distance
effect on the clump masses. Sensitivity limits might
lead to more distant clouds being systematically more massive, partly
canceling the effects of decreasing resolution.

\begin{figure}[hbt!]
\epsscale{0.70}
\rotatebox{270}{\plotone{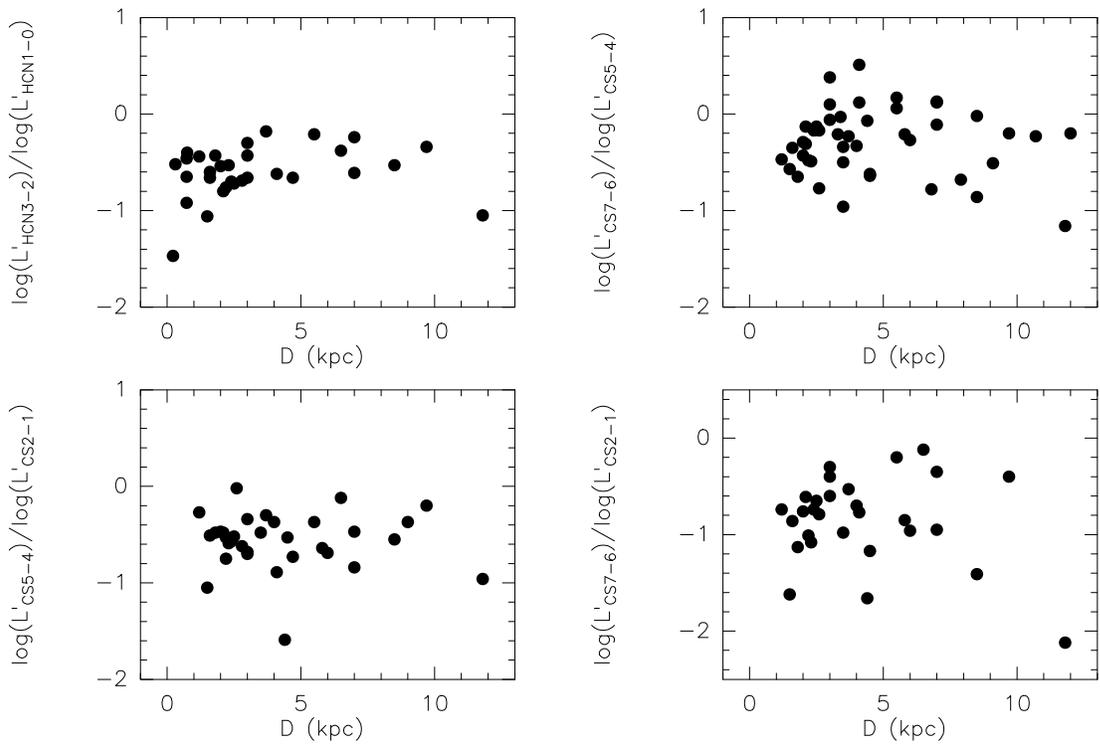}}
\caption{\label{DLratio4} Plots of the line ratios vs.
distance for massive clumps. No obvious trends are seen.}
\end{figure}

We would expect no strong trends with distance in line ratio plots,
and indeed none are seen (Figure \ref{DLratio4}).

\section{Summary}

Maps of over 50 massive, dense clumps in HCN and CS were obtained.
The main sample was drawn from the sample of Plume et al. (1992) selected
from water maser surveys and found to have CS 7-6 emission.
In addition, we mapped a smaller number of sources of lower luminosity
from other surveys. These are not included in statistics quoted for
massive, dense clumps.
We have presented maps of HCN 1-0 and 3-2, along with CS 2-1 and 7-6.
Together with previously published maps of CS 5-4, these provide the basis
for analysis of systematics of massive, dense clumps in tracers with a
wide range of excitation requirements. Upper state energies in temperature
units range from 4.3 K to 66 K, critical densities range from 3.9\ee5
\cmv\ to 6.8\ee7 \cmv, and effective densities range from 4.1\ee3 \cmv\
to 2.6\ee5 \cmv.

As excitation requirements increase, FWHM map sizes shrink, as expected
for centrally peaked density distributions. Interestingly, the linewidths
increase with excitation requirement, producing an inverse size-linewidth
relation, opposite to the one found for low density regions. The linewidths
are highly supersonic, with Mach numbers greater than 6. These facts suggest
that outflows and expanding HII regions are injecting turbulence on small
scales. An alternative would be increased inflow velocities.

Virial masses were computed for each of the tracers. Because the tracers
with higher excitation requirements have smaller FWHM sizes, the
virial masses derived from them are less. For the tracers that best
trace the whole clump, CS 2-1 and HCN 1-0, the mean and median
virial masses are 5000 and 1950 \msun\ (CS), or 5300 and 2700 \msun\ (HCN).
The mean and median infrared luminosities are 4.7\ee5 \lsun\ and
1.06\ee5 \lsun. These clumps include some of the most massive and
luminous star forming regions in the Galaxy, such as W43, W51, and W49.
For clumps with luminosity above \eten{4.5} \lsun,
$\mean{\lir} = (6.3\pm1.6)\ee5$ \lsun.

Mass surface densities are similar to those needed for high-mass
star formation, according to theoretical arguments (Krumholz \& McKee
2008). Even though only CS 7-6 indicates a mean surface density exceeding
their threshold of 1 g cm$^{-2}$ for massive star formation, the
observational and theoretical uncertainties are sufficient to suggest
that most of these clumps can form massive stars. And their high
luminosities indicate that indeed they are doing so.

Line profiles skewed to the blue, using the blue peak criterion,
are common (16 examples) in the sample,
but some (9) are skewed to the red. This excess of blue over red is less
than in the original sample of Wu \& Evans (2003), but it still suggests
the existence of large scale inflow motions in many clumps.

The line luminosities all correlate very well with virial mass, with
different offsets because of the fact that higher excitation transitions
trace smaller fractions of the core. There is no apparent correlation
of line luminosity with surface density or volume density,
contrary to some recent suggestions (e.g., Narayanan et al. 2008).
Line ratios show
little correlation with any other properties. In principle, ratios
of line luminosity
should trace density, but in a power-law density distribution,
densities range widely and line ratios tend to constant values.

The far-infrared luminosity correlates very well with all the line
luminosities. The correlations are close to linear, with some
marginal evidence for lower exponents for lines with the most
stringent excitation requirements. While that agrees qualitatively
with predictions by Narayanan et al. (2008), the effect is much
less than the predicted one.

Overall, the data support the idea that the linear relation between
\lfir\ and \lmol\ for starburst galaxies and molecular lines that are
hard to excite can be explained if such galaxies have large numbers
of massive, dense clumps similar to those in this study. In this
case, the star formation rate in starburst galaxies depends on the
total mass of gas above the threshold density or surface density needed
for massive star formation, rather than on a local, non-linear form of the
Schmidt law.

\acknowledgements
We thank D. Narayanan and M. Krumholz for lively discussions that
illuminated their models. Jingwen Wu thanks the
postdoctoral fellowship support from the Submillimeter Array of the
Smithsonian Observatory while completing this work.
This work was supported in part by NSF
Grant AST-0607793 and by McDonald Observatory.


\begin{deluxetable}{llrcccccccc}
\tabletypesize{\scriptsize}
\tablecaption{\label{sourcelist}Mapping Surveys towards Massive Clumps \tablenotemark{a} }
\tablewidth{0pt}
\tablehead{
\colhead{Source}                  &
\colhead{$\alpha$ (2000.0)}       &
\colhead{$\delta$ (2000.0)}       &
\colhead{dist}                    &
\colhead{ref. \tablenotemark{b}}  &
\colhead{\cs}                     &
\colhead{\cs \tablenotemark{c}}   &
\colhead{\cs}                     &
\colhead{HCN}                     &
\colhead{HCN}                     &
\colhead{350 $\mu$m \tablenotemark{d}}              \\
\colhead{}                        &
\colhead{}    &
\colhead{} &
\colhead{kpc}                     &
\colhead{}                        &
\colhead{2 $\rightarrow$ 1}                  &
\colhead{5 $\rightarrow$ 4}                  &
\colhead{7 $\rightarrow$ 6}                  &
\colhead{1 $\rightarrow$ 0}                  &
\colhead{3 $\rightarrow$ 2}                  &
\colhead{SHARC}
}
\startdata
G121.30$+$0.66 &  00 36 47.5 &  63 29 02.0  &1.2 & 1 & M & M & M & M  & M & M
\\
G123.07-6.31   &  00 52 25.2 &  56 33 53.2  &2.2 & 2 & M & M & M & M  & M & M
\\
W3(2)          &  02 25 40.6 &  62 05 51.1  &2.4 & 18&   &   &   &    &   & M
\\
W3(OH)         &  02 27 04.7 &  61 52 25.5  &2.4 & 3 & M & M & M & M  & M & M\\
GL490          &  03 27 38.5 &  58 46 57.9  &0.9 & 17&   &   &   &    & M & M\\
S231           &  05 39 12.9 &  35 45 54.0  &2.3 & 2 & M & M & M & M  & M & M  \\
S235           &  05 40 53.3 &  35 41 48.8  &1.6 & 2 & M & M & M & M  & M & M  \\
S241           &  06 03 53.6 &  30 14 44.0  &4.7 & 2 & M & M &   & M  & M & M\\
\enddata\\
\tablenotetext{a}{``M'' in the table indicates that the source has been mapped
in this transition}
\tablenotetext{b}{\footnotesize 1. R$_{N}$ Plume et al. 1992, 2. Blitz et al.
1982,
3. Harris \& Wynn-Williams 1976, 4.Genzel \& Downes 1977,
5. Braz \& Epchtein 1983, 6. Chini et al. 1986, 7. Hofner et al. 1996,
8. Val'tts et al. 2000, 9. Solomon et al. 1987,
10. Zinchecnko et al. 1994, 11. Hunter et al. 2000, 12. Churchwell et al. 1990,
13. Downes et al. 1980, 14. Wood \& Churchwell 1989, 15. Brand \& Blitz 1993,
16. Wink et al. 1982, 17. Zhou et al. 1996, 18. Plume et al. 1997, 19. Zhang et
al. 2005,
20. Wu et al. 2004, 21. Cernis 1990, 22. Dame \& Thaddeus 1985, 23. Kun \&
Prusti 1993.
}
\tablenotetext{c}{Mapping survey done by Shirley et al. (2003).}
\tablenotetext{d}{Mapping survey done by Mueller et al. (2002).}
\end{deluxetable}

\begin{table}[h]
\caption{\label{information}Observing Information}
\begin{tabular}{lccccc}
\hline

Transitions &  Dates &  Telescope &  frequency &$\theta_{mb}$& $\eta_{mb}$ \\
      &        &            &  (GHz)     &  (arcsec)   &
\\\hline
HCN 3-2   &  Jun 2002 & CSO  & 265886.188 & 28.1 & 0.64 \\
    &  Dec 2002 & CSO  & 265886.188 & 28.1 & 0.60 \\
    &  Jul 2003 & CSO  & 265886.434 & 28.1 & 0.76 \\
    &  Jan 2004 & CSO  & 265886.434 & 28.1 & 0.66 \\
    &  Jun 2004 & CSO  & 265886.434 & 28.1 & 0.64 \\
    &  Oct 2007 & CSO  & 265886.434 & 28.1 & 0.77 \\
CS 5-4    &  Sep 1996 & CSO  & 244935.610 & 24.5 & 0.56 \\
    &  Dec 1996 & CSO  & 244935.610 & 24.5 & 0.54 \\
    &  Apr 1997 & CSO  & 244935.610 & 24.5 & 0.56 \\
    &  Jun 1997 & CSO  & 244935.610 & 24.5 & 0.58 \\
    &  Dec 1997 & CSO  & 244935.610 & 24.5 & 0.56 \\
    &  Jul 1998 & CSO  & 244935.610 & 24.5 & 0.56 \\
    &  Jul 1999 & CSO  & 244935.610 & 24.5 & 0.60 \\
CS 7-6    &  Jul 1998 & CSO  & 342882.900 & 16.9 & 0.59 \\
    &  Dec 1998 & CSO  & 342882.900 & 20.5 & 0.61 \\
    &  Jun 2000 & CSO  & 342882.900 & 24.7 & 0.77 \\
    &  Dec 2000 & CSO  & 342882.900 & 24.7 & 0.63 \\
    &  Jun 2002 & CSO  & 342882.900 & 24.7 & 0.53 \\
HCN 1-0   &  Apr 2004 & FCRAO&  88631.847 & 58.0 & 0.50 \\
    &  Dec 2004 & FCRAO&  88631.847 & 58.0 & 0.50 \\
    &  Feb 2005 & FCRAO&  88631.847 & 58.0 & 0.50 \\
CS 2-1    &  Apr 2004 & FCRAO&  97980.968 & 50.0 & 0.50 \\
    &  Dec 2004 & FCRAO&  97980.968 & 50.0 & 0.50 \\
    &  Feb 2005 & FCRAO&  97980.968 & 50.0 & 0.50 \\
C$^{34}$S 5-4  &  Dec 1998 & CSO  & 241016.176 & 31.0 & 0.66 \\
    &  Jul 2001 & CSO  & 241016.176 & 31.0 & 0.73 \\
    &  Jan 2002 & CSO  & 241016.176 & 31.0 & 0.53 \\
    &  Jun 2002 & CSO  & 241016.176 & 31.0 & 0.59 \\
    &  Nov 2005 & CSO  & 241016.089 & 31.0 & 0.80 \\
H$^{13}$CN 3-2 &  Dec 2002 & CSO  & 259011.814 & 28.9 & 0.60 \\
    &  May 2003 & CSO  & 259011.814 & 28.9 & 0.58 \\
    &  Jun 2004 & CSO  & 259011.814 & 28.9 & 0.64 \\
    &  Jun 2006 & CSO  & 259011.814 & 28.9 & 0.80 \\\hline
\end{tabular}\\

\end{table}

\begin{deluxetable}{lrrrrrrrrrrrr}
\tabletypesize{\scriptsize}
\rotate
\tablecaption{\label{cstable}Observational Results of CS and C$^{34}$ Transitions}
\tablewidth{0pt}
\tablehead{
\multicolumn{1}{l}{SOURCE} &\multicolumn{4}{c}{CS 2 $\rightarrow$
1}&\multicolumn{4}{c}
{CS 7 $\rightarrow$ 6}&\multicolumn{4}{c}{C$^{34}$S 5 $\rightarrow$ 4}\\
\colhead{}             &
\colhead{T$_{R}^{*}$}  &
\colhead{$\int$T$_{R}^{*}$d$\nu$}  &
\colhead{V$_{LSR}$}    &
\colhead{FWHM}         &
\colhead{T$_{R}^{*}$}  &
\colhead{$\int$T$_{R}^{*}$d$\nu$}  &
\colhead{V$_{LSR}$}    &
\colhead{FWHM}         &
\colhead{T$_{R}^{*}$}  &
\colhead{$\int$T$_{R}^{*}$d$\nu$}  &
\colhead{V$_{LSR}$}    &
\colhead{FWHM}         \\
\colhead{}             &
\colhead{(K)}          &
\colhead{(Kkms$^{-1}$)}  &
\colhead{(kms$^{-1}$)}   &
\colhead{(kms$^{-1}$)}   &
\colhead{(K)}          &
\colhead{(Kkms$^{-1}$)}  &
\colhead{(kms$^{-1}$)}   &
\colhead{(kms$^{-1}$)}   &
\colhead{(K)}          &
\colhead{(Kkms$^{-1}$)}  &
\colhead{(kms$^{-1}$)}   &
\colhead{(kms$^{-1}$)}
}
\startdata
G121.30$+$0.66& 4.2(0.2)  &18.9(0.8)& -17.87(03)&  3.82(09) & 3.1(0.4)&
9.0(1.1)&-17.44(05) &2.58(13) & 0.2(.08)&   0.6(0.2)&-17.64(17) &2.87(37)  \\
G123.07-6.31  & 3.6(0.3)  &13.9(1.0)& -30.56(04)&  3.56(10) & 2.6(0.5)&
14.3(1.6)&-30.97(09) &5.33(21) & 0.2(.15)&   1.3(0.4)&-30.29(37) &5.35(108) \\
G135.28$+$2.80& ...       & ...     &   ...     &  ...      & 0.9(0.3)&
2.8(0.8)&-71.49(09) &2.47(26) & ...     &   ...     &   ...     &  ...     \\
W3(OH)        & 6.7(0.1)  &36.0(0.6)& -46.80(01)&  5.02(03) & 8.1(0.7)&
50.9(2.4)&-47.65(04) &6.05(10) & 1.0(0.1)&   5.9(0.5)&-47.49(08) &5.50(20)  \\
S231          & 3.6(0.4)  &18.4(1.5)& -16.74(06)&  4.76(16) & 3.5(0.4)&
17.3(1.5)&-15.82(04) &4.38(11) & 0.5(.08)&   1.2(0.2)&-15.77(07) &2.48(15)  \\
S235          & 4.1(0.2)  &10.9(0.8)& -16.78(02)&  2.33(06) & 5.0(0.2)&
19.5(0.8)&-17.09(02) &3.38(04) & 0.8(0.2)&   2.0(0.5)&-16.93(09) &2.35(24)  \\
S241          & 2.5(0.1)  & 7.3(0.4)&  -8.74(02)&  2.63(05) & 0.6(0.2)&
1.9(0.5)& -8.79(11) &2.98(24) & ...     &   ...     &   ...     &  ...     \\
S252A         & 3.8(0.2)  &12.8(0.8)&   9.05(03)&  3.09(07) & 2.7(0.3)&
10.5(0.9)&  9.63(05) &3.57(11) & ...     &   ...     &   ...     &  ...     \\
\enddata\\

\end{deluxetable}

\begin{deluxetable}{lrrrrrrrrlr}
\tabletypesize{\scriptsize}
\rotate
\tablecaption{\label{hcntable}Observational Results of HCN 3-2 and H$^{13}$CN 3-2 Transitions}
\tablewidth{0pt}
\tablehead{
\multicolumn{1}{l}{SOURCE} &\multicolumn{4}{c}{H$^{13}$CN 3 $\rightarrow$
2}&\multicolumn
{4}{c}{HCN 3 $\rightarrow$ 2}\\
\colhead{}             &
\colhead{T$_{R}^{*}$}  &
\colhead{$\int$T$_{R}^{*}$d$\nu$}  &
\colhead{V$_{LSR}$}    &
\colhead{FWHM}         &
\colhead{T$_{R}^{*}$}  &
\colhead{V$_{peak}$}   &
\colhead{$\int$T$_{R}^{*}$d$\nu$}  &
\colhead{V$_{LSR}$}    &
\colhead{Line$^{a}$}         &
\colhead{$\delta\upsilon$}  \\
\colhead{}             &
\colhead{(K)}          &
\colhead{(Kkms$^{-1}$)}  &
\colhead{(kms$^{-1}$)}   &
\colhead{(kms$^{-1}$)} &
\colhead{(K)}          &
\colhead{(kms$^{-1}$)} &
\colhead{(Kkms$^{-1}$)}  &
\colhead{(kms$^{-1}$)}  &
\colhead{Profile}   &
\colhead{}
}
\startdata
G121.30$+$0.66 & 0.5(.11)&   1.8(0.3) &-17.77(10)& 3.06(25) &  4.1(.30)
&-18.42(22)&  22.6(1.1) &-17.00(22) &  db & -0.37(08) \\
G123.07-6.31   & 0.5(.11)&   2.8(0.3) &-30.79(13)& 4.87(32) &  2.5(.23)
&-32.75(22)&  22.5(1.0) &-28.90(22) &  db & -0.46(12) \\
W3(2)          & 0.7(.12)&   3.3(0.4) &-38.73(10)& 4.48(25) &  5.9(.29)
&-36.97(32)&  32.8(1.4) &-39.77(32) &  dr &  0.39(04) \\
W3(OH)         & 1.8(.13)&  10.1(0.5) &-47.16(05)& 5.30(12) &  6.6(.55)
&-49.23(22)&  51.3(2.3) &-46.15(22) &  db & -0.32(03) \\
G135.28$+$2.80 & ...     &   ...      &  ...     & ...      &  0.7(.21)
&-42.18(16)&   3.3(0.7) &-42.18(19) &     &           \\
GL490          & ...     &   ...      &  ...     & ...      &  2.4(.55)
&-13.18(22)&  11.5(1.7) &-13.18(11) &     &           \\
S231           & 0.8(.10)&   3.3(0.4) &-16.41(06)& 3.74(17) &  3.5(.23)
&-14.12(22)&  20.2(1.0) & ....      &  rbs&  0.67(07) \\
S235           & 1.1(.11)&   3.4(0.3) &-17.44(05)& 2.71(13) &  5.0(.31)
&-17.36(11)&  29.1(1.0) &-17.36(03) &     & -0.18(06) \\
\enddata\\
\tablenotetext{a}{Code of line profiles: db, double peak with blue strong;
dr, double peaks with red strong; d: double peaks with neither one obviously
stronger; flat, the line has a flat top;
rbs, blue shoulder with peak skewed to red;
brs, red shoulder with peak skewed to blue; ws: wings on both sides;
rw, red wing; sr, peak skews to red; br, peak skews to blue;
two, the line has two known velocity components. }

\end{deluxetable}

\begin{deluxetable}{lrrrrrrc}
\tabletypesize{\scriptsize}
\tablecaption{\label{hcn10table}Observational Results for HCN 1-0 Transition}
\tablewidth{0pt}
\tablehead{
\colhead{Source}                  &
\colhead{T$_{R}^{*}$ (K)}  &
\colhead{T$_{R}^{*}$ (K)}  &
\colhead{T$_{R}^{*}$ (K)}  &
\colhead{$\int$T$_{R}^{*}$d$\nu$ (Kkms$^{-1}$)}  &
\colhead{V$_{LSR}$ (kms$^{-1}$)}    &
\colhead{FWHM (kms$^{-1}$}         &
\colhead{Line$^{a}$}  \\
\colhead{}                        &
\colhead{(F=1-0)} &
\colhead{(F=2-1)} &
\colhead{(F=1-1))} &
\colhead{(all hpf Fs)} &
\colhead{(F=2-1)} &
\colhead{(F=2-1)} &
\colhead{profile}
}
\startdata
G121.30$+$0.66  &  0.9(0.2)  & 4.0(0.2)  & ...       & 24.9(1.2) & -18.31(33) &
...       & ab       \\
G123.07-6.31    &  0.9(0.2)  & 2.6(0.2)  & 1.8(0.2)  & 19.6(1.0) & -31.13(04) &
3.63(05)  &          \\
W3(OH)          &  1.9(0.1)  & ...       & ...       & 41.8(0.7) &   ...      &
...       & ab,bl    \\
S231            &  0.7(0.2)  & 2.3(0.2)  & 1.3(0.2)  & 20.1(1.2) & -17.26(06) &
4.21(07)  & bl       \\
S235            &  1.2(0.2)  & 4.3(0.2)  & 1.5(0.2)  & 19.0(1.0) & -17.17(02) &
2.37(03)  &          \\
S241            &  0.3(0.1)  & 1.5(0.1)  & 0.7(0.1)  &  7.1(0.5) &  -8.90(03) &
2.66(04)  &          \\
S252A           &  1.0(0.2)  & 3.5(0.2)  & 1.6(0.2)  & 21.6(1.1) &   9.05(03) &
3.28(04)  &          \\
RCW142          &  3.1(0.3)  & 5.6(0.3)  & 3.1(0.3)  & 73.2(1.6) &  16.73(06) &
5.31(07)  & bl       \\
\enddata\\
\tablenotetext{a}{Code of line profiles: ab, absorption features on the line;
bl, three components
are blended; bl2: Only F=2-1 and F=1-1 are blended; bf, the line can not be
well fitted by 3 Gaussian
components.}
\end{deluxetable}

\begin{deluxetable}{lrrrrrrrrrrrr}
\tabletypesize{\scriptsize}
\tablecaption{\label{sizetable}The FWHM Sizes of Maps}
\tablewidth{0pt}
\tablehead{
\multicolumn{1}{c}{SOURCE} &\multicolumn{2}{c}{HCN 3 $\rightarrow$
2}&\multicolumn
{2}{c}{HCN 1 $\rightarrow$ 0}&\multicolumn{2}{c}{CS 7 $\rightarrow$
6}&\multicolumn
{2}{c}{CS 2 $\rightarrow$ 1}\\
\colhead{}             &
\colhead{R(HCN3-2)}\tablenotemark{a}    &
\colhead{$\theta_{dec}$ \tablenotemark{b}}      &
\colhead{R(HCN1-0)}\tablenotemark{a}    &
\colhead{$\theta_{dec}$ \tablenotemark{b}}      &
\colhead{R(CS7-6)}\tablenotemark{a}    &
\colhead{$\theta_{dec}$ \tablenotemark{b}}      &
\colhead{R(CS2-1)}\tablenotemark{a}    &
\colhead{$\theta_{dec}$ \tablenotemark{b}}     \\
\colhead{}            &
\colhead{(pc)}                        &
\colhead{($\arcsec$)}            &
\colhead{(pc)}                        &
\colhead{($\arcsec$)}            &
\colhead{(pc)}                        &
\colhead{($\arcsec$)}            &
\colhead{(pc)}                        &
\colhead{($\arcsec$)}
}
\startdata
G121.30$+$0.66& 0.09(01) &41.3 & 0.13(01) &  62.4& 0.05(01) &17.0 & 0.16(02)&
76.0 \\
G123.07-6.31  & 0.11(02) &20.9 & 0.41(06) &  76.2& 0.08(01) &14.2 & 0.38(05)&
70.8 \\
W3(2)         &          &     &          &      &          &     &         &
 \\
W3(OH)        & 0.22(02) &37.1 & 0.59(04) & 101.9& 0.12(01) &20.5 & 0.54(04)&
93.3 \\
G135.28$+$2.80&          &     &          &      &          &     &         &
 \\
GL490         & 0.05(01) &24.2 &          &      &          &     &         &
 \\
S231          & 0.19(03) &34.0 & 0.43(06) &  77.8& 0.11(02) &20.1 & 0.44(06)&
79.3 \\
S235          & 0.14(02) &37.3 & 0.38(04) &  96.8& 0.10(01) &24.8 & 0.32(03)&
81.7 \\
\enddata\\
\tablenotetext{a}{Numbers in the parentheses are uncertainties of FWHM size in
the unit of 0.01 pc.}
\tablenotetext{b}{The angular FWHM size of the core deconvolved from the
beam size.}
\tablenotetext{c}{Statistics do not include sources below the
horizontal line in this table.}
\tablenotemark{*}{These sources are either unresolved, or derived sizes are
less than half of the beam
size for this transition. This may cause the underestimation of the line
luminosity. We will ignore
these luminosities in later statistics.}

\end{deluxetable}

\clearpage{}
\begin{deluxetable}{lrrrrrr}
\tabletypesize{\scriptsize}
\tablecaption{\label{lumtable}Luminosities of Massive Clumps \tablenotemark{a}}
\tablewidth{0pt}
\tablehead{
\colhead{Source}                  &
\colhead{log$L_{IR}$}   &
\colhead{L$'$(HCN1-0)  }  &
\colhead{L$'$(HCN3-2) }   &
\colhead{L$'$(CS7-6) }  &
\colhead{L$'$(CS5-4) \tablenotemark{b}}  &
\colhead{L$'$(CS2-1) }  \\
\colhead{}                            &
\colhead{(L$_{\odot}$)}                         &
\colhead{K kms$^{-1}$pc$^{2}$}         &
\colhead{K kms$^{-1}$pc$^{2}$}         &
\colhead{K kms$^{-1}$pc$^{2}$}         &
\colhead{K kms$^{-1}$pc$^{2}$}         &
\colhead{K kms$^{-1}$pc$^{2}$}
}
\startdata
G121.30+0.66 & 2.85 &   0.53(12) &   0.09(17) &  -0.29(26) & 0.18(13) &
0.45(15)\\
G123.07-6.31 & 3.76 &   1.36(16) &   0.60(15) &   0.15(18) & 0.63(14) &
1.16(14)\\
W3(OH)       & 5.09 &   1.96(06) &   1.26(08) &   1.07(08) & 1.24(10) &
1.81(09)\\
GL490        & 3.37 &            &  -0.41(17) &            &          &
\\
S231         & 4.03 &   1.49(16) &   0.96(17) &   0.29(17) & 0.78(15) &
1.37(15)\\
S235         & 3.72 &   1.36(12) &   0.70(13) &   0.31(15) & 0.66(12) &
1.17(10)\\
S241         & 4.14 &   1.49(14) &   0.83(27) &            & 0.66(26) &
1.39(22)\\
S252A        & 4.56 &   1.41(65) &   0.35(15) &  -0.43(19) & 0.14(15) &
1.19(23)\\
\enddata\\
\tablenotetext{a}{Values in this table are in log scale, except the mean and
median at the bottom. Numbers in the parentheses are uncertainties of the line
luminosity in the log scale, in the unit of 0.01. }
\tablenotetext{b}{Converted from Shirley et al. (2003).  }
\tablenotetext{c}{Statistics on massive dense clumps only, which does not
include the sources below the horizontal line. Values and uncertainties (in the
parentheses) are in the unit of Kkms$^{-1}$pc$^{2}$. }

\end{deluxetable}

\begin{table}
\caption{\label{ratiotable}Ratios of Sizes and Luminosities for Different Tracers$^{a}$}
\begin{tabular}{lccccc}
\hline
 & $\frac{R_{HCN 1-0}}{R_{HCN 3-2}}$ & $\frac{R_{CS 2-1}}{R_{CS 7-6}}$
&$\frac{R_{HCN 1-0}}{R_{CS 2-1}}$  &
$\frac{R_{CS 5-4}}{R_{CS 7-6}}$ & $\frac{R_{HCN 3-2}}{R_{CS 5-4}}$  \\\hline
Mean$^{b}$ & 2.97$\pm$0.19  & 4.12$\pm$0.45 & 1.11$\pm$0.03 & 1.42$\pm$0.09 &
1.11$\pm$0.04 \\
Median& 2.71          &  3.36         & 1.07          & 1.31          &   1.05
  \\
Sdv$^{c}$ & 1.06&  2.40 & 0.23 & 0.64 & 0.21 \\\hline
\hline
 & $\frac{L'_{HCN 1-0}}{L'_{HCN 3-2}}$ & $\frac{L'_{CS 2-1}}{L'_{CS 7-6}}$ &
$\frac{L'_{CS 2-1}}{L'_{CS 5-4}}$  &$\frac{L'_{CS 5-4}}{L'_{CS 7-6}}$ &
$\frac{L'_{HCN 3-2}}{L'_{CS 5-4}}$  \\\hline
mean$^{b}$& 4.3$\pm$0.4    & 9.4$\pm$2.1   & 4.0$\pm$0.4   & 2.6$\pm$0.4 &
1.6$\pm$0.1 \\
median   & 3.8            & 5.9          & 3.4            &1.7  &1.5 \\
Sdv$^{c}$& 2.5&  11.0 & 2.3 & 2.6 & 0.7 \\\hline
\end{tabular}\\
\tablenotetext{a}{Sizes of CS 5-4 maps come from Shirley et al. (2003). }
\tablenotetext{b}{The uncertainty is the sigma of the mean}
\tablenotetext{c}{The standard deviation of the ratios}

\end{table}

\begin{table}
\caption{\label{property}Properties of Different  Transitions}
\begin{tabular}{lccccccc}
\hline
Transition & n$_{crit}$ $^{a}$& E$_{up}$ $^{a}$ &
n$_{eff}$  $^{a}$& Mean L$^{'}$& Mean L$^{'}_{sub}$$^{c}$ &
Mean R & Mean R$_{sub}$$^{c}$\\
 & (100K)   &   (K)  &(100K) & Kkm/spc$^{2}$ & Kkm/spc$^{2}$ & (pc) &
(pc)\\\hline
CS 7-6 & 2.0$\times 10^{7}$ & 66  & 2.6$\times 10^{5}$ & 64 & 42 & 0.28
& 0.24\\
CS 5-4 $^{b}$ & 6.9$\times 10^{6}$ & 35  & 6.0$\times 10^{4}$ & 77 & 68 & 0.37
& 0.31 \\
HCN 3-2& 6.8$\times 10^{7}$ & 26  & 3.6$\times 10^{4}$ & 79 & 101 & 0.32
& 0.35 \\
CS 2-1 & 3.9$\times 10^{5}$ & 7.1  & 4.1$\times 10^{3}$ & 272 & & 1.03 &\\
HCN 1-0& 4.5$\times 10^{6}$ & 4.3  & 5.1$\times 10^{3}$ & 359 & & 1.13
&\\\hline
\end{tabular}\\
\tablenotetext{a}{From Evans (1999). }
\tablenotetext{b}{From Shirley et al.( 2003).}
\tablenotetext{c}{Statistics on a subsample of the sources only, which have
been mapped by
all the 3 high-J transitions CS 7-6, CS 5-4, and HCN 3-2}
\end{table}

\begin{table}
\caption{\label{fwhm}Statistics of linewidth for different tracers}
\begin{tabular}{lccccc}
\hline
Tansitions & $\Delta V$(CS7-6)& $\Delta V$(CS5-4) $^{a}$&$\Delta V$(CS2-1)&
$\Delta V$(HCN 1-0)&$\Delta V$(C$^{34}$S5-4)\\
 & (km s$^{-1}$) & (km s$^{-1}$)  & (km s$^{-1}$)   & (F=2-1)(km s$^{-1}$)
& (km s$^{-1}$) \\\hline
Mean & 5.9$\pm$2.3 & 5.6$\pm$2.2 & 4.8$\pm$1.8 & 3.9$\pm$1.0 & 4.7$\pm$1.9 \\
Median & 5.7   &  5.1  & 4.8 & 3.7 & 4.2 \\\hline
\end{tabular}\\
\tablenotetext{a}{From Shirley et al.( 2003).}
\end{table}

\begin{deluxetable}{lrccrccrccrcc}
\tabletypesize{\scriptsize}
\rotate
\tablecaption{\label{virialtable}The Virial Mass, Surface Density and Average
Volume Density of Dense Clumps \tablenotemark{a}}
\tablewidth{0pt}
\tablehead{
\multicolumn{1}{c}{SOURCE} &\multicolumn{3}{c}{HCN 3 $\rightarrow$
2}&\multicolumn
{3}{c}{HCN 1 $\rightarrow$ 0}&\multicolumn{3}{c}{CS 7 $\rightarrow$
6}&\multicolumn
{3}{c}{CS 2 $\rightarrow$ 1}\\
\colhead{}             &
\colhead{log(M$_{vir}$)}     &
\colhead{$\Sigma$ }      &
\colhead{log(\nbar)}                     &
\colhead{log(M$_{vir}$)}    &
\colhead{$\Sigma$ }      &
\colhead{log(\nbar)}                     &
\colhead{log(M$_{vir}$)}  &
\colhead{$\Sigma$ }      &
\colhead{log(\nbar)}                     &
\colhead{log(M$_{vir}$)}    &
\colhead{$\Sigma$ }    &
\colhead{log(\nbar)}                      \\
\colhead{}            &
\colhead{(M$_{\odot}$)}                        &
\colhead{(g cm$^{-2}$)}            &
\colhead{}                                     &
\colhead{(M$_{\odot}$)}                        &
\colhead{(g cm$^{-2}$)}            &
\colhead{}                                     &
\colhead{(M$_{\odot}$)}                        &
\colhead{(g cm$^{-2}$)}            &
\colhead{}                                     &
\colhead{(M$_{\odot}$)}                        &
\colhead{(g cm$^{-2}$)}            &
\colhead{}
}
\startdata
G121.30+0.66   & 2.17(12) &  0.69 & 5.54 &  2.35(10) & 0.45 & 5.19 & 1.79(16)& 1.65 & 6.31 & 2.44(11) &  0.37 & 5.01 \\
G123.07-6.31   & 2.68(16) &  2.58 & 6.15 &  3.24(16) & 0.71 & 5.03 & 2.52(17)& 3.75 & 6.48 & 3.21(15) &  0.76 & 5.09 \\
W3(OH)         & 2.99(05) &  1.40 & 5.60 &  3.43(03) & 0.51 & 4.72 & 2.74(05)& 2.54 & 6.11 & 3.39(05) &  0.56 & 4.80 \\
S231           & 2.24(10) &  0.33 & 5.03 &  2.60(09) & 0.14 & 4.30 & 2.02(10)& 0.55 & 5.47 & 2.61(09) &  0.14 & 4.29 \\
S235           & 2.08(09) &  0.38 & 5.21 &  2.49(09) & 0.15 & 4.38 & 1.90(10)& 0.58 & 5.57 & 2.42(08) &  0.18 & 4.53 \\
S255           & 2.25(08) &  0.71 & 5.53 &           &      &      & 2.15(10)& 0.91 & 5.74 &          &       &      \\
RCW142         & 2.97(06) &  1.60 & 5.70 &  3.45(06) & 0.53 & 4.74 & 2.73(03)& 2.73 & 6.16 & 3.23(05) &  0.87 & 5.17 \\
W28A2(1)       &          &       &      &  3.28(04) & 0.95 & 5.20 & 2.70(01)& 3.66 & 6.37 & 3.20(04) &  1.15 & 5.36 \\
\enddata\\
\tablenotetext{a}{The virial mass, surface density and volume density are calculated for
mass within the FWHM contour of each transitions. Values for virial mass and volume density
are in log scale, except the mean and median at bottom. Numbers in the parentheses are
uncertainties in the log scale, in the unit of 0.01. }
\tablenotetext{b}{Statistics on massive dense clumps only, which does not
include the sources
below the horizontal line in the table. }
\tablenotetext{c}{The $\sigma$ of the virial mass is the $\sigma$ of the mean.}

\end{deluxetable}

\begin{table}
\caption{\label{ratio7654} Statistics of Line Ratio $log\frac{L'_{CS 7-6}}
{L'_{CS 5-4}}$ by Evolutionary Stages $^{a}$}
\begin{tabular}{lcccc}
\hline
 & No HII region & UC HII  & Comp. HII & HII   \\
\hline
mean$^{b}$ & -0.45 $\pm$0.08    & -0.21$\pm$0.08   & -0.40$\pm$0.15   &
-0.21$\pm$0.08  \\
median   & -0.47            &  -0.20          & -0.28            &-0.21  \\
Std. dev.$^{c}$& 0.34&  0.37 & 0.41 & 0.25 \\\hline
\end{tabular}\\
\tablenotetext{a}{ Luminosities of CS 5-4 come from Shirley et al. (2003). }
\tablenotetext{b}{The uncertainty is the sigma of the mean}
\tablenotetext{c}{The standard deviation of the ratios}

\end{table}

\clearpage{}
\begin{deluxetable}{lrccccc}
\tabletypesize{\scriptsize}
\tablecaption{\label{evoltable} Evolutionary Analysis of Massive Clumps}
\tablewidth{0pt}
\tablehead{
\colhead{Source}                                    &
\colhead{HII? \tablenotemark{a}}                                      &
\colhead{log$\frac{L_{IR}}{M_{vir}(R_{CS5-4})}$ \tablenotemark{b,c} } &
\colhead{log$\frac{L'(HCN3-2)}{L'(HCN1-0)}$ \tablenotemark{b} }  &
\colhead{log$\frac{L'(CS7-6)}{L'(CS5-4)}$ \tablenotemark{b,c}} &
\colhead{Outflow? \tablenotemark{d} }               &
\colhead{Infall? \tablenotemark{e} }   \\
\colhead{}                            &
\colhead{}                            &
\colhead{(M$_{\odot}$)}                 &
\colhead{}                            &
\colhead{}                            &
\colhead{}                         &
\colhead{}
}
\startdata
G121.30+0.66&       &    0.34 & -0.44 & -0.47 &   Y &    Y \\
S231        &       &    1.77 & -0.53 & -0.49 &   Y &      \\
S241        &       &    1.99 & -0.66 &       &   Y &      \\
12.89+0.49  &       &    2.13 &       & -0.96 &          &      \\
W33A        &       &    2.53 &       & -0.62 &          &      \\
24.49-0.04  &       &    2.26 &       & -0.34 &          &      \\
28.86+0.07  &       &    2.68 &       &       &          &      \\
W43 Main3   &       &         &       & -0.78 &          &      \\
\enddata\\
\tablenotetext{a}{The classification of evolutionary status of HII regions are
judged from the size
 of the HII region from 2cm survey. Refer to Shirley at al. 2003 and references
in there for details.}
\tablenotetext{b}{Values are in log scale.}
\tablenotetext{c}{CS 5-4 data from Shirley et al. 2003.}
\tablenotetext{d}{Refer to outflow catalogue, Wu et al. (2004).  }
\tablenotetext{e}{Judged from HCN 3-2 profile for $\delta\upsilon$ $<$ -0.25.}
\tablenotemark{f}{The mean in log scale. The uncertainty is the sigma of the mean.}
\end{deluxetable}

\begin{table}
\caption{\label{ratiomvdl} Statistics of Ratio $\frac{M_{Vir}(R_{molecule})}{L'_{molecule}}$}
\begin{tabular}{lccccc}
\hline
 & {\rm log} $\frac{M_{Vir}(R_{CS2-1})}{L'_{CS2-1}}$& {\rm log}
$\frac{M_{Vir}(R_{CS7-6})}{L'_{CS7-6}}$&  {\rm log}
$\frac{M_{Vir}(R_{HCN1-0})}{L'_{HCN1-0}}$
&{\rm log} $\frac{M_{Vir}(R_{HCN3-2})}{L'_{HCN3-2}}$ \\
\hline
mean$^{a}$ & 1.34 $\pm$0.07    & 1.57$\pm$0.07   & 1.30$\pm$0.10   &
1.35$\pm$0.07  \\
median   & 1.30            &  1.59         & 1.20            &1.29  \\
Std. dev.$^{b}$& 0.39&  0.43 & 0.54 & 0.37 \\\hline
\end{tabular}\\
\tablenotetext{a}{The uncertainty is the sigma of the mean}
\tablenotetext{b}{The standard deviation of the ratios}
\end{table}

\begin{table}
\caption{\label{statishm}Statistics of Properties of Massive Dense Clumps
$^{a}$ }
\begin{tabular}{lccccc}
\hline
 & M$_{vir(HCN1-0)}$ & M$_{vir(HCN3-2)}$ & M$_{vir(CS2-1)}$ &
M$_{vir(CS7-6)}$ & L$_{IR}$  \\
 & (\msun) & (\msun) &(\msun) & (\msun) &(\lsun) \\\hline
Mean$^{b}$ & 5300$\pm$1760 & 1555$\pm$445 & 5000$\pm$1870 & 1300$\pm$310 & 470000$\pm$120000\\
Median& 2700       &  890       & 1950          & 650 &106000         \\
Sdv$^{c}$ & 9600&  2310 & 10300 & 1930 & 920000\\\hline
\hline
 & $\frac{L_{IR}}{M_{vir(HCN1-0)}}$ & $\frac{L_{IR}}{M_{vir(HCN3-2)}}$ &
$\frac{L_{IR}}{M_{vir(CS2-1)}}$ & $\frac{L_{IR}}{M_{vir(CS7-6)}}$ &\\
 &(\lsun\ /\msun) &(\lsun\ /\msun) &(\lsun\ /\msun) &(\lsun\ /\msun)&\\\hline
mean$^{b}$& 117$\pm$20    & 371$\pm$102   & 139$\pm$26   & 592$\pm$195 & \\
median   & 98           & 182        & 129            &229 & \\
Sdv$^{c}$& 105&  487 & 134 & 1085 & \\\hline
\end{tabular}\\
\tablenotetext{a}{Statistics on massive dense clumps only, which does not
include the sources below the horizontal line in Table 1. }
\tablenotetext{b}{The uncertainty is the sigma of the mean.}
\tablenotetext{c}{The standard deviation.}

\end{table}

\begin{table}
\caption{\label{statisgt45}Statistics of Properties of Massive
Dense Clumps with $\lir > \eten{4.5}$ \lsun\ $^{a}$ }
\begin{tabular}{lccccc}
\hline
 & M$_{vir(HCN1-0)}$ & M$_{vir(HCN3-2)}$ & M$_{vir(CS2-1)}$ &
M$_{vir(CS7-6)}$ & L$_{IR}$  \\
 & (\msun) & (\msun) &(\msun) & (\msun) & (\lsun) \\\hline
Mean$^{b}$ & 6650$\pm$2350  & 2070$\pm$630 & 6300$\pm$2500 & 1480$\pm$420 &
630000$\pm$160000 \\
Median& 2950        &  1170        & 2500          & 710          &  200000
  \\
Sdv$^{c}$ & 10800&  2600 & 11500 & 2030 & 1020000 \\\hline
\hline
 & $\frac{L_{IR}}{M_{vir(HCN1-0)}}$ & $\frac{L_{IR}}{M_{vir(HCN3-2)}}$ &
$\frac{L_{IR}}{M_{vir(CS2-1)}}$ & $\frac{L_{IR}}{M_{vir(CS7-6)}}$ &  \\
 &(\lsun\ /\msun) &(\lsun\ /\msun) &(\lsun\ /\msun) &(\lsun\ /\msun)&\\\hline
mean$^{b}$& 137$\pm$23    & 473$\pm$130   & 169$\pm$30   & 726$\pm$246 & \\
median   & 120           & 324        & 155            &439  & \\
Sdv$^{c}$& 106&  532 & 137 & 1204 & \\\hline
\end{tabular}\\
\tablenotetext{a}{Statistics on massive dense clumps only, which does not
include the sources below the horizontal line in Table 1. }
\tablenotetext{b}{The uncertainty is the sigma of the mean.}
\tablenotetext{c}{The standard deviation.}

\end{table}

\clearpage{}

\clearpage{}

\clearpage{}

\clearpage{}

\clearpage{}

\clearpage{}

\clearpage{}

\clearpage{}

\clearpage{}

\clearpage{}

\end{document}